\documentclass[iop]{emulateapj}
\usepackage{graphicx}
\usepackage{color}

\shorttitle{KAPPA: Synthesis of spectra for $\kappa$-distributions}
\shortauthors{Dzif\v{c}\'akov\'a et al.}

\begin{document}

\title{KAPPA: A Package for Synthesis of optically thin spectra for the non-Maxwellian $\kappa$-distributions based on the CHIANTI database}

\author{Elena Dzif\v{c}\'akov\'a} \author{Jaroslav Dud\'ik\altaffilmark{1}} \author{Pavel Kotr\v{c}} \author{Franti\v{s}ek F\'arn\'ik} \author{Alena Zemanov\'{a}}
\affil{Astronomical Institute of the Academy of Sciences of the Czech Republic, Fri\v{c}ova 298, 251 65 Ond\v{r}ejov, Czech Republic}
\email{elena@asu.cas.cz}

\altaffiltext{1}{RS Newton International Fellow, DAMTP, CMS, University of Cambridge, Wilberforce Road, Cambridge CB3 0WA, United Kingdom}

\begin{abstract}
The non-Maxwellian $\kappa$-distributions have been detected in the solar transition region and flares. These distributions are characterized by a high-energy tail and a near-Maxwellian core and are known to have significant impact on the resulting optically thin spectra arising from collisionally dominated astrophysical plasmas. We developed the KAPPA package\footnote{http://kappa.asu.cas.cz} for synthesis of such line and continuum spectra. The package is based on the freely available CHIANTI database and software, and can be used in a similar manner. Ionization and recombination rates together with the ionization equilibria are provided for a range of $\kappa$ values. Distribution-averaged collision strengths for excitation are obtained by an approximate method for all transitions in all ions available within CHIANTI. The validity of this approximate method is tested by comparison with direct calculations. Typical precisions of better than 5\% are found, with all cases being within 10\%. Tools for calculation of synthetic line and continuum intensities are provided and described. Examples of the synthetic spectra and \textit{SDO}/AIA responses to emission for the $\kappa$-distributions are given.
\end{abstract}


%
\section{Introduction}
\label{Sect:1}

In astrophysics, the emitted radiation is usually the only source of information about the physical conditions in the emitting medium. Physical properties of the emitting plasma are then derived by analysis and modeling of the observed spectra. For a long time, this has been done under the assumption of a local, equilibrium Maxwellian distribution. This is done even if the emitting medium is optically thin and therefore perhaps not dense enough for the equilibrium to be always ensured locally. Such assumption is at best difficult in dynamic situations with particle acceleration, as e.g., a  high-energy tail is difficult to equilibrate collisionally, since the collision frequency scales inversely with $E^{3/2}$, where $E$ is the particle energy \citep[e.g.,][]{Meyer-Vernet07}. \citet{Scudder13} argue that, in the case of stellar coronae, the assumption of the Maxwellian distribution should always be violated at heights above 1.05 of the stellar radius. If long-range interactions are induced, e.g., by reconnection, wave-particle interaction, or shocks, the particles in the system can become correlated and do not have a Maxwellian distribution \citep[e.g.,][]{Collier04,Vocks03,Vocks08,Drake06,Livadiotis09,Livadiotis10,Livadiotis13,Pierrard10,Gontikakis13,Laming13}. Rather, the distribution exhibits a high-energy power-law tail. The $\kappa$-distributions are a class of particle distributions having a near-Maxwellian core and a high-energy power-law tail, both of which are described by an analytic expression \citep[][, see also Sect. 2]{Vasyliunas68,Owocki83}. The $\kappa$ index has been shown to be an independent thermodynamic index \citep{Livadiotis09,Livadiotis10,Livadiotis11a,Livadiotis13} in the generalized Tsallis statistical mechanics \citep[e.g.,][]{Tsallis88,Tsallis09,Leubner02,Leubner04a}.

The $\kappa$-distributions can be derived analytically in case of a turbulent velocity diffusion coefficient inversely proportional to velocity. This has been shown for the plasma in a suprathermal radiation field \citep{Hasegawa85}, for electrons heated by lower hybrid waves \citep{Laming07} and for solar flare plasmas where the distribution function arises as a consequence of balance between diffusive acceleration and collisions \citep{Bian14}.

Indeed, in solar flares, the $\kappa$-distributions provide a good fit to some of the X-ray spectra of coronal sources observed during partially occulted flares \citep{Kasparova09,Oka13}, although a second, near-Maxwellian distribution is also present \citep{Oka13}. \citet{Battaglia13} used the AIA and RHESSI observations of flares to derive the distribution function in the range of 0.1 to tens of keV. These authors shown that the distribution derived in the low-energy range from AIA does not match the high-energy tail observed by RHESSI. A possible cause of this mis-match is the assumption of Maxwellian distribution in the calculation of AIA differential emission measures (DEMs), which may compromise the analysis, especially if the high-energy tail is present and observed by RHESSI. 

\citet{Dzifcakova11} have shown that the $\kappa$-distributions can explain the \ion{Si}{3} transition-region line intensities observed by the \textit{SOHO}/SUMER instrument \citep{Wilhelm95}, especially in the active region spectra (see also \citet{DelZanna14}). \citet{Testa14} inferred presence of high-energy tails from the analysis of \ion{Si}{4} spectra observed by the IRIS spectrometer \citep{DePontieu14}. The $\kappa$-distributions are also routinely detected in the solar wind \citep[e.g.,][]{Collier96,Maksimovic97a,Maksimovic97b,Livadiotis10,LeChat11}. The high-energy tails at keV energies can arise as a consequence of coronal nanoflares \citep{Gontikakis13} that are also able to produce the ``halo'' in the solar wind electron distribution \citep{Che14}. Furthermore, a claim has been made that the $\kappa$-distributions were detected also in the spectra of planetary nebulae \citep{Binette12,Nicholls12,Nicholls13,Dopita13}, although this has been challenged as a possible effect of atomic data uncertainties \citep{Storey13,Storey14}. The kappa-distributions are also one of the possible explanations of the non-Maxwellian H$\alpha$ profiles detected in the Tycho's supernova remnant \citep{Raymond10}.

Although the $\kappa$-distributions were detected in solar flares, transition region and solar wind, their presence in the solar corona is currently unknown despite numerous attempts at their diagnostics. A diagnostics of the high-energy electrons have been attempted by \citet{Feldman07} and \citet{Hannah10}. \citet{Feldman07} investigated whether the He-like intensities observed by SUMER could correspond to a bi-Maxwellian distribution with the second Maxwellian having a temperature of 10\,MK. These authors argued that no such second Maxwellian is necessary. However, this analysis was limited to Maxwellian distribution and did not include the effect of a proper high-energy power-law tail. \citet{Hannah10} used the X-ray off-limb observations of the quiet-Sun performed by the RHESSI instrument \citep{Lin02} to obtain upper-limits on the emission measures as a function of $\kappa$. However, for temperatures of several MK corresponding to the solar corona, these upper limits are large and increase with increasing $\kappa$. Direct attempts at spectroscopic diagnostics using EUV line intensities observed by \textit{Hinode}/EIS \citep{Culhane07} were performed by \citet{Dzifcakova10} and \citet{Mackovjak13}. Indications of non-Maxwellian distributions were found using the \ion{O}{4}--\ion{O}{5} and \ion{S}{10}--\ion{S}{11} lines. However, such analysis was problematic due to large photon noise uncertainties affecting weak lines, atomic data uncertainties and the possible presence of multi-thermal effects that would complicate the analysis. Therefore, even diagnostics using only strong lines will have be supplemented by a DEM analysis under the assumption of a $\kappa$-distribution. Under the constraints of the current EUV instrumentation, such DEM analysis typically involves many different elements and ionization stages \citep[see, e.g.,][]{Warren12,Mackovjak14}.

All this leads to a requirement of reliable calculation of synthetic spectra involving many different elements and ionization stages. In this paper, we describe the KAPPA package for calculation of optically thin astrophysical spectra that arise due to collisional excitation by electrons with a $\kappa$-distribution. This package, allowing for fast calculation of line and continuum spectra for $\kappa$-distributions, is based on the freely available CHIANTI atomic database and software, currently in version 7.1 \citep{Dere97,Landi13}. The manuscript is organized as follows. The $\kappa$-distributions are described in Sect. \ref{Sect:2}. Synthesis of line spectra and continua are described in Sect. \ref{Sect:3} and Sect. \ref{Sect:4}, respectively. Section \ref{Sect:5} describes the database and the software implementation. Examples of the synthetic spectra and the AIA filter responses fo $\kappa$-distributions are provided in Sect. \ref{Sect:6}. Summary is given in Sect. \ref{Sect:7}.

\vspace{0.5cm}
%
   \begin{figure}[!t]
	\centering
	\includegraphics[width=8.8cm]{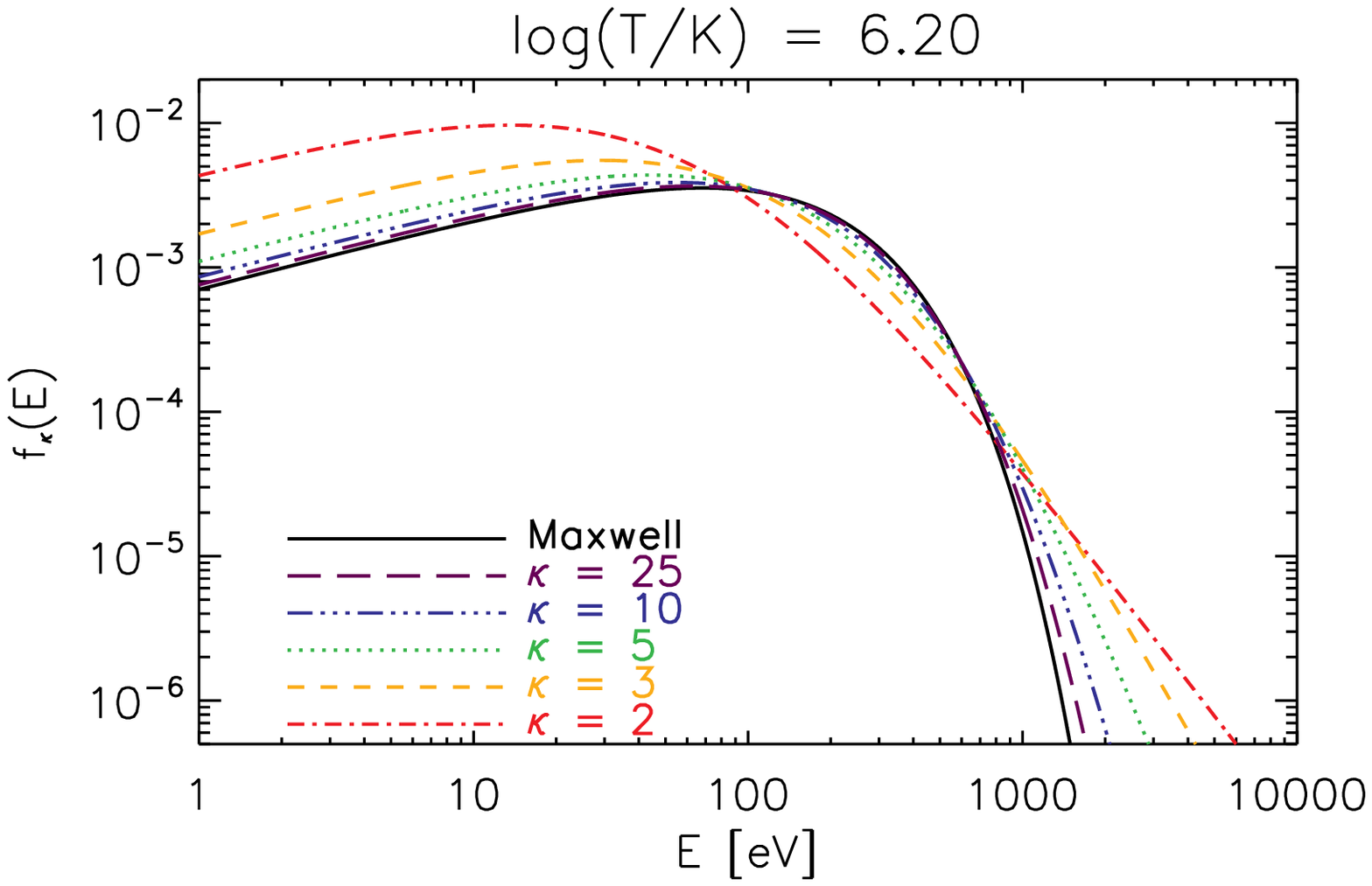}
	\includegraphics[width=8.8cm]{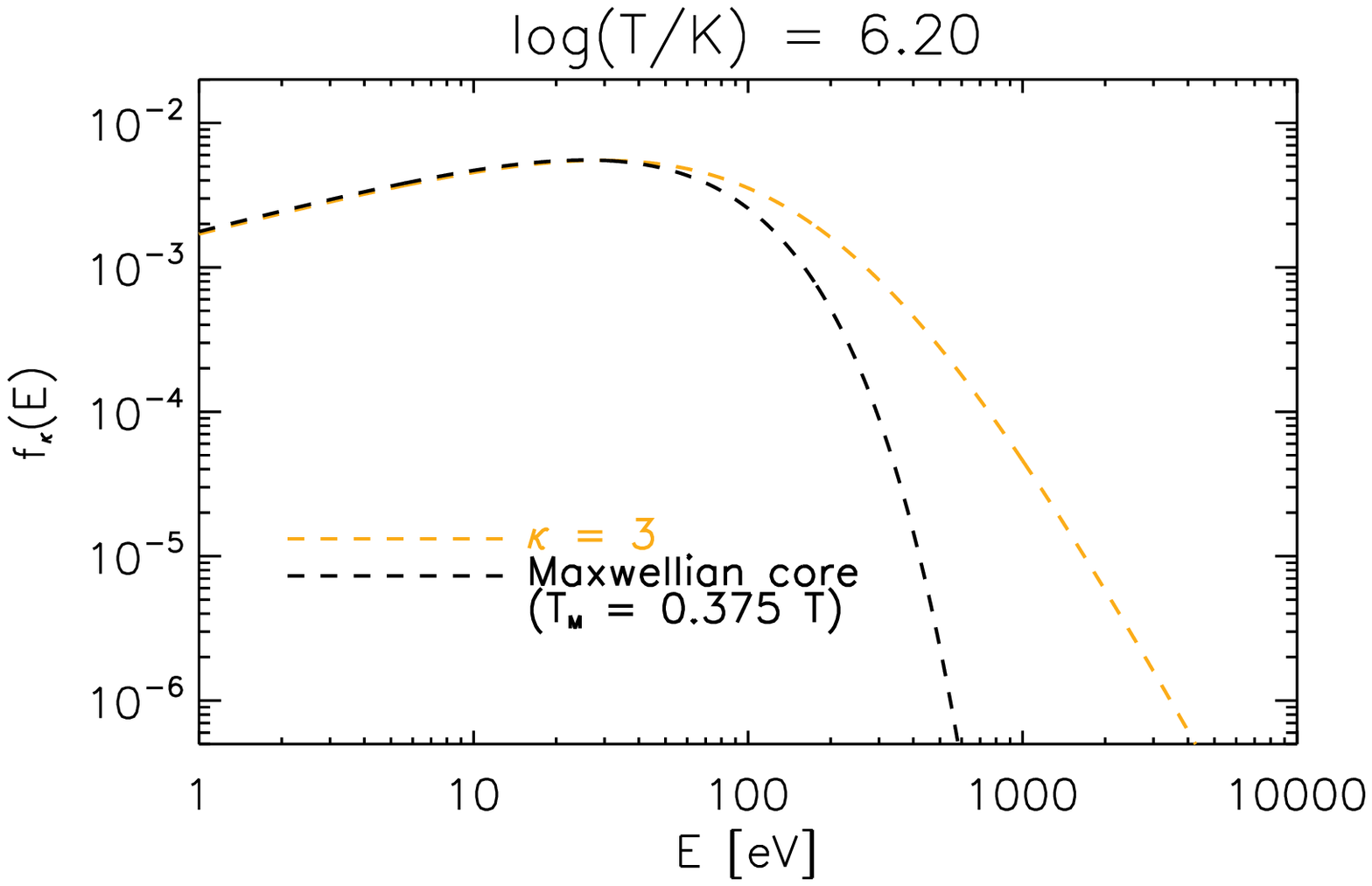}
	\includegraphics[width=8.8cm]{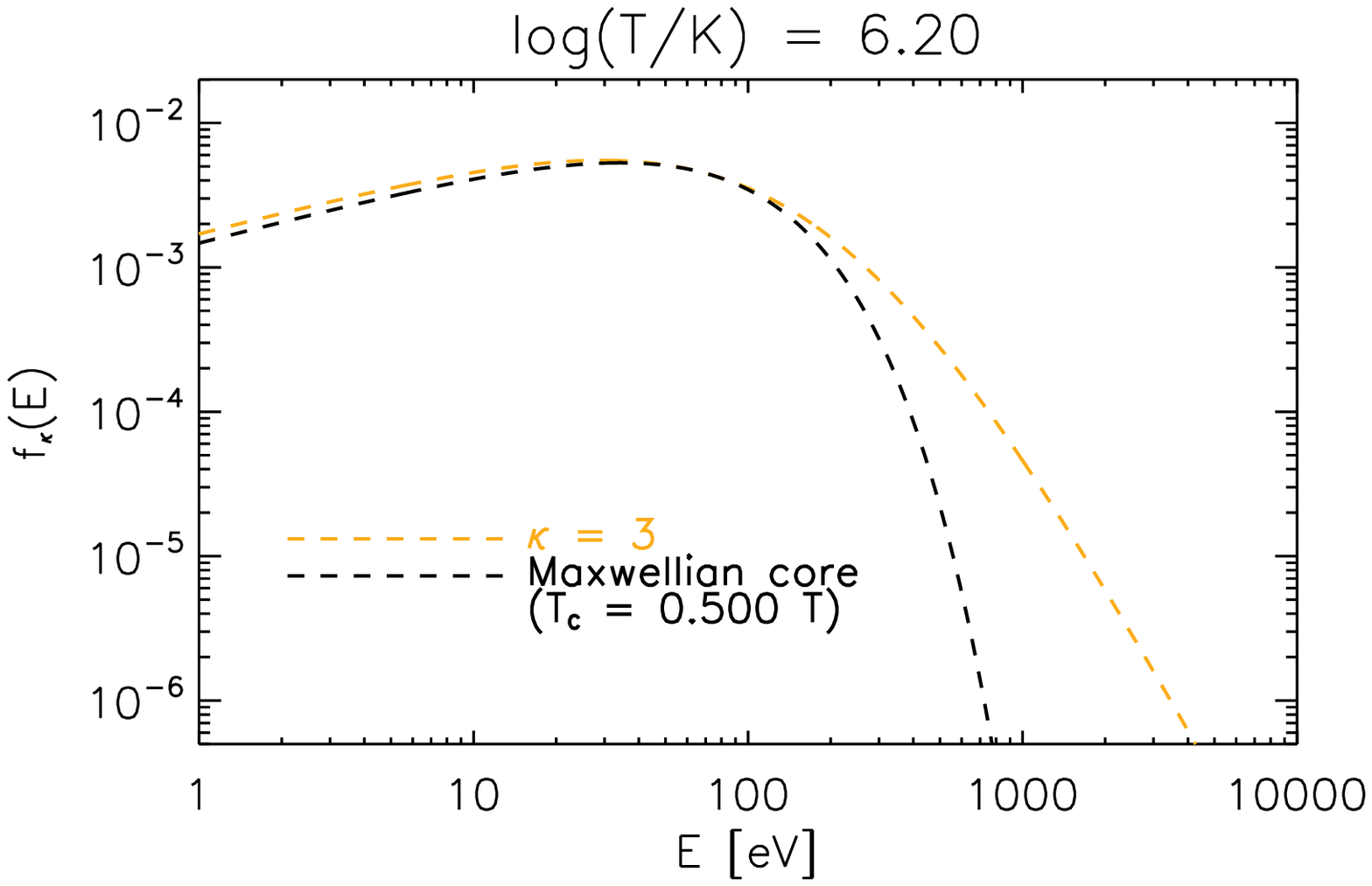}
	\caption{The $\kappa$-distributions with $\kappa$\,=\,2, 3, 5, 10, 25 and the Maxwellian distribution plotted for log($T$/K)\,=\,6.20 (\textit{top}). Colors and linestyles denote the different values of $\kappa$. Approximations of the $\kappa$\,=\,3 distribution in the low-energy range with a Maxwellian distribution according to \citet{Livadiotis09} and \citet{Oka13} are shown in the \textit{middle} and \textit{bottom} panels, respectively. \\ A color version of this image is available in the online journal.}
       \label{Fig:Kappa}
   \end{figure}
%

%
\section{The Non-Maxwellian $\kappa$-distributions}
\label{Sect:2}

\subsection{Definition and Basic Properties}
\label{Sect:2.1}

The $\kappa$-distribution of electron energies (Fig. \ref{Fig:Kappa}) is defined as \citep[e.g.,][]{Owocki83,Livadiotis09}
        \begin{equation}
                f_\kappa(E) \mathrm{d}E = A_{\kappa} \frac{2}{\sqrt{\pi} (k_\mathrm{B}T)^{3/2}} \frac{E^{1/2}\mathrm{d}E}{\left 			  (1+ \frac{E}{(\kappa - 3/2) k_\mathrm{B}T} \right)^{\kappa+1}}\,,
                \label{Eq:Kappa}
        \end{equation}
where the $A_{\kappa}$\,=\,$\Gamma(\kappa+1)$/$\left(\Gamma(\kappa-1/2) (\kappa-3/2)^{3/2}\right)$ is the normalization constant and $k_\mathrm{B}$\,=\,1.38 $\times 10^{-16}$ erg\,s$^{-1}$ is the Boltzmann constant. The $\kappa$-distribution has two parameters, $T$\,$\in$\,$\left(0,+\infty\right)$ and $\kappa$\,$\in$\,$\left(3/2,+\infty\right)$. The Maxwellian distribution at a given $T$ corresponds to $\kappa$\,$\to$\,$\infty$. The departure from the Maxwellian distribution increases with decreasing $\kappa$, with the maximum departure occurring for $\kappa$\,$\rightarrow$\,3/2.

While the most probable energy $E_\mathrm{max}$\,=\,$(\kappa-3/2)k_\mathrm{B}T/\kappa$ is a decreasing function of $\kappa$, the mean energy $\left< E \right> = {3k_\mathrm{B}T}/{2}$ of a $\kappa$-distribution does not depend on $\kappa$ and is only a function of $T$. Because of this, the parameter $T$ has the same physical meaning for the $\kappa$-distributions as the (kinetic) temperature for the Maxwellian distribution. Additionally, \citet{Livadiotis09} and \citet{Livadiotis10} show that the $T$ also corresponds to the definition of physical temperature in the framework of the generalized Tsallis statistical mechanics \citep{Tsallis88,Tsallis09}, and permits the generalization of the zero-th law of thermodynamics. Note that this fact permits e.g. the definition of electron kinetic pressure $p$\,=\,$n_\mathrm{e}k_\mathrm{B}T$ in the usual manner.

Note also that the $\kappa$-distribution is not the only possible representation of a non-Maxwellian distribution with a high-energy tail \citep[e.g.,][]{Dzifcakova11,Che14}. Nevertheless, its analytical expression and a single additional parameter $\kappa$ make it a useful special case of an equilibrium particle distribution associated with turbulence \citep{Hasegawa85,Laming07,Bian14}, offering a rather straightforward evaluation of various rate coefficients associated with radiative processes (Sects. \ref{Sect:3} and \ref{Sect:4}).

%
\subsection{Approximation by Maxwellian Core and a Power-Law Tail}
\label{Sect:2.2}

It is straightforward to see from Eq. (\ref{Eq:Kappa}) that in the high-energy limit, the $\kappa$-distribution approaches a power-law with the index of $-(\kappa+1/2)$. On the other hand, \citet{Meyer-Vernet95} and \citet{Livadiotis09} showed that, in the low-energy limit, the $\kappa$-distribution behaves as a Maxwellian with 
	\begin{equation}
	  T_\mathrm{M} = \frac{\kappa-3/2}{\kappa+1}T\,.
	  \label{Eq:T_M}
	\end{equation}
The low-energy end of a $\kappa$-distribution can indeed be well approximated by a Maxwellian, if this Maxwellian is scaled by a constant
	\begin{equation}
	  c_\mathrm{M}(\kappa) =  A_\kappa \frac{\left(\frac{\kappa-3/2}{\kappa+1}\right)^{3/2}  \mathrm{exp}\left(\frac{\kappa+1}{2\kappa+1}\right) } {\left(1+\frac{1}{2\kappa+1}\right)^{(\kappa+1)}}\,,
	  \label{Eq:C_M}
	\end{equation}
so that the two distributions match at the most probable energy $E_\mathrm{max}$\,=\,$(\kappa-3/2)k_\mathrm{B}T/\kappa $ (Fig. \ref{Fig:Kappa}, \textit{middle}; see also e.g., \citet{Dzifcakova02}, Fig. 1 therein).

\citet{Oka13} attempted to approximate the core of a $\kappa$-distribution with a Maxwellian at temperature $T_\mathrm{C}$ 
	\begin{equation}
		T_\mathrm{C} = \frac{\kappa-3/2}{\kappa} T\,.
		\label{Eq:T_C}
	\end{equation}
Such Maxwellian core has to be adjusted (Fig. \ref{Fig:Kappa}, \textit{bottom}) by a scaling constant \citep[][Eq. (3) therein]{Oka13}
	\begin{equation}
		c(\kappa) = \mathrm{exp(1)} \frac{\Gamma(\kappa+1)}{\Gamma(\kappa -1/2)} \kappa^{-3/2} \left(1 +\frac{1}{\kappa}\right)^{-(\kappa+1)}\,,
		\label{Eq:C_kappa}
	\end{equation}
so that the two distributions match at $E$\,=\,$k_\mathrm{B}T_\mathrm{C}$. This approximation by a Maxwellian core leads to a worse match at very low energies $E$\,$\to$\,0 (Fig. \ref{Fig:Kappa}, \textit{bottom}).

These approximations suggest that the $\kappa$-distribution can be thought of as a Maxwellian core at a lower temperature with a power-law tail.

%
\begin{figure*}[!ht]
	\centering
	\includegraphics[width=17.4cm]{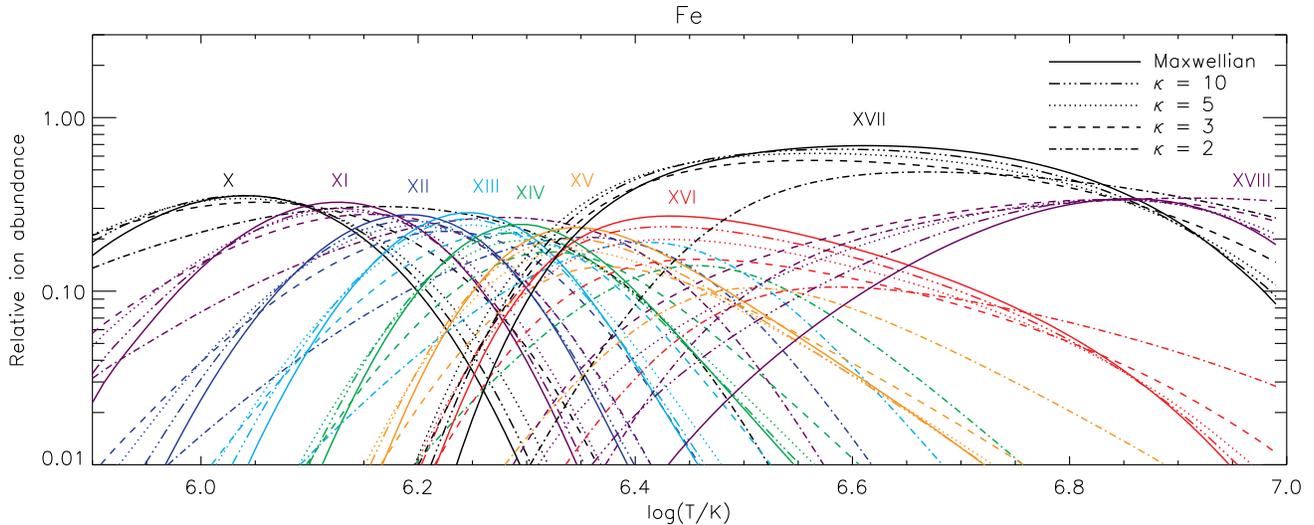}
	\caption{Example of the ionization equilibrium for iron and the $\kappa$-distributions. Relative ion abundances for \ion{Fe}{10}--\ion{Fe}{18} are plotted as a function of $T$ and $\kappa$ and compared to the relative ion abundances for the Maxwellian distribution from CHIANTI v7.1 (full lines). \\ A color version of this image is available in the online journal.}
\label{Fig:Ioneq}
\end{figure*}
%

%
\section{Line intensities for the $\kappa$-distributions}
\label{Sect:3}

In the optically thin solar and stellar coronae, as well as the associated transition regions and flares, spectral lines arise as a consequence of particle collisions exciting the ions in the highly ionized plasma. The total emissivity $\varepsilon_{ji}$ of a spectral line $\lambda_{ji}$ corresponding to a transition $j \to i$, $j > i$, in a $k$-times ionized ion of the element $X$ is usually expressed as \citep[e.g.,][]{Mason94,Phillips08}
	\begin{eqnarray}
		\nonumber \varepsilon_{ji} &=& \frac{hc}{\lambda_{ji}} A_{ji} n(X_j^{+k}) = \frac{hc}{\lambda_{ji}} \frac{A_{ji}}{n_\mathrm{e}} \frac{n(X_j^{+k})}{n(X^{+k})} \frac{n(X^{+k})}{n(X)} A_X n_\mathrm{e} n_\mathrm{H}\\
		 &=& A_X G_{X,ji}(T,n_\mathrm{e},\kappa) n_\mathrm{e} n_\mathrm{H}\,,
		\label{Eq:line_emissivity}
	\end{eqnarray}
where $h$\,$\approx$\,6.62\,$\times$\,10$^{-27}$ erg\,s is the Planck constant, $c$\,=\,3\,$\times$10$^{10}$\,cm\,s$^{-1}$ is the speed of light, $A_{ji}$ the Einstein coefficient for spontaneous emission, and $n(X_j^{+k})$ the density of the ion $+k$ with electron on the excited upper level $j$. In Eq. \ref{Eq:line_emissivity}, the latest quantity is usually expanded in terms of the ionization fraction $n(X^{+k})/n_X$ (Sect. \ref{Sect:3.1}) and the excitation fraction $n(X_j^{+k})/n(X^{+k})$ (Sect. \ref{Sect:3.2}). There, $n(X^{+k})$ denotes the total density of the ion $+k$, and $n(X)$\,$\equiv$\,$n_X$ corresponds to the total density of element $X$ whose abundance is $A_X$, with $n_\mathrm{H}$ being the hydrogen density. The function $G_{X,ji}(T,n_\mathrm{e},\kappa)$ is the contribution function for the line $\lambda_{ji}$. The intensity $I_{ji}$ of the spectral line is then given by the emissivity integral of emissivity along a path $l$ corresponding to the line of sight
	\begin{equation}
		I_{ji} = \int A_X G_{X,ji}(T,n_\mathrm{e},\kappa) n_\mathrm{e} n_\mathrm{H} \mathrm{d}l\,,
		\label{Eq:line_intensity}
	\end{equation}
where $EM$\,=\,$\int n_\mathrm{e} n_\mathrm{H} \mathrm{d}l$ is the emission measure of the emitting plasma.

The CHIANTI atomic database provides the observed wavelengths $\lambda_{ji}$ and the corresponding Einstein coefficients $A_{ji}$, while the electron density $n_\mathrm{e}$ is a free parameter. To complete the synthesis of line intensities for the $\kappa$-distributions, the relative ion abundance $n(X^{+k})/n_X$ and the relative level population $n(X_j^{+k})/n(X^{+k})$ must be calculated. This is detailed in the remainder of this section.

\subsection{Ionization Equilibrium}
\label{Sect:3.1}

A common assumption in calculation of the relative ion abundance $n(X^{+k})/n_X$ is that of ionization equilibrium, i.e., that the relative ion abundance is not a function of time. Then, the relative ion abundance is given by the equilibrium between the ionization and recombination rates. In coronal conditions, the dominating ionization processes are the direct ionization and the autoionization \citep[e.g.,][]{Phillips08}, while the dominant recombination processes are radiative and dielectronic recombination. Since these processes involve free electrons, all of these rates depend on $T$ and $\kappa$ \citep[e.g.,][]{Dzifcakova92,Anderson96,Dzifcakova02,Wannawichian03,Dzifcakova13}. We note that in the non-equilibrium ionization conditions, the $n(X^{+k})/n_X$ depends on the specific evolution of the system, in particular on the energy sources, sinks, and the resulting flows \citep[e.g.,][]{Bradshaw03,Bradshaw04,Bradshaw09}. Since radiation is an energy sink, the system is then coupled.

\citet{Dzifcakova13} provide the latest available ionization equilibria for $\kappa$-distributions for all ions of the elements with $Z$\,$\leqq$\,30, i.e., H to Zn. These calculations use the same atomic data for ionization and recombination as the ionization equilibrium for the Maxwellian distribution available in the CHIANTI database, v7.1 \citep{Landi13,Dere07,Dere97}.

Figure \ref{Fig:Ioneq} shows examples of the behaviour of the relative ion abundances of \ion{Fe}{10}--\ion{Fe}{18} with $\kappa$. The ionization peaks are in general wider for lower $\kappa$. Compared to the Maxwellian distribution, ionization peaks of the transition-region ions are in general shifted to lower log$(T/$K), while the coronal ions are generally shifted to higher $T$, especially for low $\kappa$\,=\,2--3 \citep{Dzifcakova13}. Exceptions from these rules of thumb occur. E.g., the ionization peak of \ion{Fe}{17} is shifted to lower $T$ for $\kappa$\,=\,5, while for $\kappa$\,=\,2, it is shifted to higher $T$ compared to the Maxwellian distribution. The shifts of the ionization peaks are typically $\Delta$log($T$/K)\,$\approx$\,0.10--0.15 for $\kappa$\,=\,2, although much larger shifts can also occur, e.g., for \ion{Fe}{7} \citep{Dzifcakova13}.

This behaviour of the individual ionization peaks with $\kappa$ strongly influence the resulting line intensities (Eqs. \ref{Eq:line_emissivity} and \ref{Eq:line_intensity}). Therefore, the approximate temperatures determined from the observed lines in the spectrum can be different for a $\kappa$-distribution and the Maxwellian distribution. For a plasma where the high-energy tail or electron beams can be expected, the $T$ is related to the mean energy of the distribution (Sect. \ref{Sect:2.1}) including the high-energy tail. Notably, $T$ can be very different from the Maxwellian ``bulk'' temperature $T_\mathrm{M}$ or $T_\mathrm{C}$ (Sect. \ref{Sect:2.2}). Strong changes in the ionization equilibrium for the $\kappa$-distributions mainly in the transition region result e.g. in the \ion{O}{4} being formed at log$(T/K$)\,$\approx$\,5.15 for the Maxwellian distribution, but at $\approx$\,5.0 for $\kappa$\,=\,5 and $\approx$\,4.8 for $\kappa$\,=\,2 \citep{Dudik14a}. The core of the distribution can have even lower temperatures -- for log$(T/K$)\,$\approx$\,5.15, the log($T_\mathrm{C}$/K)\,=\,5.0 for $\kappa$\,=\,5, but only 4.55 for $\kappa$\,=\,2. The $T_\mathrm{M}$ are even lower (Sect. \ref{Sect:2.2}). Therefore, without a diagnostics of $\kappa$ in situations where the high-energy tail can exist, one has to be very careful in the estimation of the plasma temperature from the fact that a particular line is observed. The situation is furthermore complicated by the dependence of line emission on the differential emission measure of the emitting plasma \citep[e.g.,][]{Warren12,Teriaca12b}, which is itself a function of $\kappa$ \citep{Mackovjak14}.

\subsection{Excitation Equilibrium and Rates}
\label{Sect:3.2}

The relative level populations $n(X_j^{+k})/n(X^{+k})$ can be obtained under an assumption of excitation equilibrium \citep[][Eqs. (4.24) and (4.25) therein]{Phillips08}. In equilibrium, the total number of transitions to and from any given level $j$ is balanced by transitions both from all other levels $m$ to the level $j$, as well as from the level $j$ to any other level $m$. In the conditions of the solar and stellar coronae, ion-electron collisions are the dominant excitation mechanism, while deexcitations are facilitated either by spontaneous radiative decay (with the rates $A_{jm}$) and/or collisional deexcitation during ion-electron collisions. The rates of electron excitation and deexcitation, $C_{jm}^\mathrm{e}$ and $C_{jm}^\mathrm{d}$, can be expressed as \citep{Bryans06,Dudik14b}
	\begin{eqnarray}
		C_{ij}^\mathrm{e} &=& \frac{2 \sqrt{2} a_0^2 I_H   }{\sqrt{m_\mathrm{e}}\omega_i} \left(\frac{\pi}{k_\mathrm{B}T}\right)^{1/2}  \mathrm{e}^{-\frac{\Delta E_{ij}}{k_\mathrm{B}T}}
\Upsilon_{ij}(T,\kappa)\,, \label{Eq:Excit_rate_Upsilon} \\
		C_{ji}^\mathrm{d} &=& \frac{2 \sqrt{2} a_0^2 I_H   }{\sqrt{m_\mathrm{e}}\omega_j}\left(\frac{\pi}{k_\mathrm{B}T}\right)^{1/2} \, \rotatebox[origin=c]{180}{$\Upsilon$}_{ji}(T,\kappa)\,, 
		\label{Eq:Deexcit_rate_Downsilon}
	\end{eqnarray}
where $a_0$\,=\,5.29\,$\times$10$^{-9}$ cm is the Bohr radius, $m_\mathrm{e}$\,=\,9.1\,$\times$10$^{-28}$ is the electron rest mass, $I_H$\,$\approx$\,13.6\,eV\,$\equiv$\,1\,Ryd is the hydrogen ionization energy, $\omega_i$ and $\omega_j$ are the statistical weights of the levels $i$ and $j$, respectively, $\Delta E_{ij}$\,=\,$E_i - E_j$ is the energy of the transition, and $E_i$ and $E_j$ are the incident and final electron energies. The $\Upsilon_{ij}(T,\kappa)$ and \rotatebox[origin=c]{180}{$\Upsilon$}$_{ji}(T,\kappa)$ denote the distribution-averaged collision strengths, given by
	\begin{eqnarray}
		\Upsilon_{ij} &=& A_\kappa \frac{\Delta E_{ij}}{k_\mathrm{B}T} \mathrm{e}^{\frac{\Delta E_{ij}}{k_\mathrm{B}T}}  \int\limits_{\Delta E_{ij}}^{+\infty} \frac{\Omega_{ji}(E_i)}{ \left(1+ \frac{E_i}{(\kappa-3/2)k_\mathrm{B}T}\right)^{\kappa+1}} \,\frac{\mathrm{d}E_i}{\Delta E_{ij}}\,,
		\label{Eq:Upsilon_kappa} \\
		\rotatebox[origin=c]{180}{$\Upsilon$}_{ji} &=& A_\kappa \frac{\Delta E_{ij}}{k_\mathrm{B}T} \int\limits_0^{+\infty} \frac{\Omega_{ji}(E_j)}{\left(1+ \frac{E_j}{(\kappa-3/2)k_\mathrm{B}T}\right)^{\kappa+1}} \,\frac{\mathrm{d}{E_j}}{\Delta E_{ij}}\,.
		\label{Eq:Downsilon_kappa}
	\end{eqnarray}
In these expressions, $\Omega_{ji}(E_j)$\,=\,$\Omega_{ij}(E_i)$ is the collision strength, i.e., the non-dimensionalised cross-section
	\begin{equation}
		\Omega_{ji}(E_j) = \omega_j \frac{E_j}{I_H} \frac{\sigma_{ji}^\mathrm{d}(E_j)}{\pi a_0^2} = \omega_i \frac{E_i}{I_H} \frac{\sigma_{ij}^\mathrm{e}(E_i)}{\pi a_0^2}\,, \\
		\label{Eq:Omega}
	\end{equation}
where the $\sigma_{ji}^\mathrm{e}$ and $\sigma_{ij}^\mathrm{d}$ are the electron impact excitation and deexcitation cross-sections, respectively. Note that with $\kappa$\,$\to$\,$\infty$, the $\Upsilon_{ij}(T,\kappa)$ and \rotatebox[origin=c]{180}{$\Upsilon$}$_{ji}(T,\kappa)$ revert to the $\Upsilon_{ij}(T)$ commonly used for the Maxwellian distribution \citep[][]{Seaton53,Burgess92,Mason94,Bradshaw13}, with the property of $\Upsilon_{ij}(T)$\,$\equiv$\,\rotatebox[origin=c]{180}{$\Upsilon$}$_{ji}(T)$ being recovered. The $\Upsilon_{ij}(T,\kappa)$ and \rotatebox[origin=c]{180}{$\Upsilon$}$_{ji}(T,\kappa)$, together with the Eqs. (\ref{Eq:Excit_rate_Upsilon}) and (\ref{Eq:Deexcit_rate_Downsilon}) and the equations of statistical equilibrium \citep[Eqs. (4.24) and (4.25) in][]{Phillips08}, can then be used to synthesize the spectra for the $\kappa$-distributions in the same manner as for the Maxwellian distribution (Sect. \ref{Sect:5.2}).

%
%
%
\begin{figure*}
	\centering
	\includegraphics[width=5.9cm]{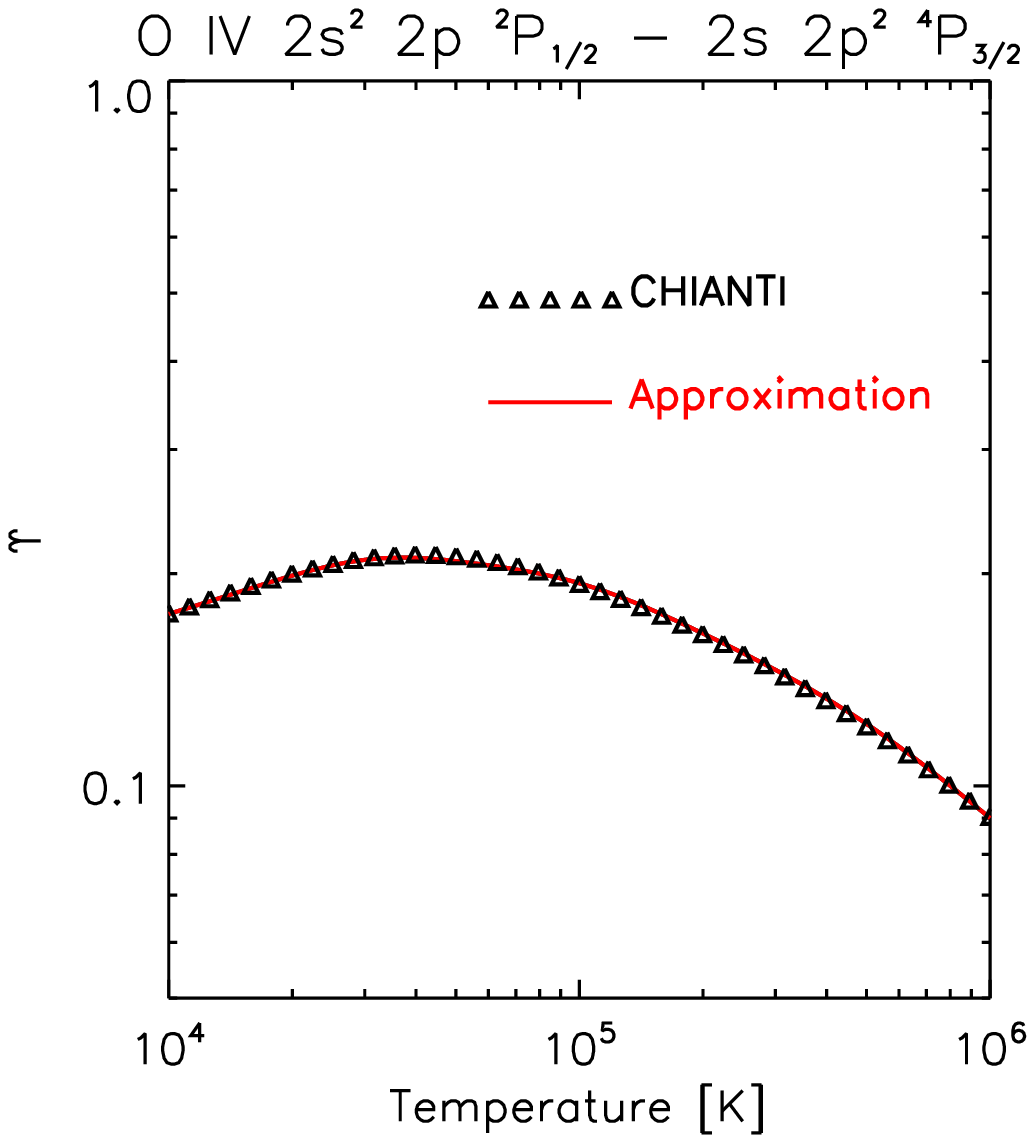}
	\includegraphics[width=5.9cm]{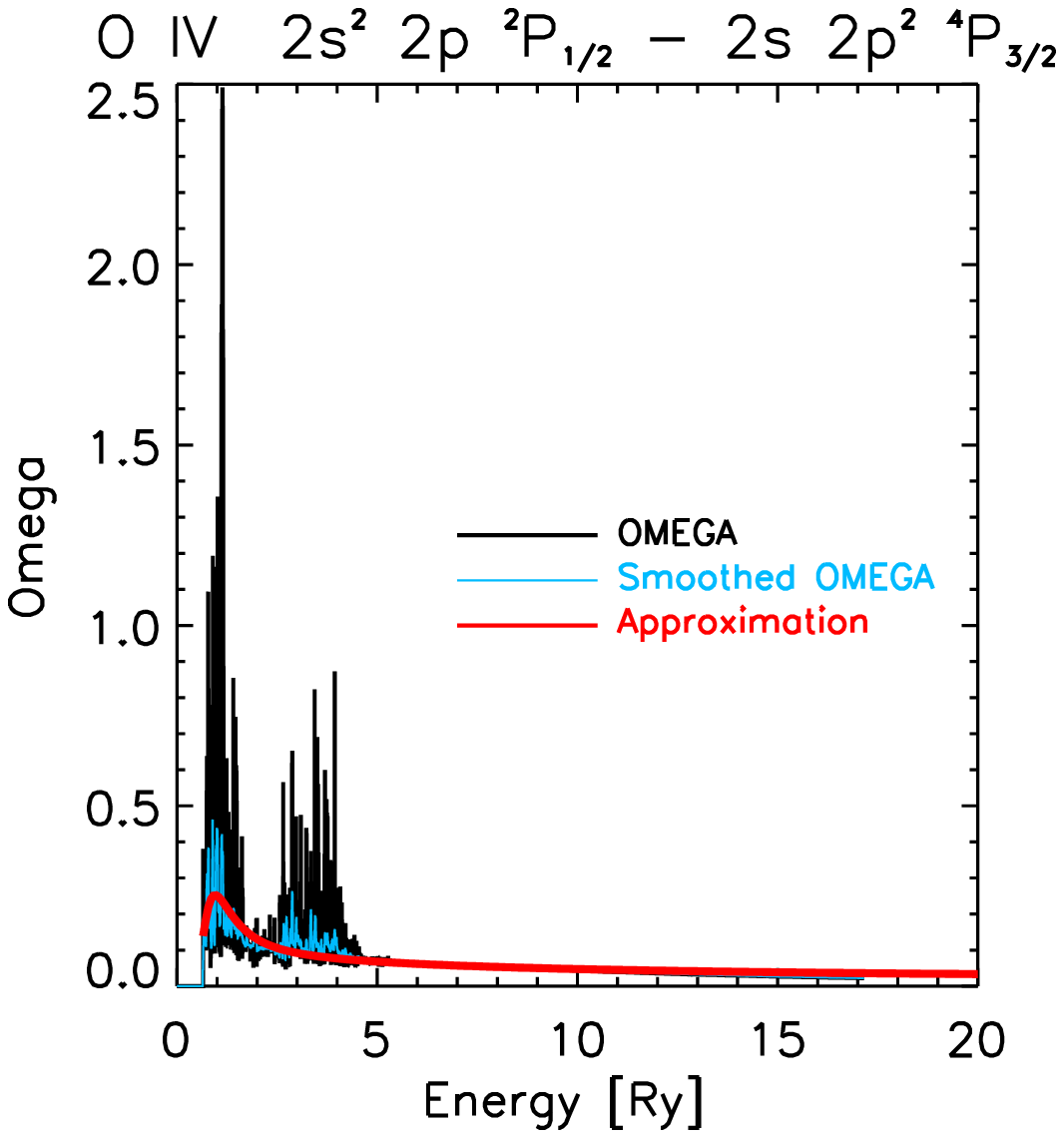}
	\includegraphics[width=5.9cm]{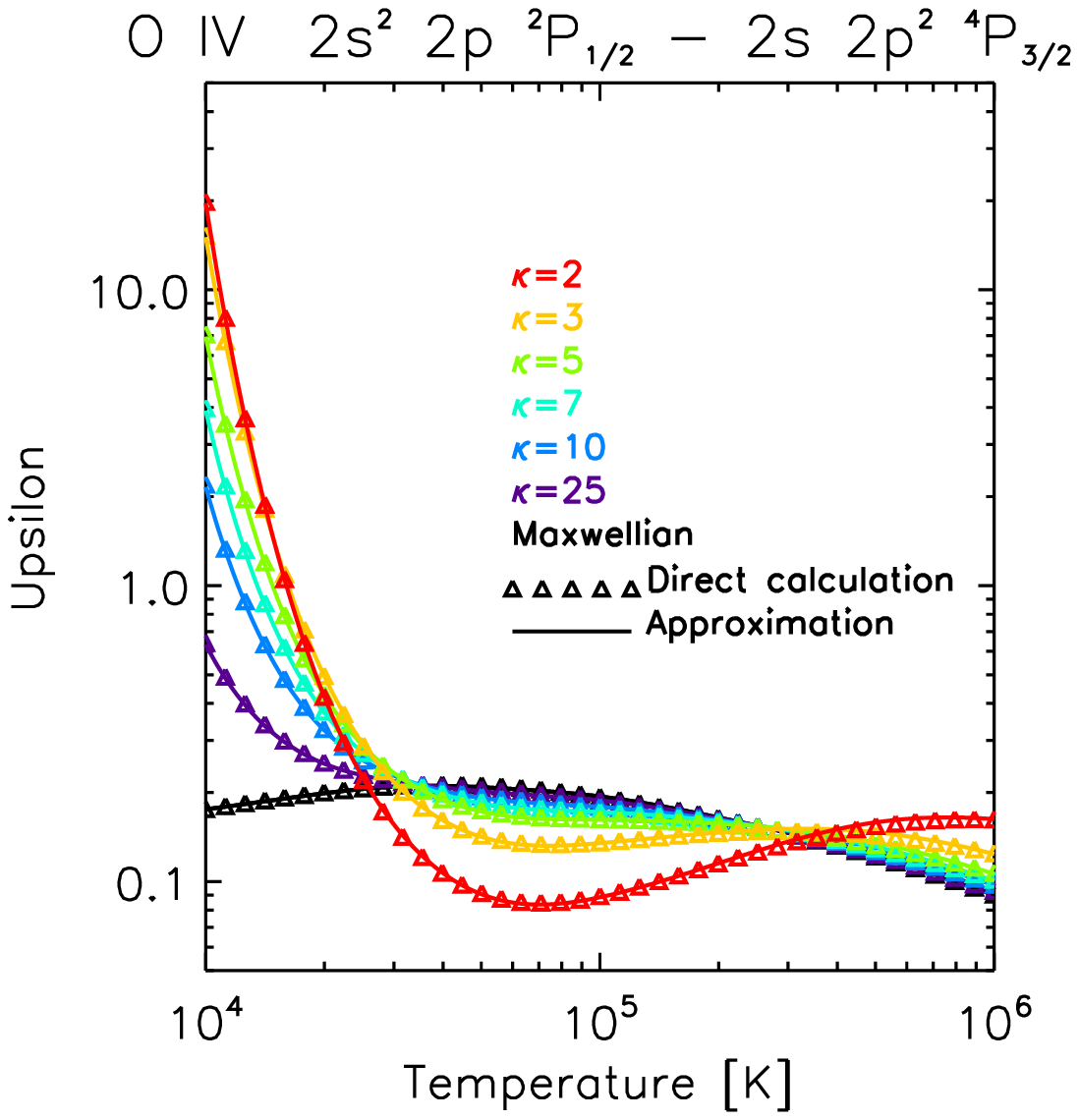}
	\caption{\textit{Left:} Approximation of $\Upsilon$ from CHIANTI. \textit{Middle:} Comparison of the approximation of $\Omega$ (red line) with atomic data of \citet{Liang12} without (black) and with smoothin (blue). \textit{Right:} Comparison of the $\Upsilon$s calculated using our approximation (full lines) with direct calculations for the Maxwellian distribution (triangles) for different $\kappa$ distributions with $\kappa$\,=\, 25 (violet), 10 (blue), 7 (turquoise), 5 (green), 3 (yellow), and 2 (red). }
\label{Fig:O4}
\end{figure*}
\begin{figure*}[!ht]
	\centering
	\includegraphics[width=5.9cm,bb=20 0 453 340,clip]{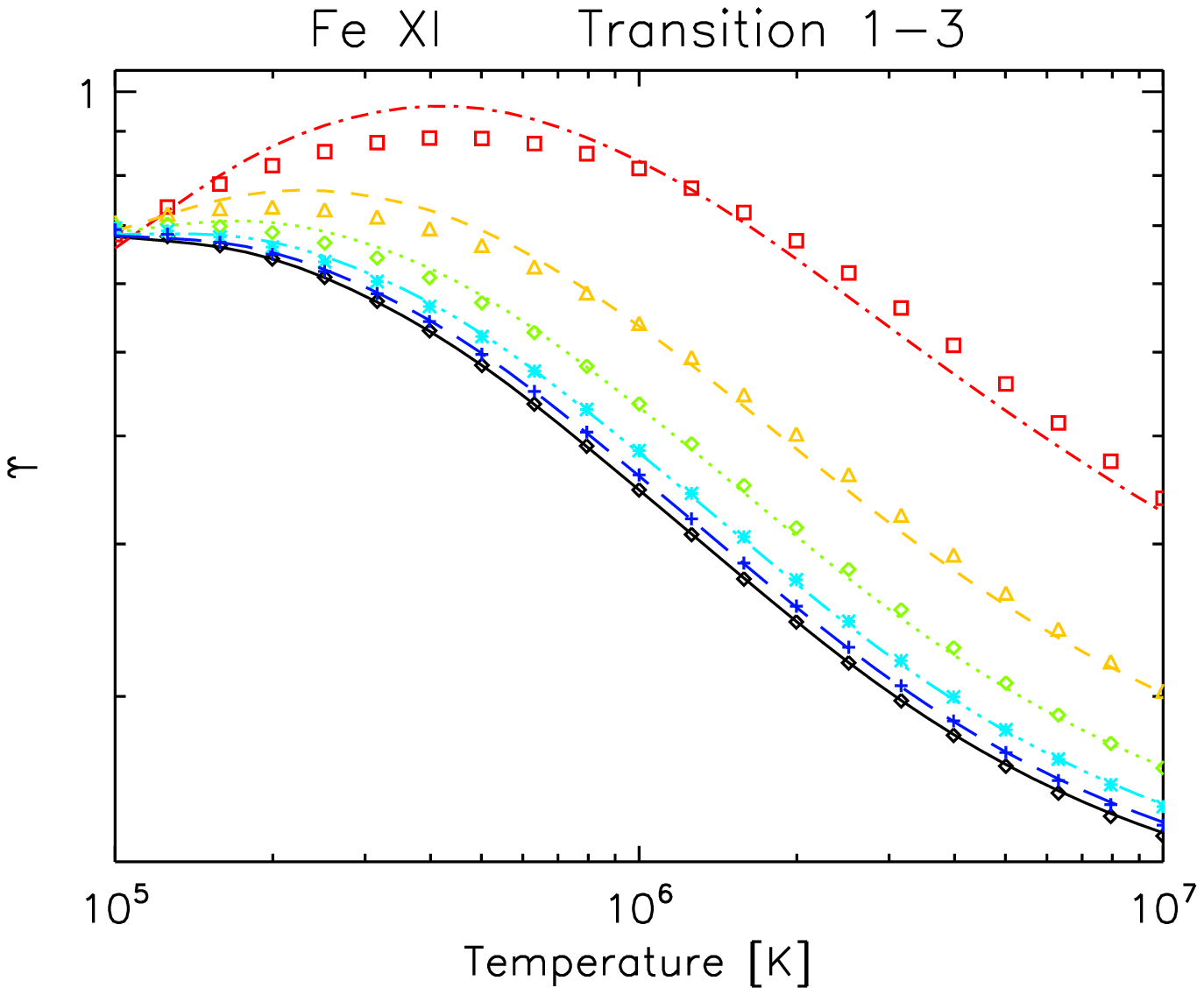}
	\includegraphics[width=5.9cm,bb=20 0 453 340,clip]{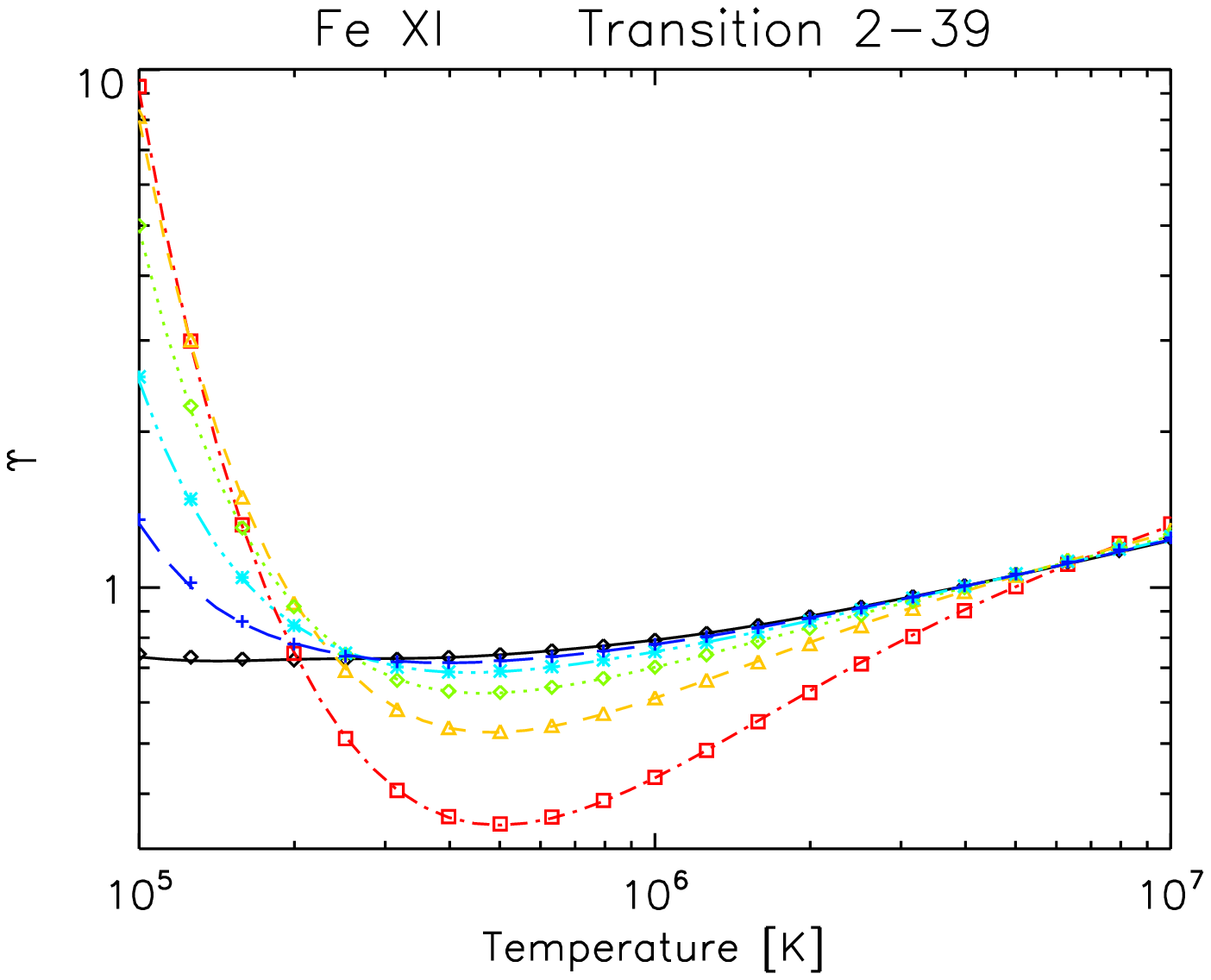}
	\includegraphics[width=5.9cm,bb=20 0 453 340,clip]{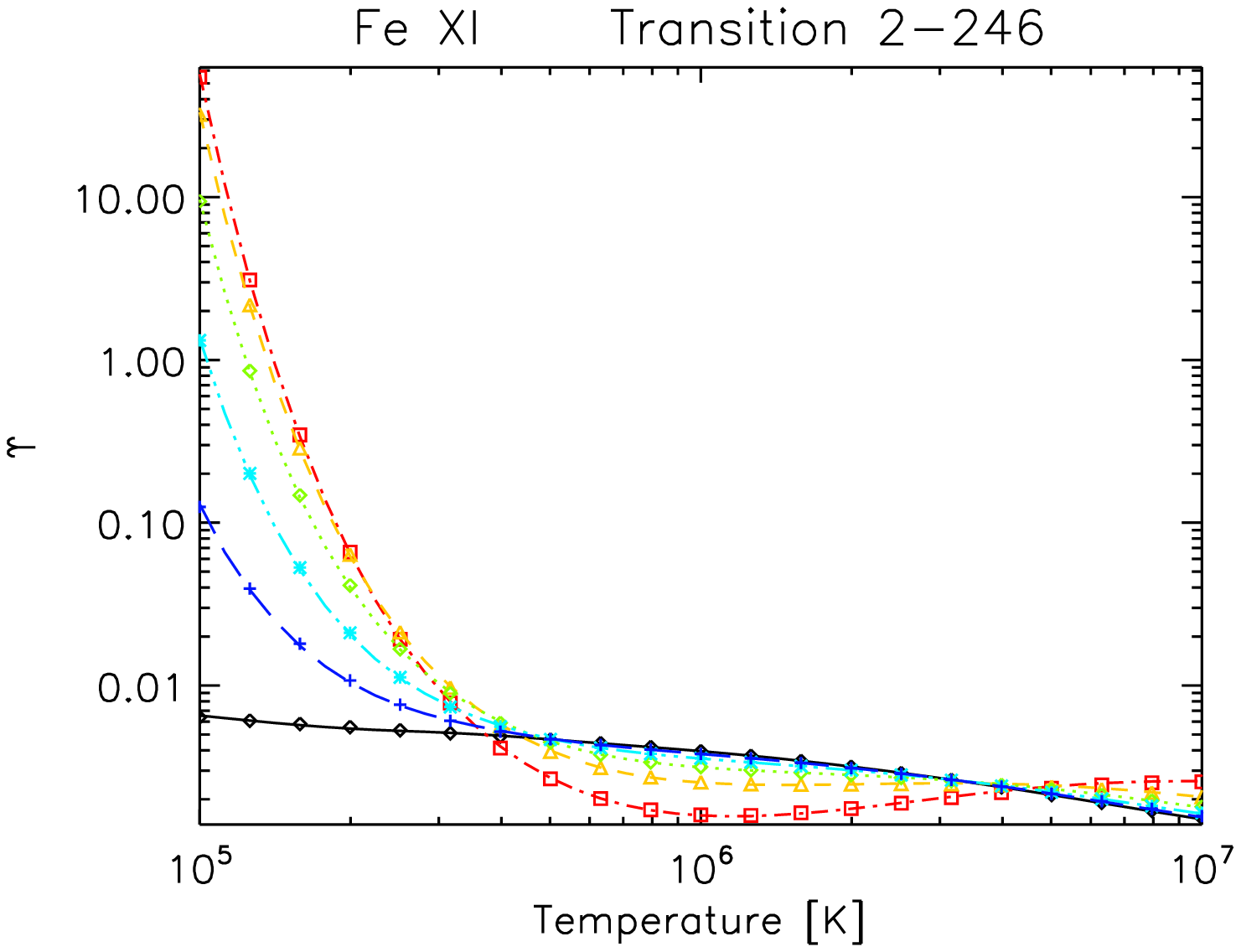}
	\includegraphics[width=5.9cm,bb=20 0 453 340,clip]{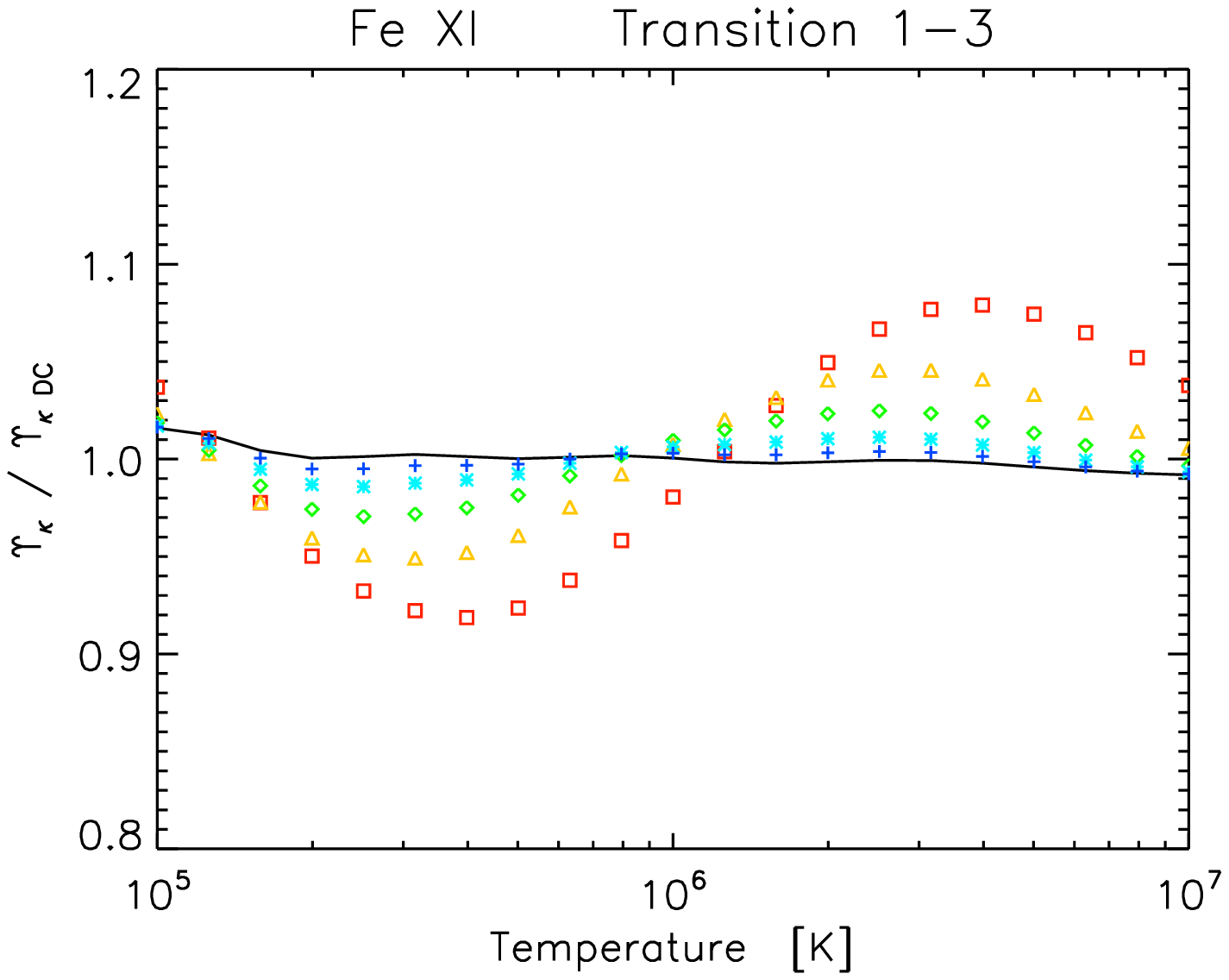}
	\includegraphics[width=5.9cm,bb=20 0 453 340,clip]{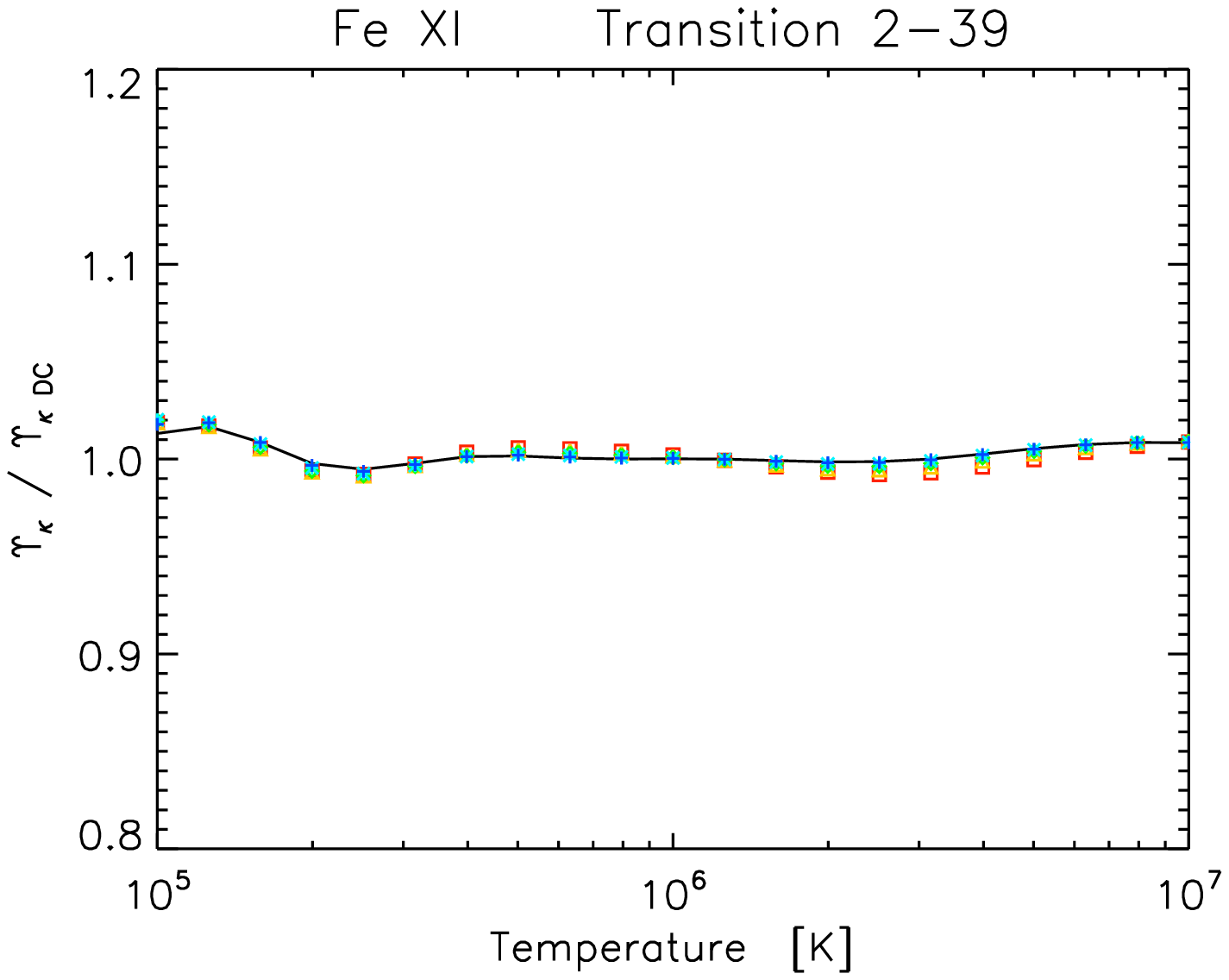}
	\includegraphics[width=5.9cm,bb=20 0 453 340,clip]{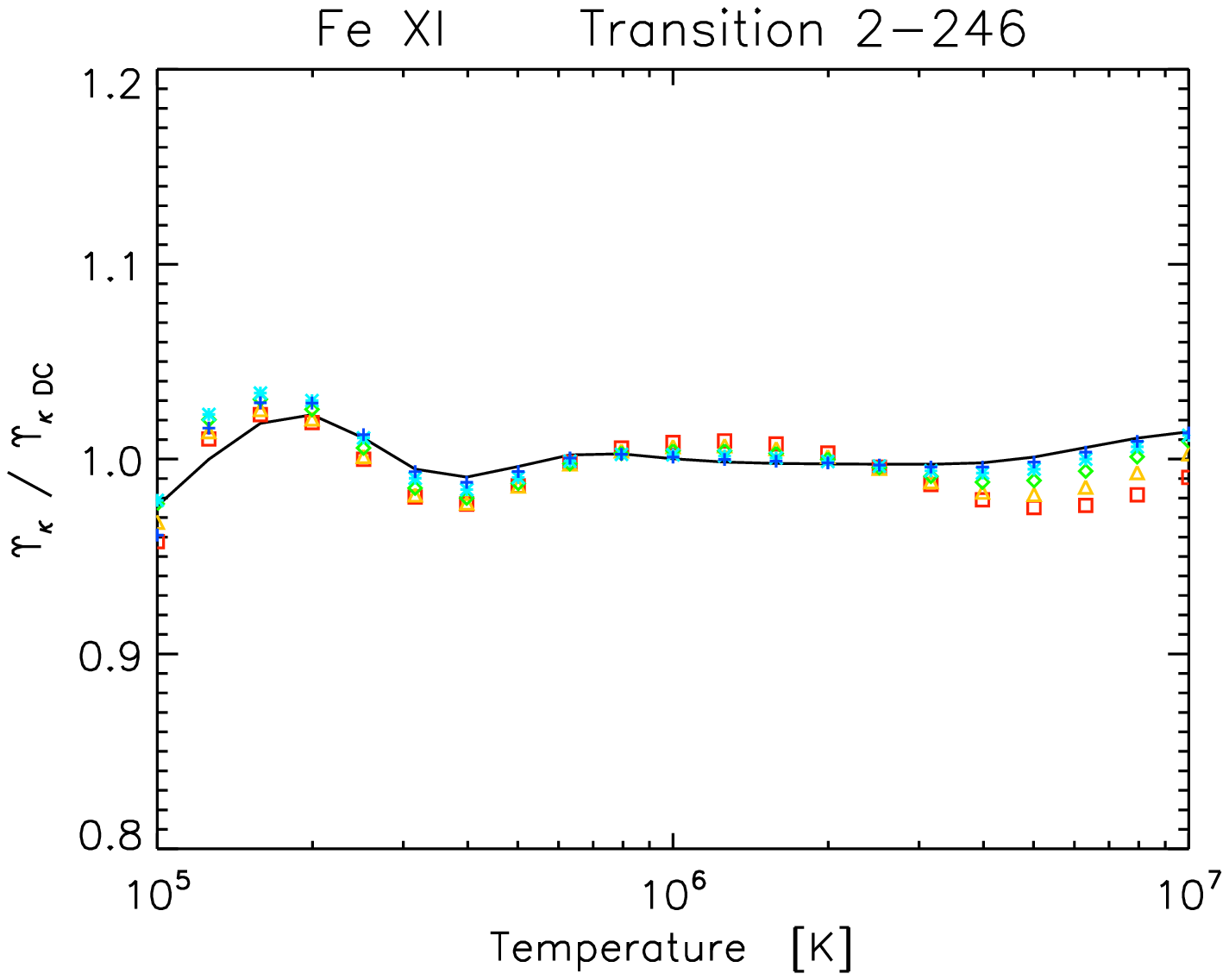}
	\caption{$\Upsilon_{\kappa}$ (\textit{top}) and their relative errors (\textit{below}) for the \ion{Fe}{11} 3s$^2$3p$^4$ $^3$P$_2$\,--\,3s$^2$3p$^4$ $^3$P$_0$ \textit{(left)}, 3s$^2$3p$^4$ $^3$P$_1$ -- 3s$^2$3p$^4$($^2$D) $^3$S$_0$ (\textit{middle}), and 3s$^2$3p$^4$ $^3$P$_1$ -- 3p$^5$ 3d  $^3$F$_3$ \textit{(right)} transitions. Black lines show comparison of the CHIANTI's approximation with the direct calculations for the Maxwellian distribution. Colors show the comparison of our approximation to direct calculations for the $\kappa$-distribution with $\kappa$\,=\,25 (blue), 10 (turquoise), 5 (green), 3 (yellow), and 2 (red).}
\label{Fig:Errors_fe11}
\end{figure*}
%
%
\begin{figure*}[!ht]
	\centering
	\includegraphics[width=5.9cm,bb=20 0 453 340,clip]{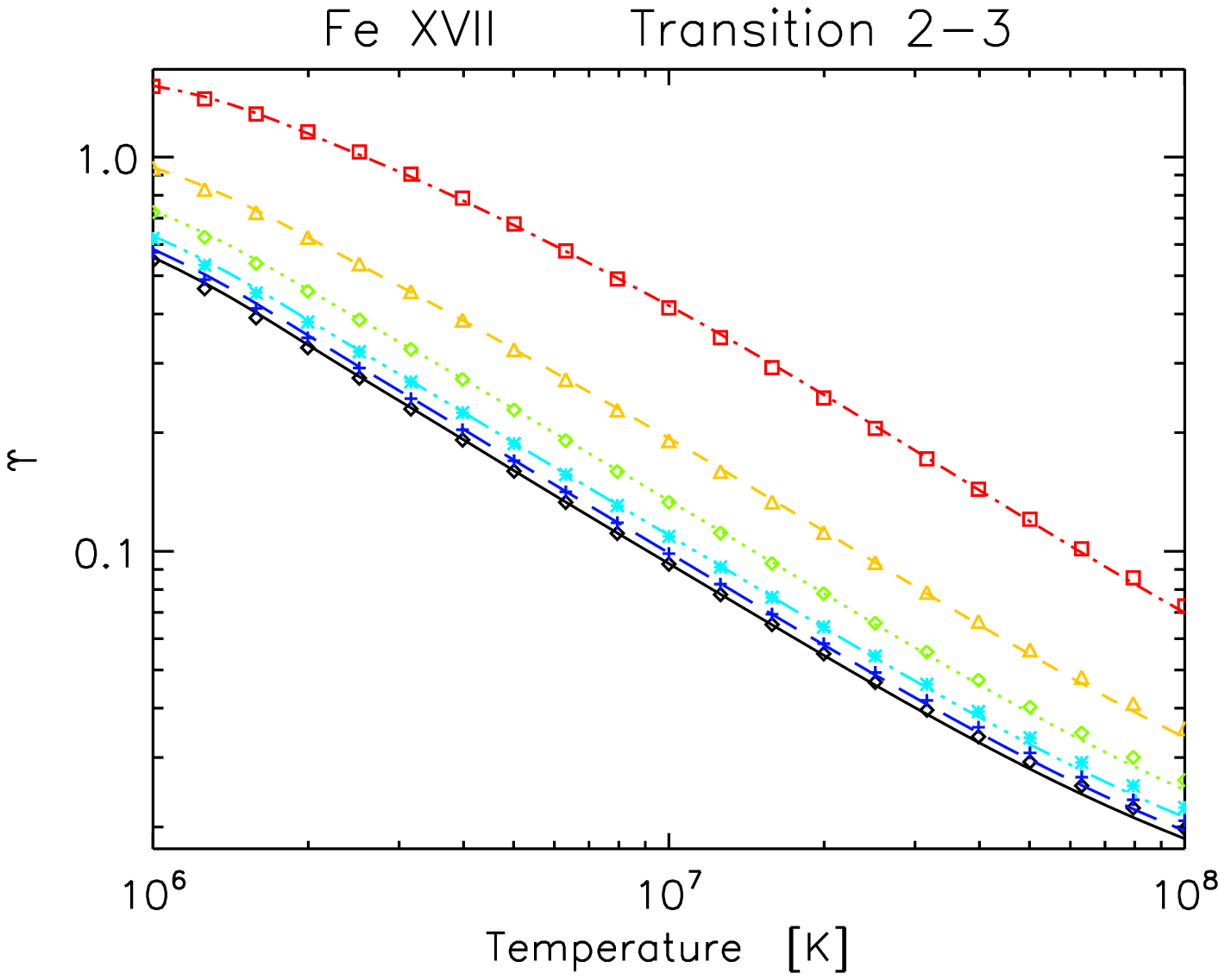}
	\includegraphics[width=5.9cm,bb=20 0 453 340,clip]{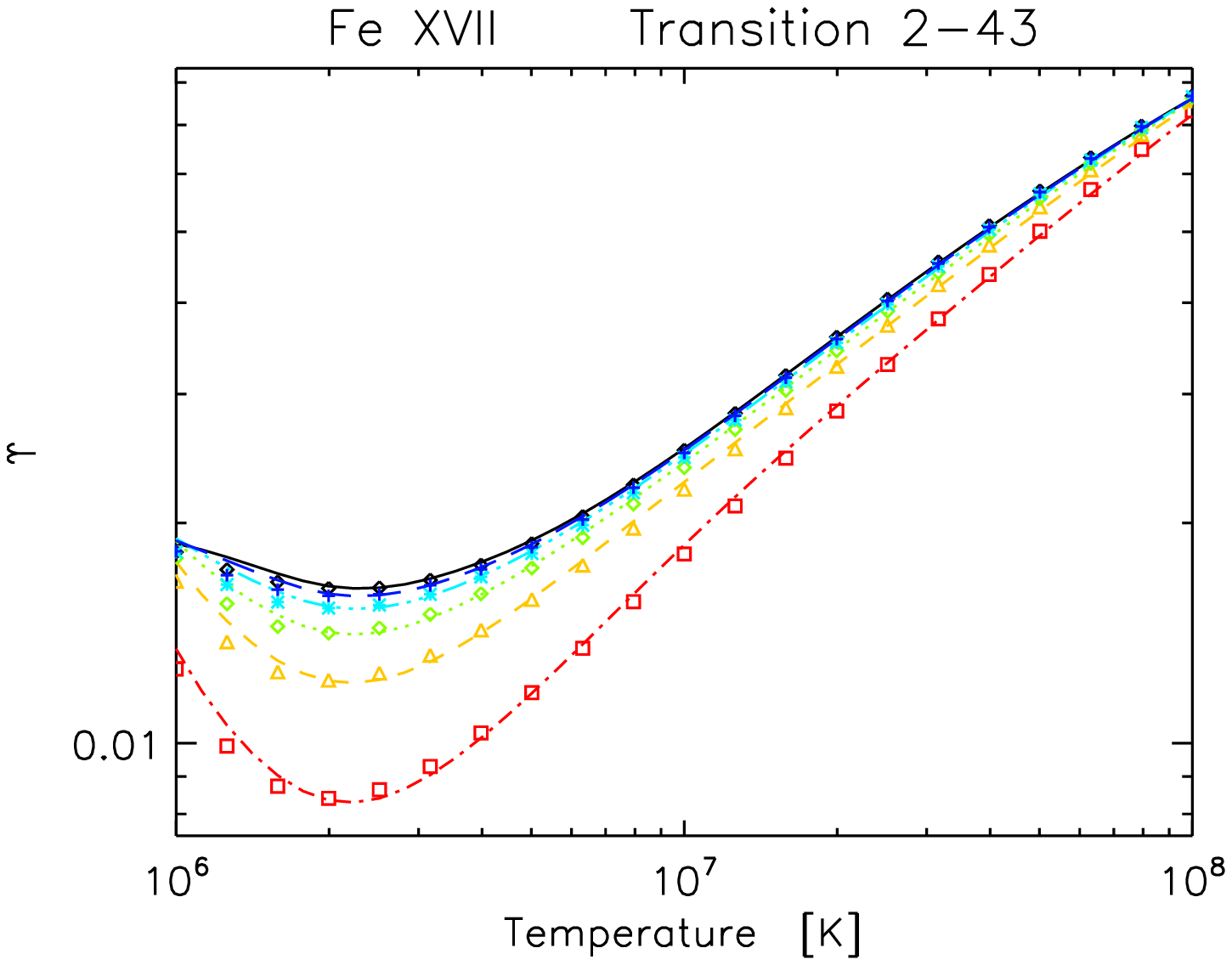}
	\includegraphics[width=5.9cm,bb=20 0 453 340,clip]{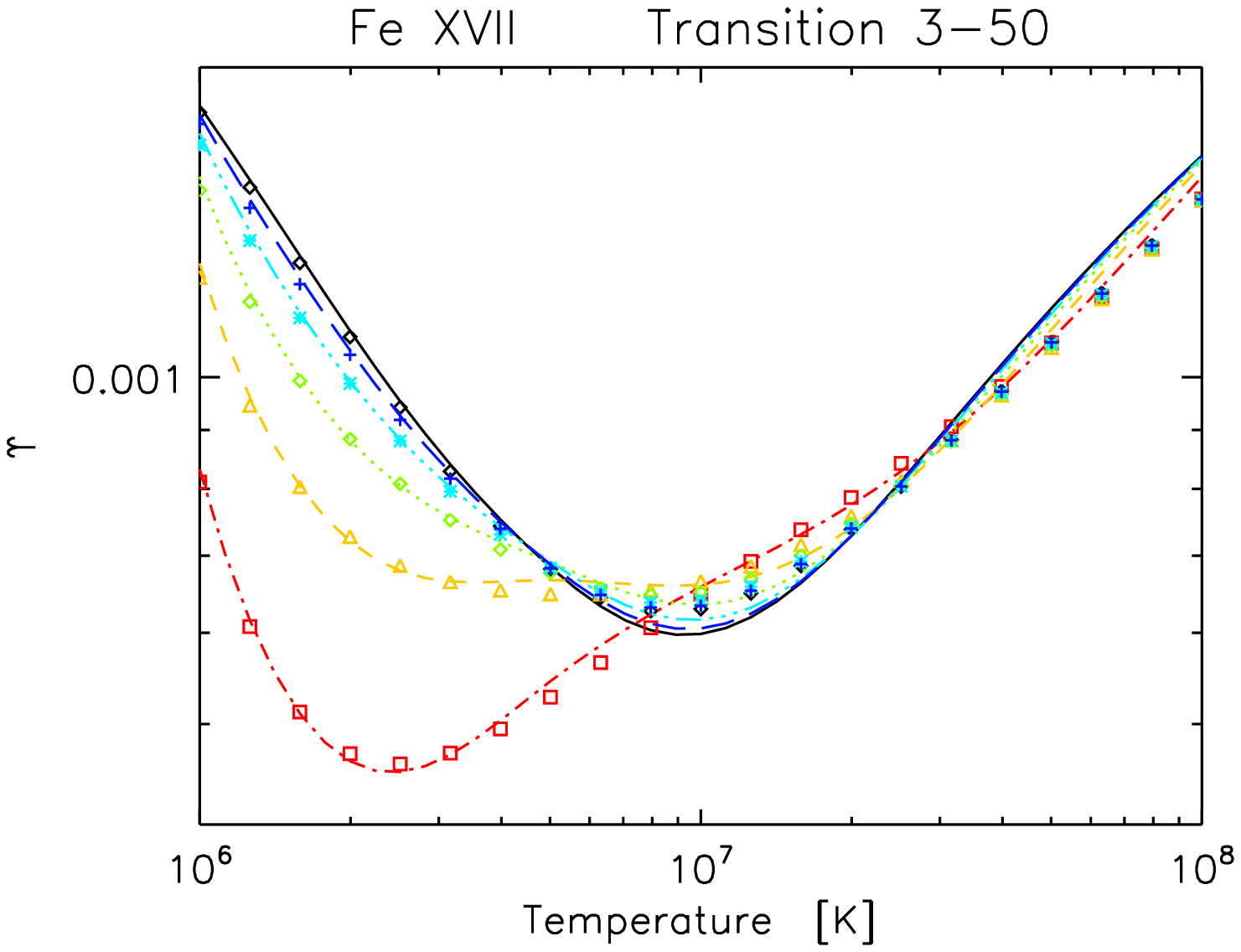}
	\includegraphics[width=5.9cm,bb=20 0 453 340,clip]{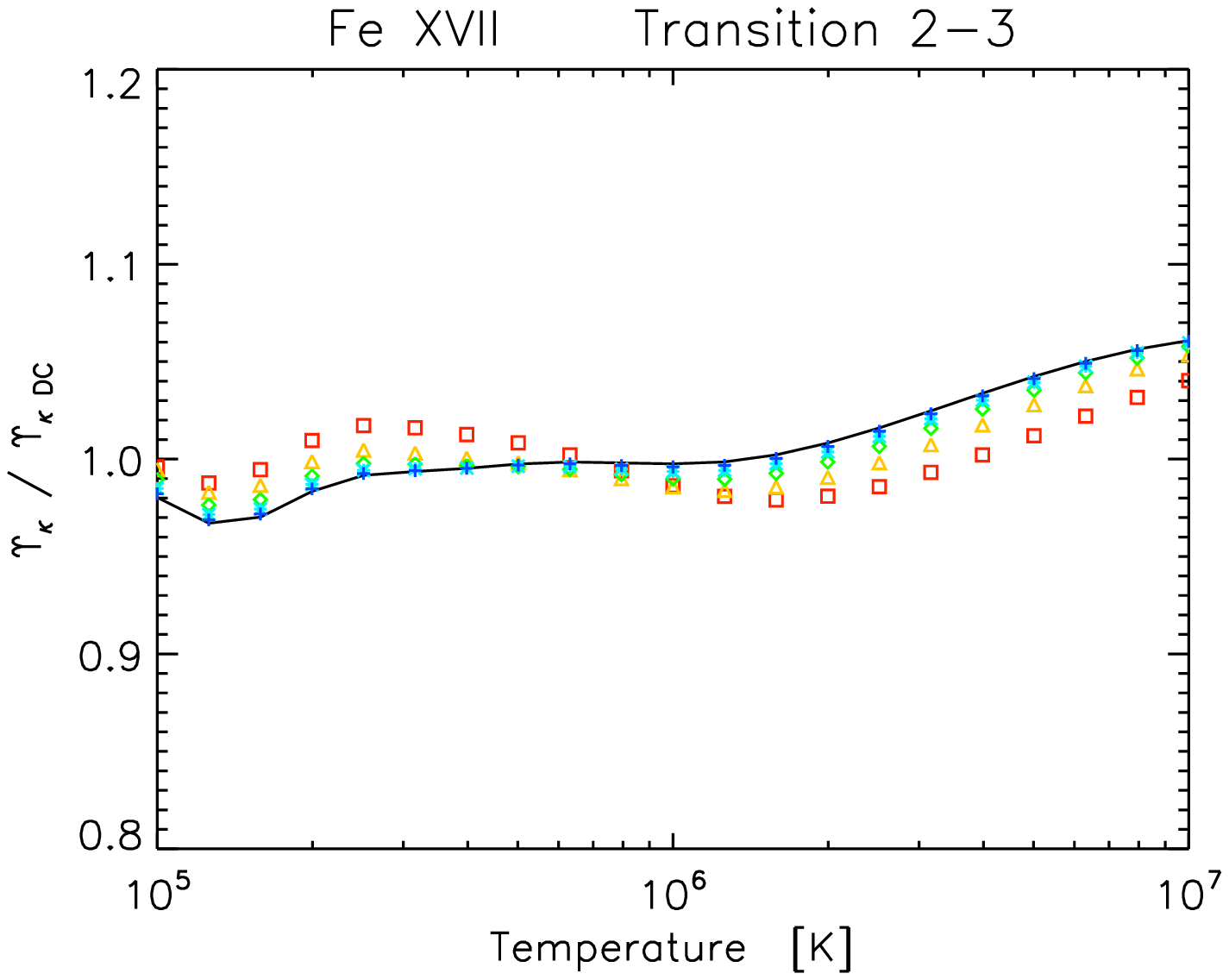}
	\includegraphics[width=5.9cm,bb=20 0 453 340,clip]{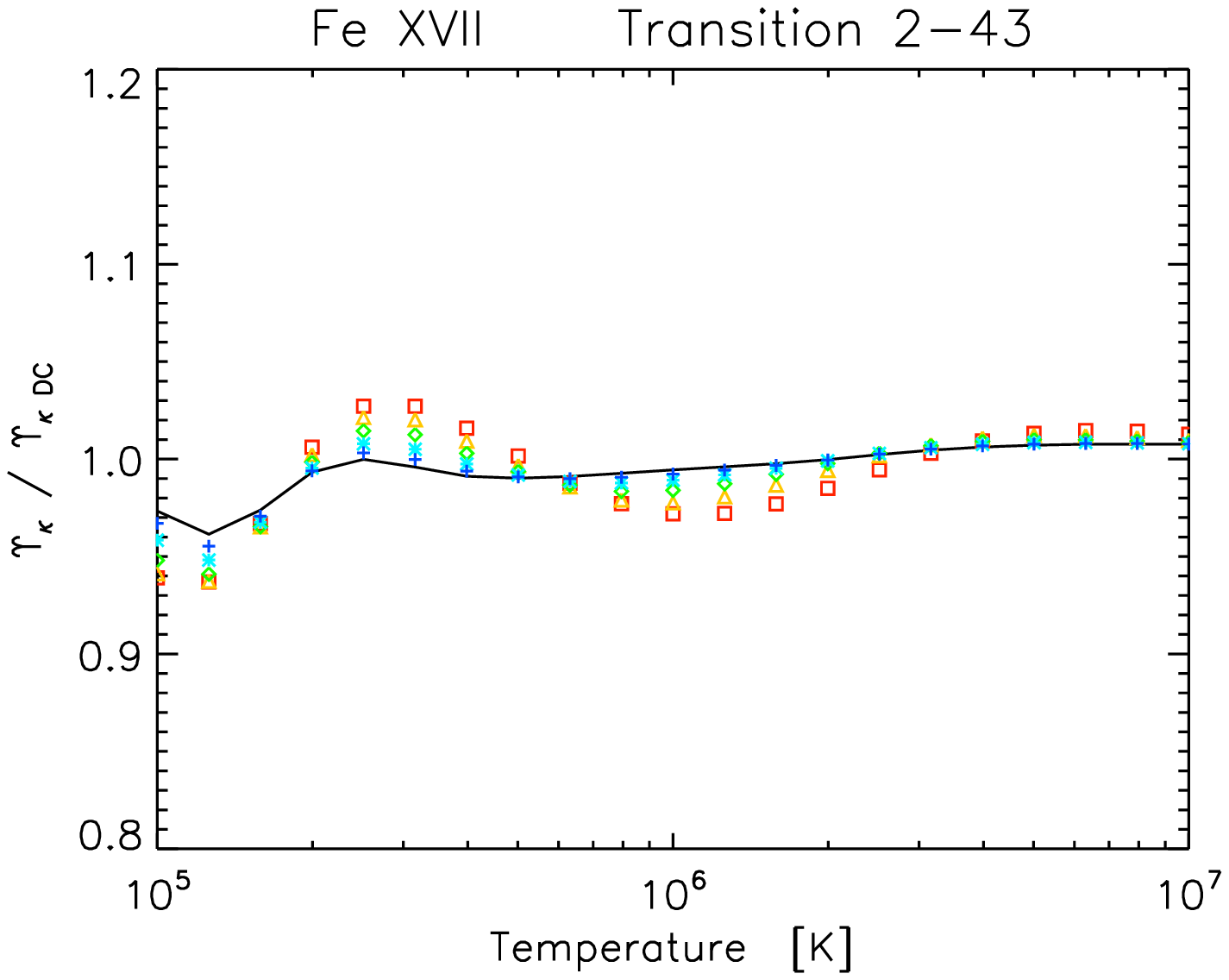}
	\includegraphics[width=5.9cm,bb=20 0 453 340,clip]{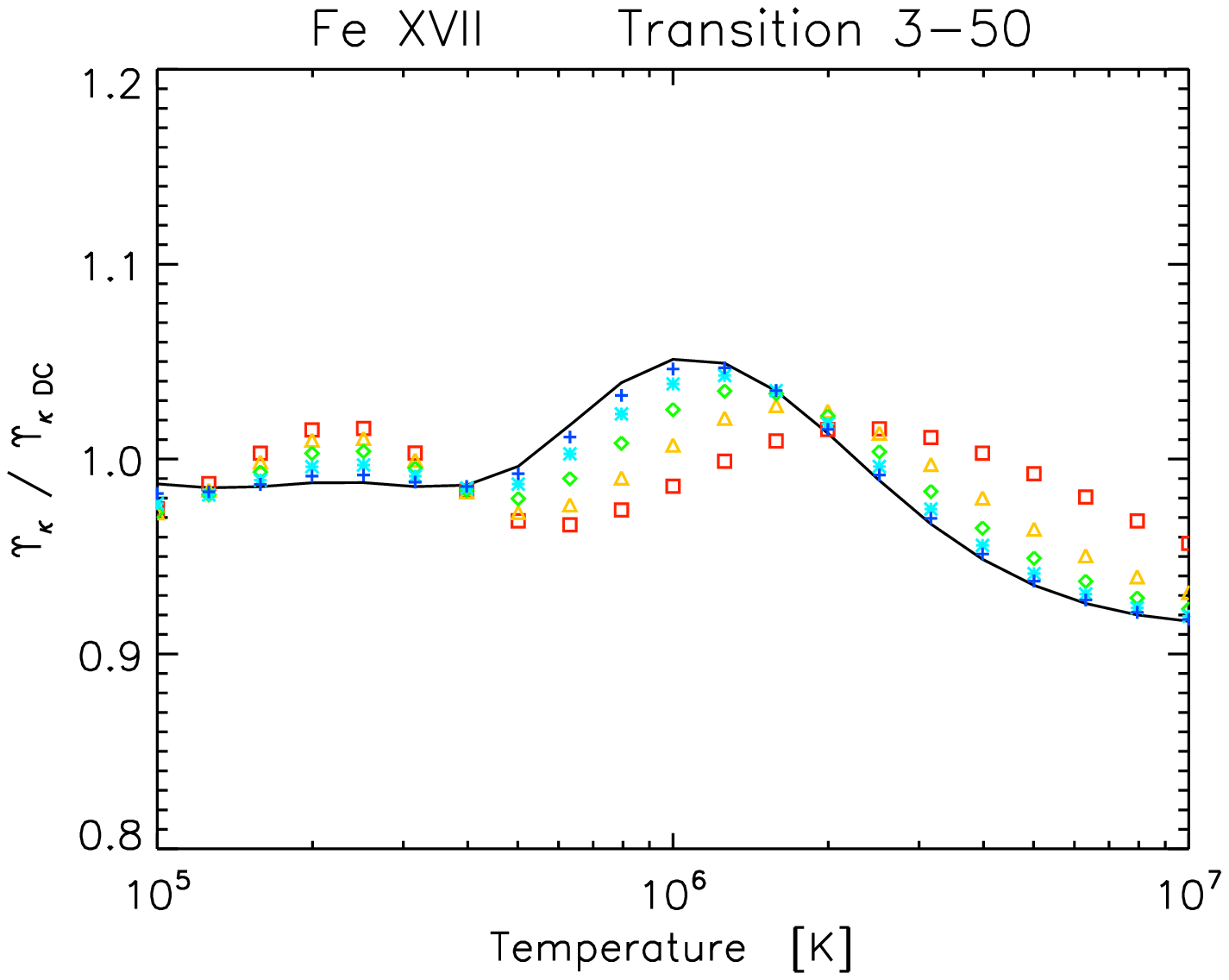}
	\caption{Same as in Fig. \ref{Fig:Errors_fe11}, but for the following transitions in \ion{Fe}{17}: 2s$^2$2p$^5$3s $^1$P$_2$ -- 2s$^2$2p$^5$3s $^1$P$_1$ (\textit{left}), 2s$^2$2p$^5$3s $^1$P$_2$ -- 2s$^2$2p$^5$4p $^2$D$_2$ (\textit{middle}), and 2s$^2$2p$^5$3s $^1$P$_1$ -- 2s$^2$2p$^5$4p $^2$D$_2$ (\textit{right}).}
\label{Fig:Errors_fe17}
\end{figure*}
%
%
\begin{figure*}[!ht]
	\centering
	\includegraphics[width=8.8cm]{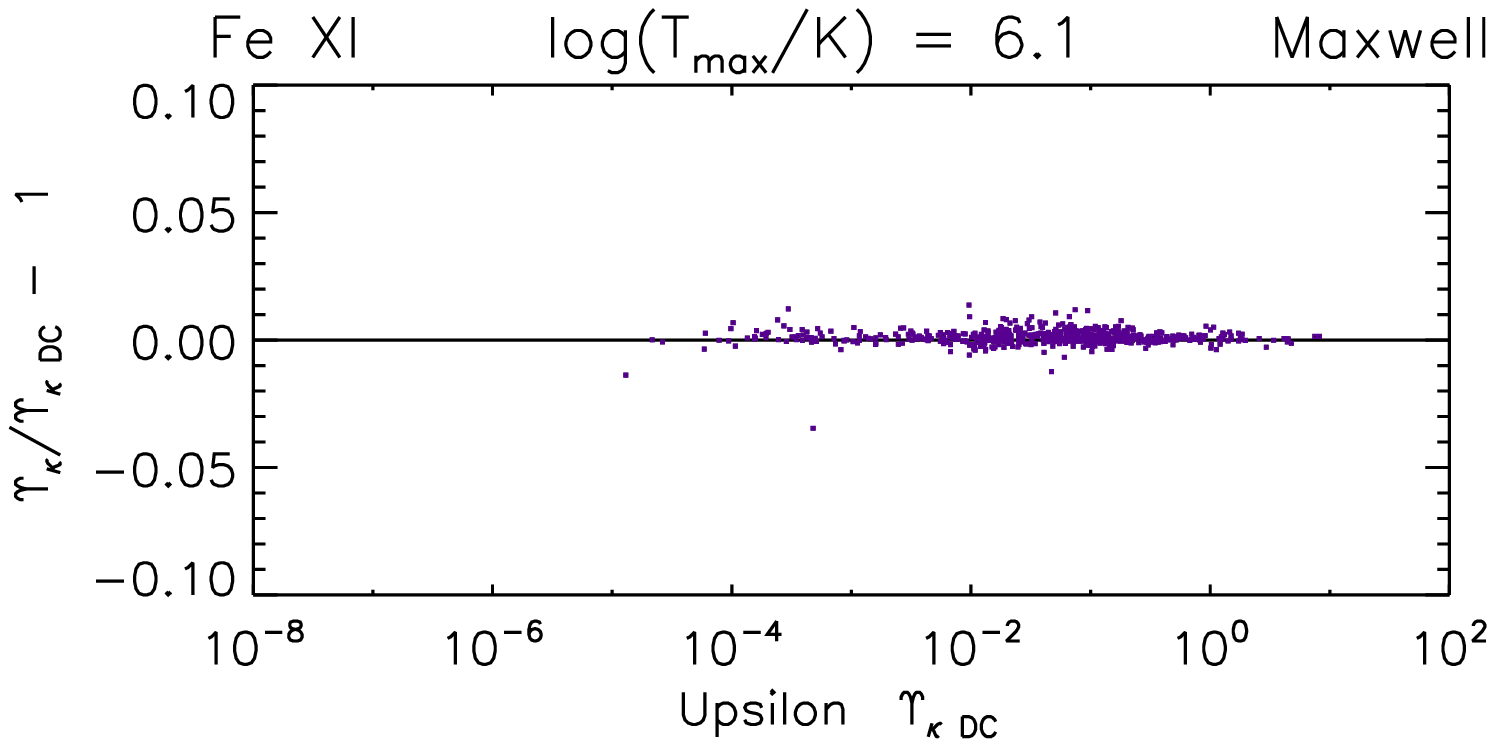}
	\includegraphics[width=8.8cm]{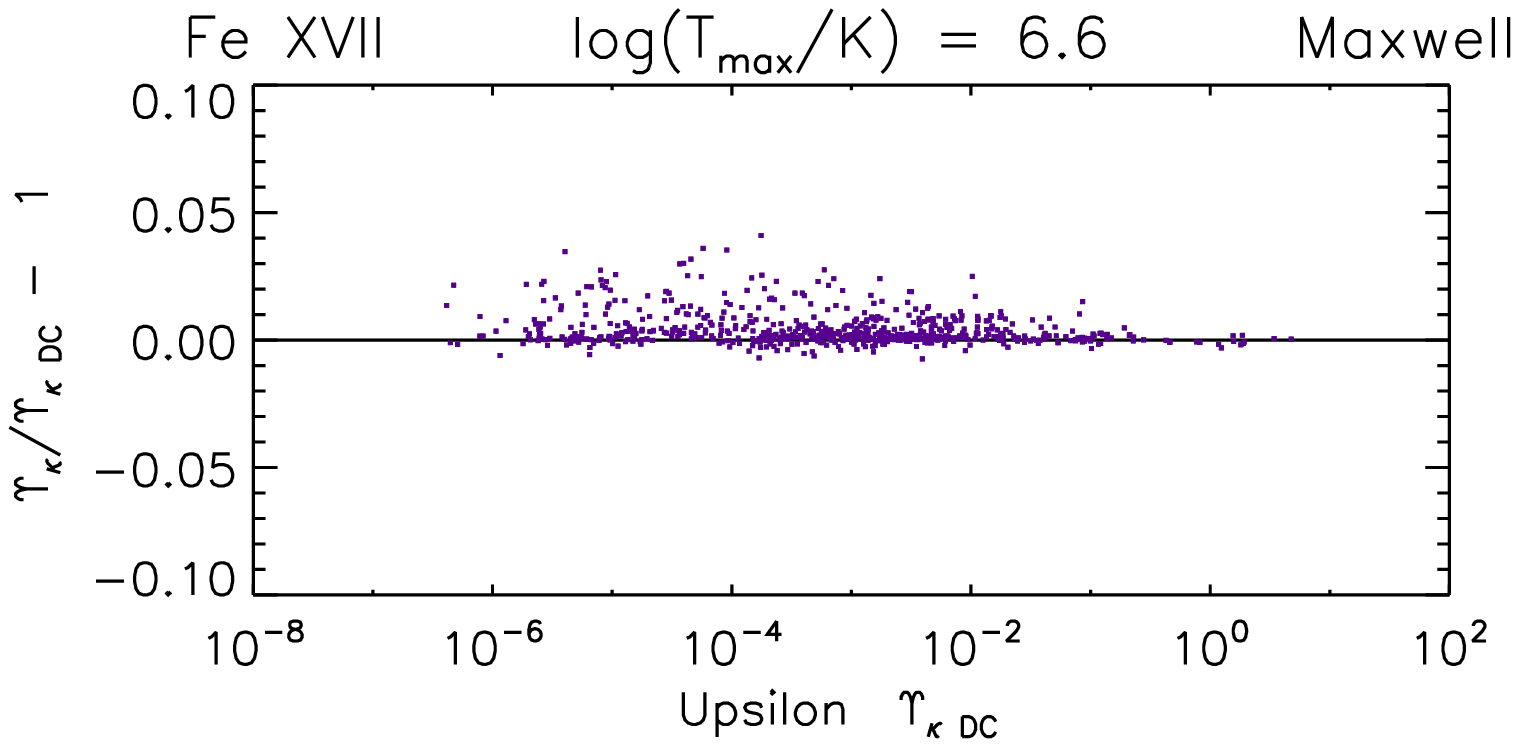}
	\includegraphics[width=8.8cm]{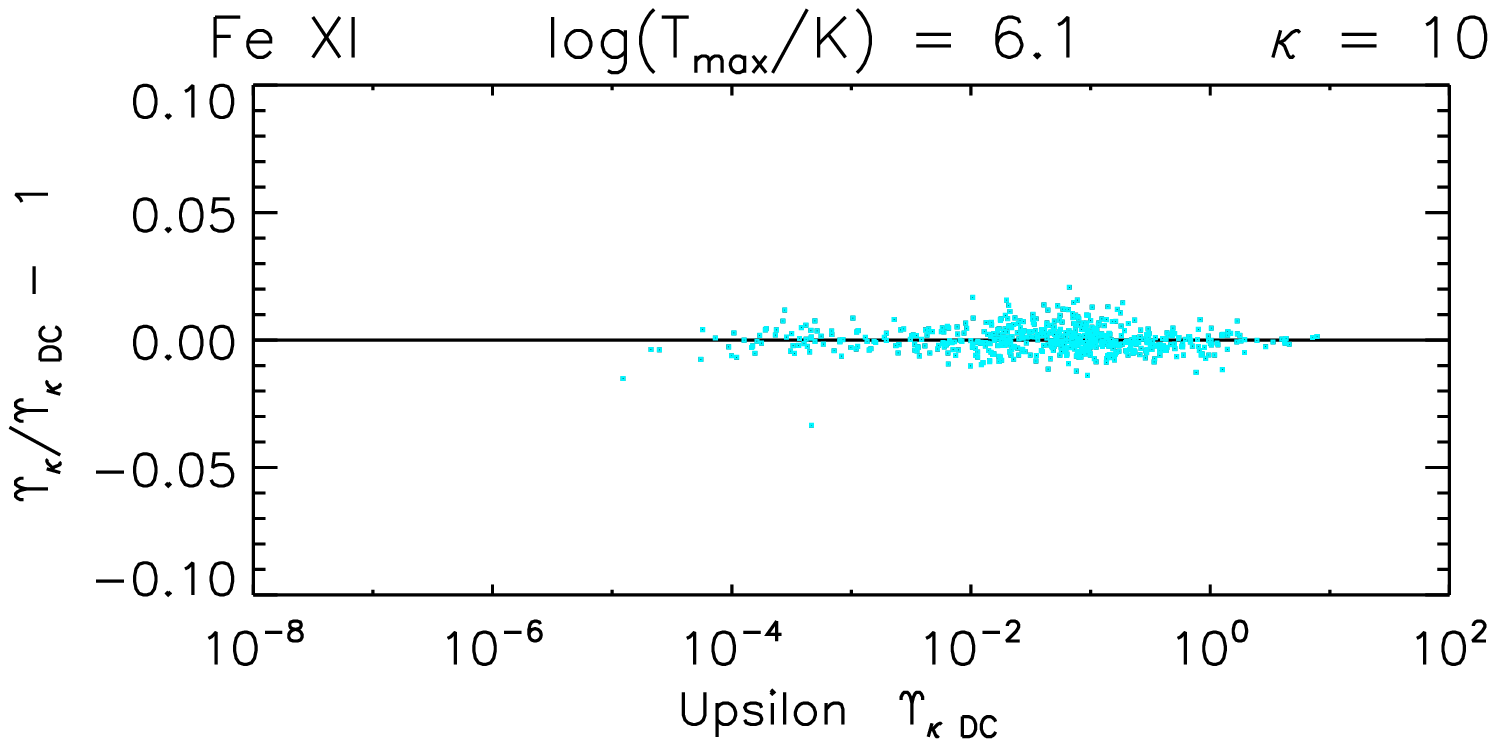}
	\includegraphics[width=8.8cm]{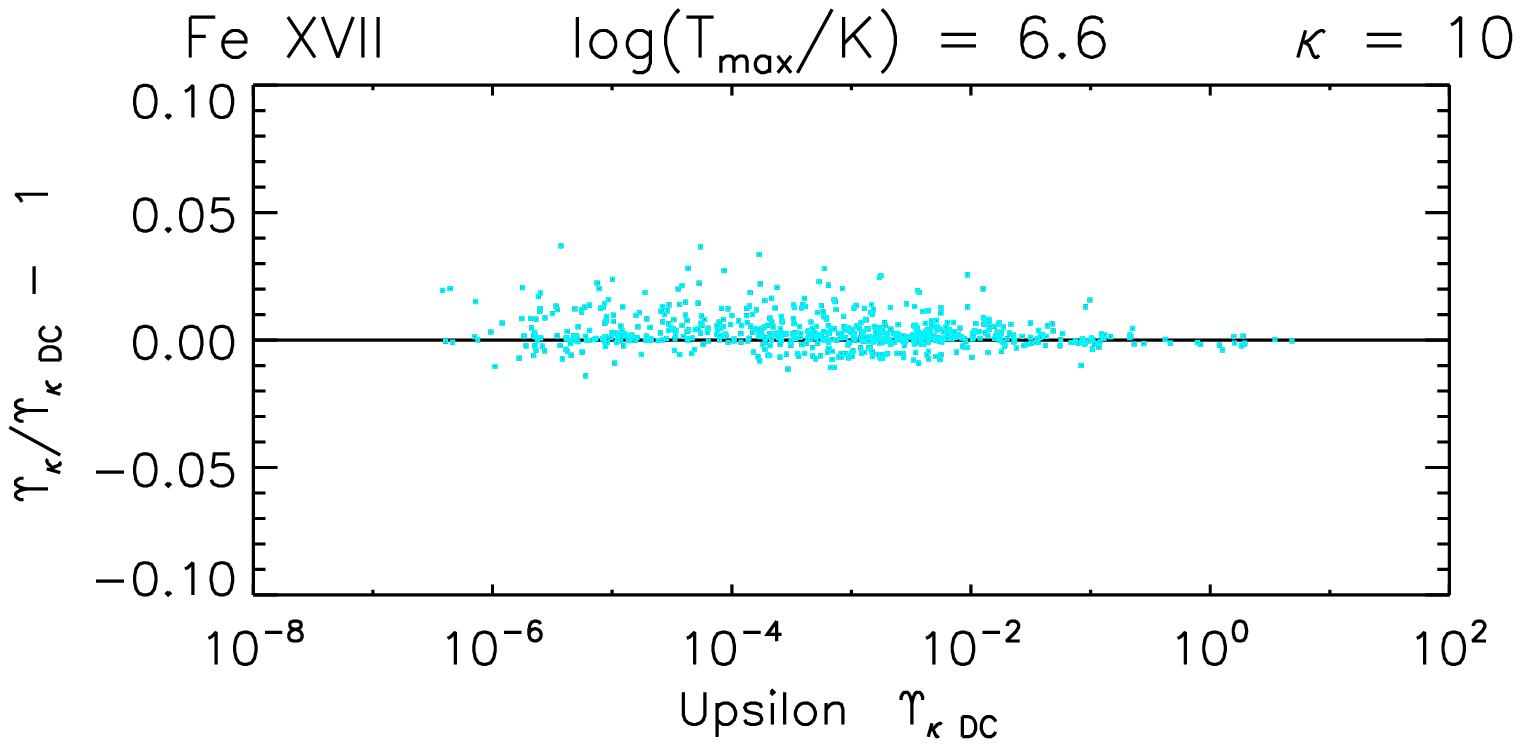}
	\includegraphics[width=8.8cm]{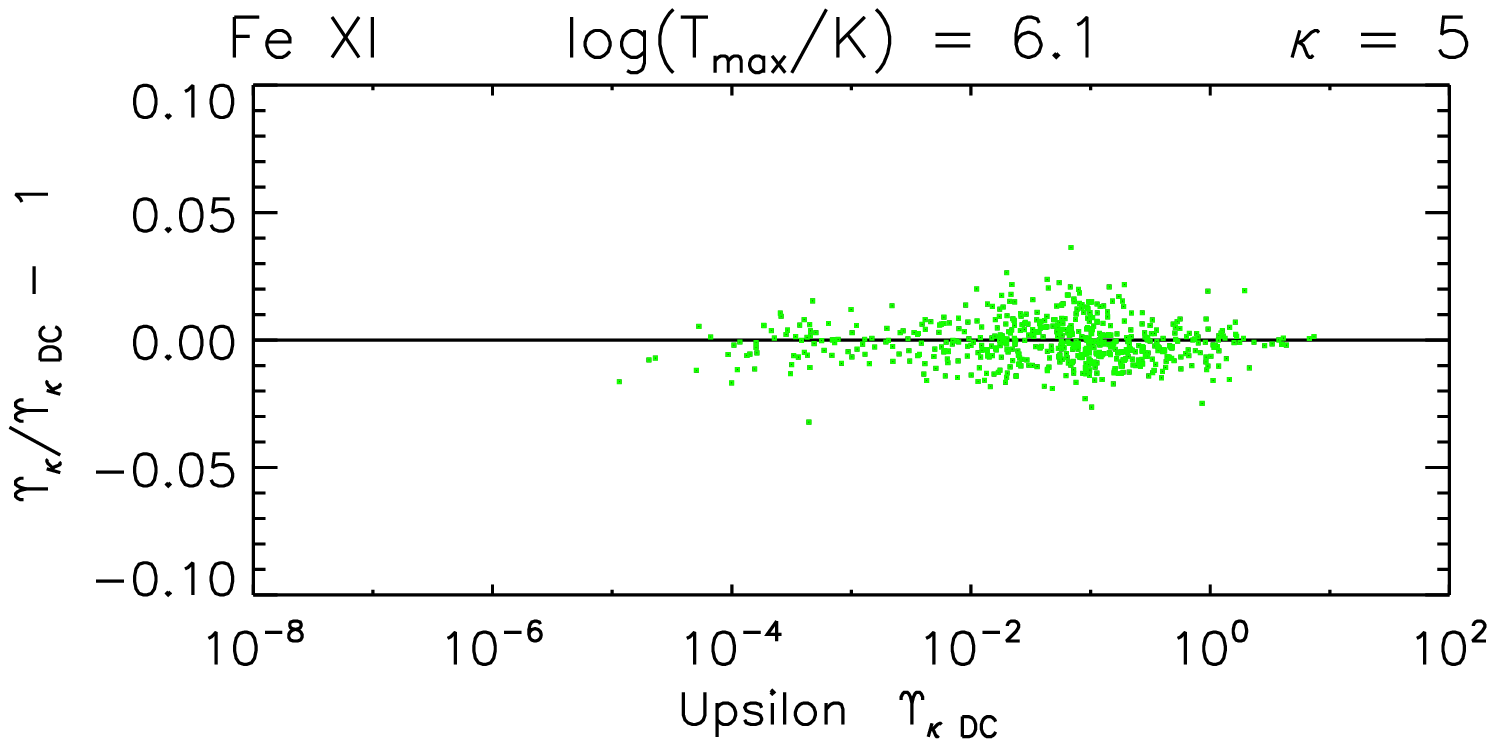}
	\includegraphics[width=8.8cm]{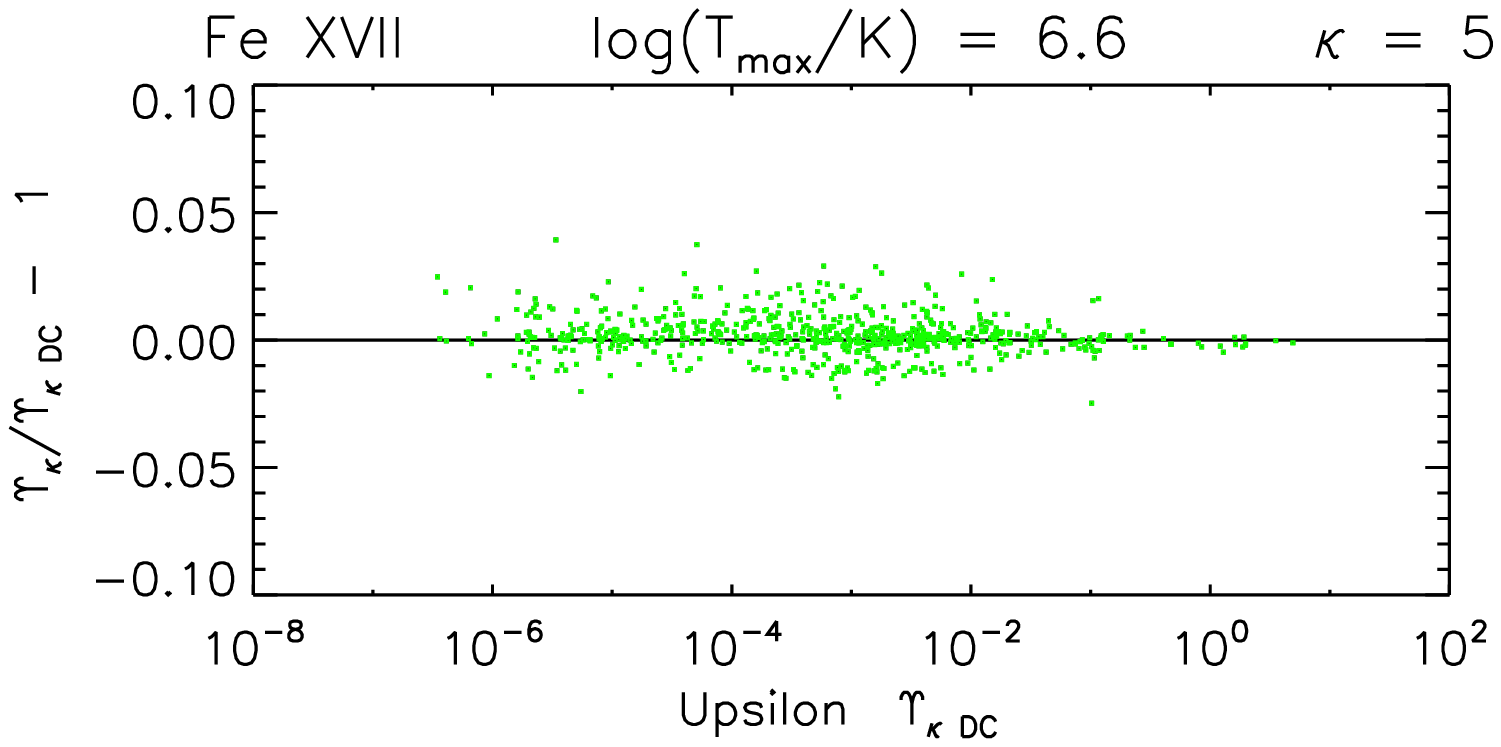}
	\includegraphics[width=8.8cm]{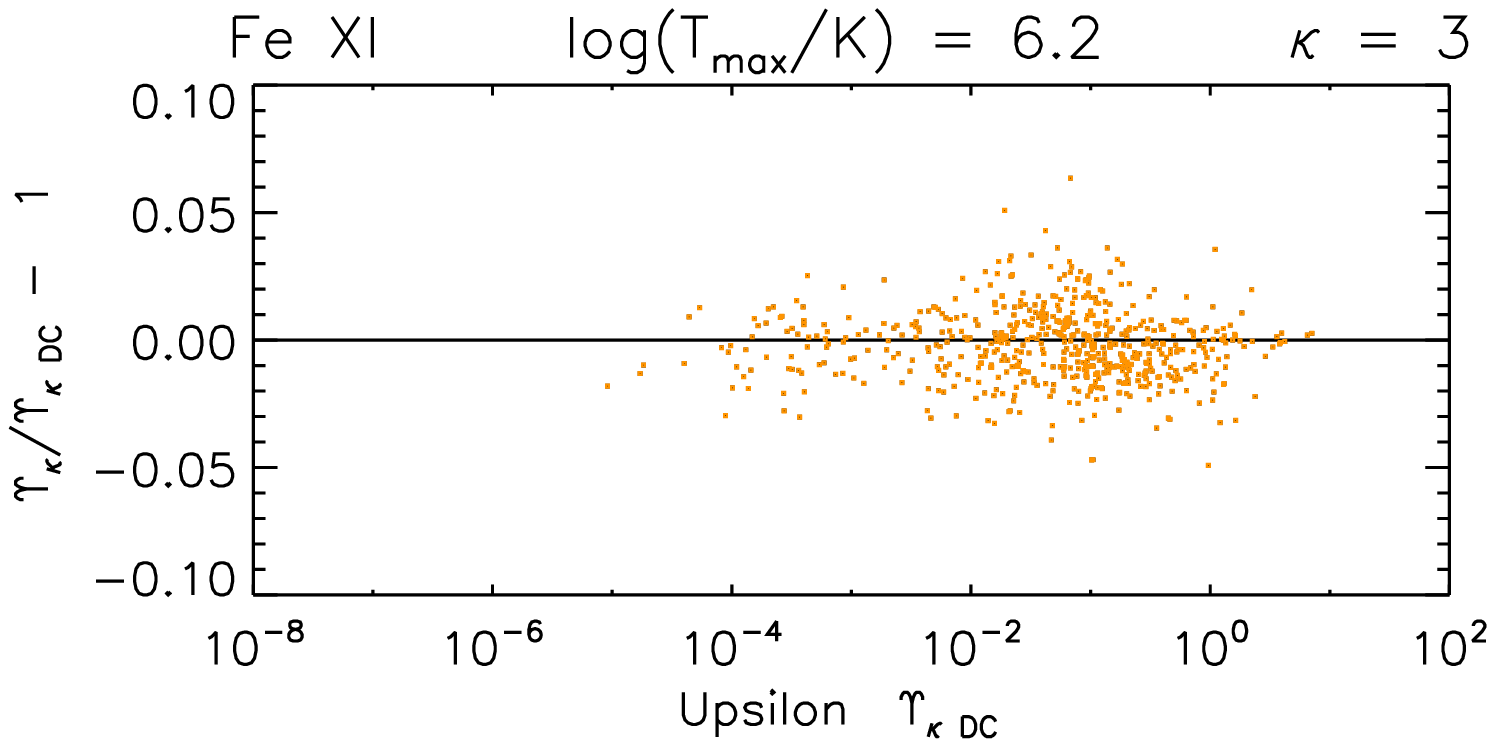}
	\includegraphics[width=8.8cm]{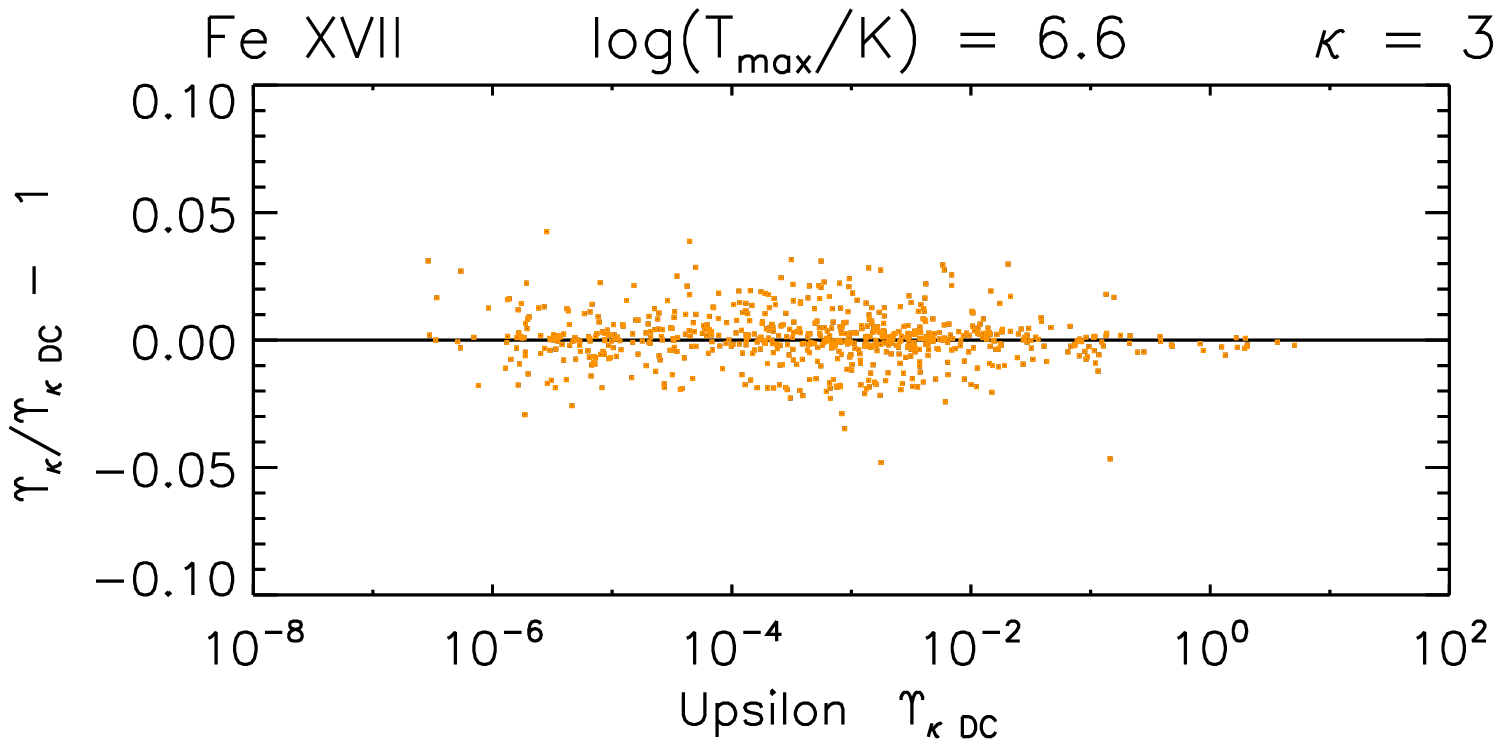}
	\includegraphics[width=8.8cm]{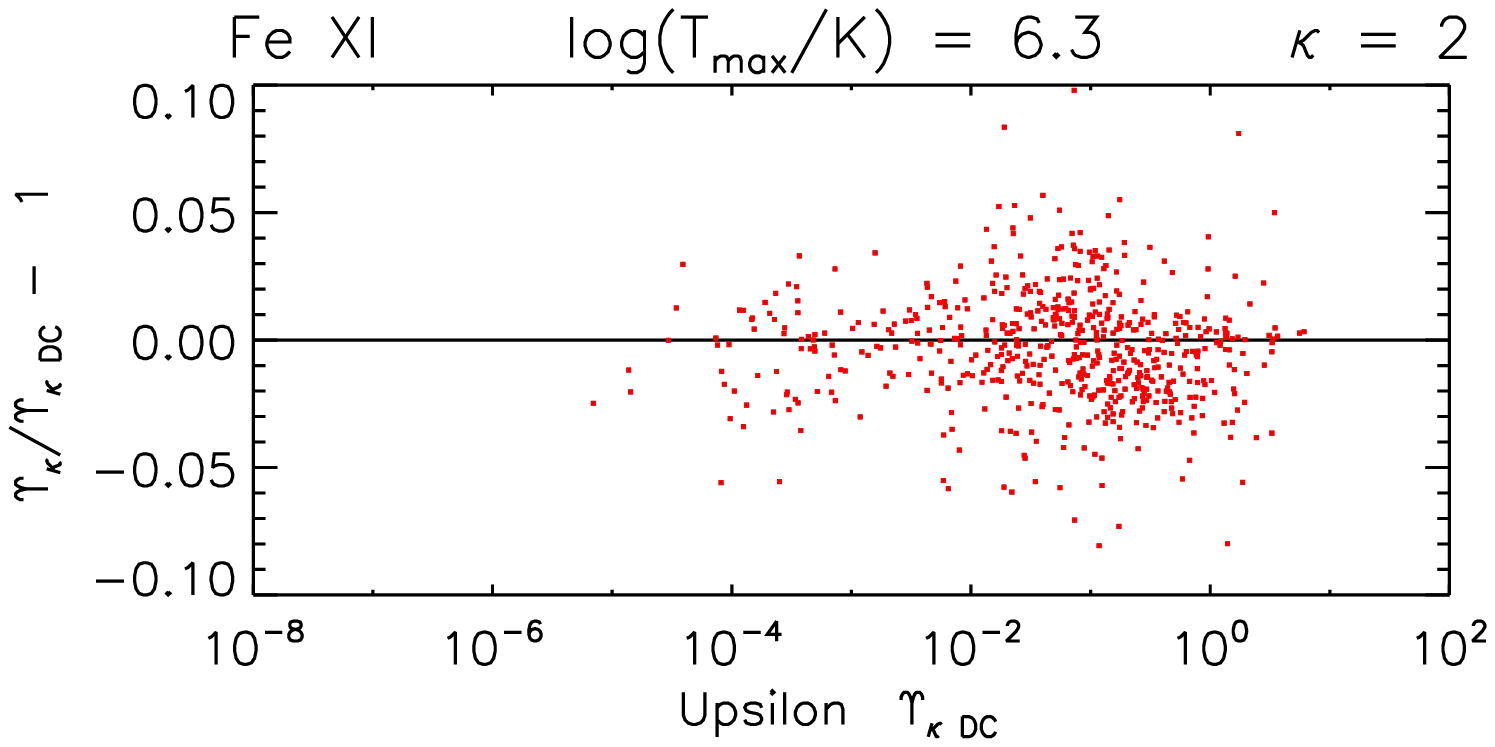}
	\includegraphics[width=8.8cm]{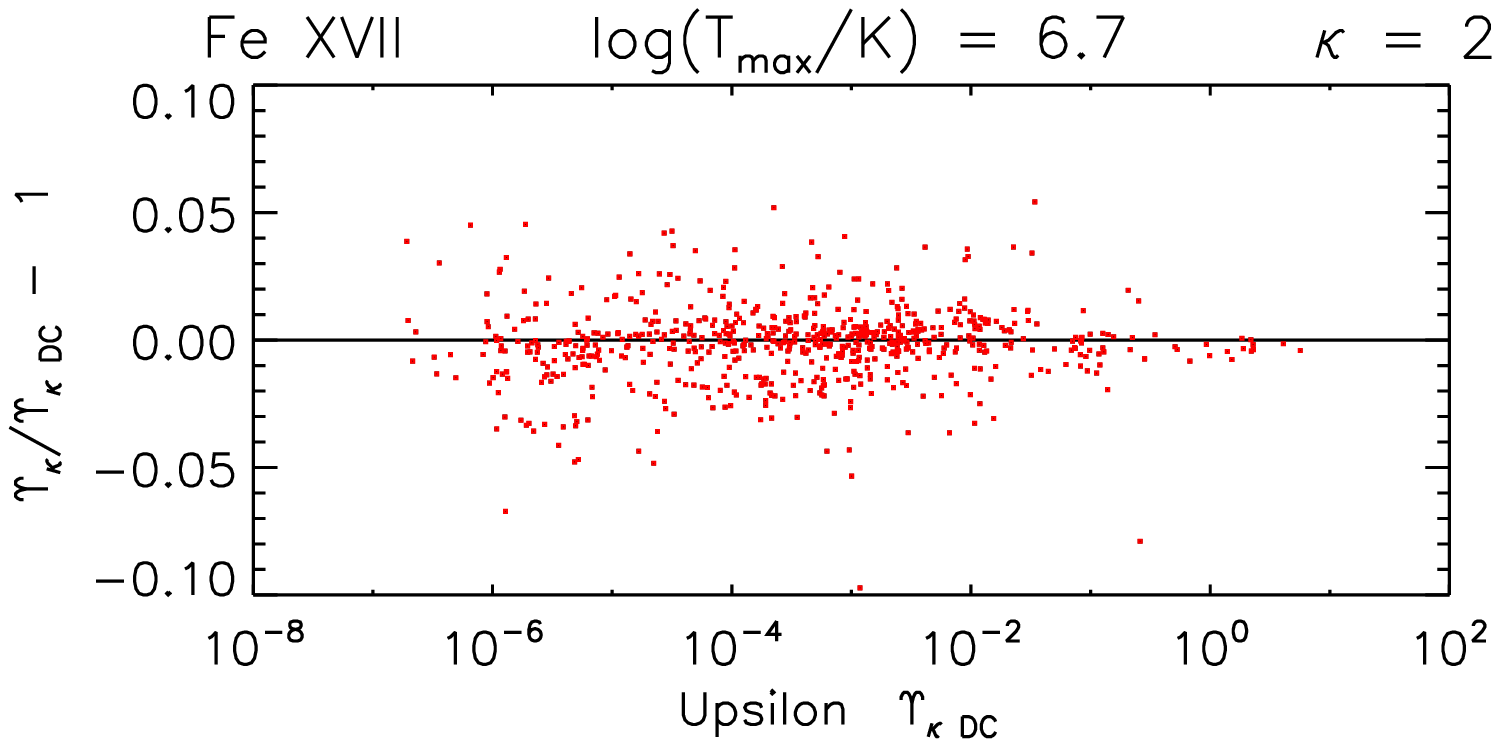}
	\caption{The relative error of $\Upsilon_{\kappa}$ to $\Upsilon_{\kappa DC}$ for \ion{Fe}{11} (\textit{left}) and \ion{Fe}{17}  (\textit{right}) as a function of $\Upsilon_{\kappa DC}$ at temperatures corresponding to the maximum of the ion abundance $\Upsilon_{\kappa,(T_{\mathrm{max}})}$. Black points are for the CHIANTI approximation (Eq. \ref{Eq:Ups_k_approx}). Different colors stand for results for $\kappa$\,=\,2 (red), 3 (orange), 5 (green), and 10 (blue). }
\label{Fig:Scatterplots}
\end{figure*}
%
%
\subsection{Collision Strength Approximation}
\label{Sect:3.3}

The calculation of the collision strengths for excitation and deexcitation averaged over $\kappa$-distributions for large number of transitions introduces a problem of accessibility of atomic cross-setions $\Omega_{ji}(E_j)$. Only a few database contain these data, and typically only for a small number of transitions. The CHIANTI database and software \citep{Dere97,Landi13} contains spline approximations to the Maxwellian-averaged collision strengths for the majority of the astronomically interesting ions of elements H to Zn. CHIANTI allows for computation and analysis of solar spectra and is an important tool of the diagnostics of the solar plasma under the assumption of a Maxwellian distribution.

We used the CHIANTI database to calculate the approximate cross-sections $\Omega$ and subsequently approximate excitation and de-excitation rates $\Upsilon_{ij}(T,\kappa)$ and \rotatebox[origin=c]{180}{$\Upsilon$}$_{ji}(T,\kappa)$ for the $\kappa$-distributions. This approximate method was described e.g. in \citet{Dzifcakova06} and tested for \ion{Fe}{15} by \citet{DzifcakovaMason08}. Here, we use this method to obtain the $\Upsilon_{ij}(T,\kappa)$ and \rotatebox[origin=c]{180}{$\Upsilon$}$_{ji}(T,\kappa)$ for all transitions in all the elements and ions available within CHIANTI.

The approximation works as follows: A functional form for the approximation of $\Omega$ is assumed \citep{Abramowitz65} 
\begin{equation}
	\Omega = \sum_{n=0}^{n_\mathrm{max}}{\cal C}_{n}u^{-n}+D~\mathrm{ln}(u),
	\label{Eq:Omega_approx}
\end{equation}
where ${\cal C}_{k}$ and $D$ are coefficients and $u=E_{i}/\Delta E_{ij}$. The advantage of this approximation is a simple analytical evaluation of its integral over the distribution function. This approximation was often used for expression of the collision strength e.g. by \citet{Mewe72}.

The $\Upsilon_{ij}$ for the Maxellian distribution can then be written as:
\begin{equation}
	\Upsilon_{ij}=\frac{\Delta E_{ij}}{k_\mathrm{B}T} \mathrm{e}^{\frac{\Delta E_{ij}}{k_\mathrm{B}T}} \int_{1}^{\infty}\Omega_{ij} \mathrm{e}^{\left(-\frac{E_{i}}{k_\mathrm{B}T}\right)} \mathrm{d}\left(\frac{E_{i}}{\Delta E_{ij}}\right)\,,
	\label{Eq:Ups_mxw_approx}
\end{equation}
which after integration leads to
\begin{equation}
	\Upsilon_{ij}={\cal C}_{0}+\left( \sum_{k=1}^{n_\mathrm{max}}y{\cal C}_{n}{\cal E}_{n}(y)+D {\cal E}_{1}(y) \right) e^{y},
	\label{Eq:Ups_mxw_approx_expr}
\end{equation}
where $y=\Delta E_{ij}/k_\mathrm{B}T$ and ${\cal E}_{n}(y)$ is an $n$-th order exponential integral.

%
%
%
%
%
%
The behaviour of $\Omega$ in the high-energy limit and the corresponding behaviour of $\Upsilon_{ij}$ provide following conditions for the coefficients ${\cal C}_{n}$ and $D$ for the electric dipole transitions
\begin{equation}
	D=\frac{4\omega_{i}f_{ij}}{\Delta E_{ij}},~
	\Upsilon_{ij}(\rightarrow\infty)=\sum\limits_{n=0}^{n_\mathrm{max}}{\cal C}_{n}=\Omega(u=1)\,,
	\label{Eq:D-Ups_conditions_type1}
\end{equation}
while for the non electric dipole, non exchange transitions
\begin{eqnarray}
	D&=&0,~\Upsilon_{ij}(\rightarrow0)={\cal C}_{0},\nonumber \\
	\Upsilon_{ij}(\rightarrow\infty)&=&\sum_{n=0}^{n_{max}}{\cal C}_{n}=\Omega(u=1)
	\label{Eq:D-Ups_conditions_type2}
\end{eqnarray}
and finally for the exchange transitions
\begin{eqnarray}
	{\cal C}_{0}&=&{\cal C}_{1}=D=0,~\nonumber \\ \Upsilon_{ij}(\rightarrow 0)&=&y\int_{1}^{\infty}\Omega d(u),~\nonumber \\
	\Upsilon_{ij}(\rightarrow\infty)&=&\sum_{n=0}^{n_\mathrm{max}}{\cal C}_{n}=\Omega(u=1).
	\label{Eq:D-Ups_conditions_type3}
\end{eqnarray}
The low- and high-energy limits $\Upsilon_{ij}(\rightarrow0)$ and $\Upsilon_{ij}(\rightarrow\infty)$ can be found in the CHIANTI database. The coefficients ${\cal C}_{n}$ and $D$ can be evaluated from the collisional strengths in CHIANTI, averaged over the Maxwellian distribution by the least square method. To achieve the higher precision, we used approximations (Eq. \ref{Eq:Omega_approx}) up to $n_\mathrm{max}=7$.

The approximate method described here is also used to calculate the distribution-averaged collision strengths for deexcitation \rotatebox[origin=c]{180}{$\Upsilon$}$_{ji}(T,\kappa)$.

%
\subsection{Validity of the Approximate Method}
\label{Sect:3.4}

Figure \ref{Fig:O4} demonstrates the approximation of $\Omega$ and calculation of $\Upsilon_{\kappa}$ for the \ion{O}{4} transition 2s$^2$\,2p\,$^2$P$_{1/2}$\,--\,2s\,2p$^2$\,$^4$P$_{3/2}$ at 1401.16\AA. The atomic data for this transition are taken from \citet{Liang12}. We find a typical precision in the approximation of CHIANTI $\Upsilon$'s of a few percent. This is the case for the \ion{O}{4} 1401.16\AA~transition shown, for which we find a precision of 1--2\%. However, for a small part of transitions the precision can be significantly worse, up to approximately  15\%.

Fulfilling the conditions (\ref{Eq:D-Ups_conditions_type1})--(\ref{Eq:D-Ups_conditions_type3}) for the coefficients guarantees correct behaviour of $\Omega$ for high and threshold energies. It is however difficult to compare data for all transitions of each of ion. Occasional errors in the approximation of $\Omega$ (Eq. \ref{Eq:Omega_approx}) cannot be excluded at present. Their propagation to the calculated of $\Upsilon_{\kappa}$ are further minimized by adopting
\begin{equation}
	\Upsilon_{\kappa}=\Upsilon_\mathrm{Maxwell}^\mathrm{CHIANTI} \frac{\Upsilon_{\kappa}^\mathrm{approx}}{\Upsilon_\mathrm{Maxwell}^\mathrm{approx}}\,,
	\label{Eq:Ups_k_approx}
\end{equation}
where $\Upsilon_{\kappa}$ is the final $\Upsilon(\kappa,T)$ for the $\kappa$-distributions, $\Upsilon_\mathrm{Maxwell}^\mathrm{CHIANTI}$ is ${\Upsilon}(T)$ taken from CHIANTI for the Maxwellian distribution, and $\Upsilon_{\kappa}^\mathrm{approx}$ and $\Upsilon_\mathrm{Maxwell}^\mathrm{approx}$ are $\Upsilon$'s calculated from our approximations of the cross sections for the $\kappa$-distributions and Maxwellian distribution, respectively. 

First tests of the precision of the approximate method desribed in Sect. \ref{Sect:3.3} were performed by \citet{DzifcakovaMason08}. These authors used $n_\mathrm{max}=5$ and tested the validity of the approximation of the cross-section $\Omega$ for some of the \ion{Fe}{15} transitions. An overall precision better than 10\% was found. The approximation worked almost perfectly for the alowed transitions. Worse results were found for the forbidden transitions. It was also found that transitions with strong resonance contributions and a low ratio of the excitation energy to temperature can also be problematic. However, all the $\Omega$s for all transitions were reproduced to an accuracy better than 15\% \citep{DzifcakovaMason08}.

To supplement this analysis, we used $n_\mathrm{max}=7$ (Sect. \ref{Sect:3.3}) and tested the approximate method on two ions, \ion{Fe}{11} and \ion{Fe}{17}. We used the original atomic cross sections from \citet{DelZanna10a} for \ion{Fe}{11} and \citet{DelZanna11b} for \ion{Fe}{17}. These Maxwellian-averaged $\Upsilon_{ij}(T)$ are implemented in the CHIANTI database, version 7.1 \citep{Landi13}. Here, we compare our approximation based on these Maxwellian data in CHIANTI with the $\Upsilon_{ij}(T,\kappa)$ and \rotatebox[origin=c]{180}{$\Upsilon$}$_{ji}(T,\kappa)$ calculated directly from the $\Omega$s using the method of \citet{Dudik14b}.

Figures. \ref{Fig:Errors_fe11} and \ref{Fig:Errors_fe17} show several examples of the comparison of the direct calculation (hereafter, DC) with the approximate method for \ion{Fe}{11} (Fig. \ref{Fig:Errors_fe11}) and \ion{Fe}{17} (Fig. \ref{Fig:Errors_fe17}). The DC are denoted by squares and the approximate $\Upsilon_\kappa$ by the full lines. Left columns in these figures show typical worst cases for strong transitions. We see that the error of the approximation depends on $\kappa$ and $T$; it typically increases with decreasing $\kappa$. The worst cases are however still within 10\% even for the extreme value of $\kappa$\,=\,2 considered here. Typical cases are shown in the middle columns of Figs. \ref{Fig:Errors_fe11} and \ref{Fig:Errors_fe17}. Here, the approximations are valid to within a few per cent for all $\kappa$s. Finally, typical approximations for the weak transitions are shown in the right columns of Figs. \ref{Fig:Errors_fe11} and \ref{Fig:Errors_fe17}. We again find that the approximations are valid to within $\approx$10\% for all $\kappa$s.

Figure \ref{Fig:Scatterplots} contain scatterplots of the relative error $\Upsilon_\kappa$/$\Upsilon_\kappa,\mathrm{DC}$\,$-$1 plotted for each $\kappa$ at the peak of the corresponding relative ion abundance. These scatterplots contain 447 transitions in \ion{Fe}{11} and 1050 transitions in \ion{Fe}{17} that we were able to unambiguously indentify both in both the CHIANTI database and the atomic data themselves. The plots in Fig. \ref{Fig:Scatterplots} confirm that the approximate $\Upsilon_\kappa$ do not depart from the directly calculated one $\Upsilon_\kappa,\mathrm{DC}$ by more than 10\%. Typically, the relative errors increase with decreasing $\kappa$; smallest errors are found for the Maxwellian distribution. Strong transitions typically have higher accuracy than the weaker ones, in agreement with the results of \citet{DzifcakovaMason08}.

Finally we note that the approximation of $\Upsilon(\kappa,T)$ to within 10\% is considered satisfactory given the uncertainties in the atomic data themselves, which are typically of the same order of magnitude, and the uncertainties of the spline-fits of the Maxwellian $\Upsilon(T)$ contained in CHIANTI, which are typically $<5$\%.

%
\subsection{Dielectronic Satellite Lines}
\label{Sect:3.5}

The rate coefficient for the dielectronic excitation from level $i$ to level $j$ and for an arbitrary electron distribution funtion $f(E)$ can be expressed as \citep{Seely87}
\begin{equation}
	C^\mathrm{diel}=\left( \frac{2}{m_\mathrm{e} \Delta E_{ji}} \right)^{1/2}\frac{h^3g_j}{16\pi m_\mathrm{e} g_i}f(\Delta E_{ji})A_a,
	\label{Eq:C_diel}
\end{equation}
where $g_j$ and $g_i$ are statistical wieghts of double excited state and lower level, respectively; $A^\mathrm{a}$ is the autoionization (Auger) rate. The transition occurs at discrete energy $\Delta E_{ji}$ which corresponds to the energy difference between energy of states $j$ and $m$.
For the Maxwellian distribution, this equation leads to the well-known expression \citep[e.g.,][Eq. (4.19) therein]{Phillips08}
\begin{equation}
	C^\mathrm{diel}_\mathrm{Maxw} = \frac{h^3}{2(2\pi m_\mathrm{e} k_\mathrm{B}T)^{3/2}} \frac{g_j}{g_i} \mathrm{e}^{-\frac{\Delta E_{ji}}{k_\mathrm{B}T}}A^\mathrm{a} \,.
	\label{Eq:C_maxw}
\end{equation}
For the $\kappa$-distribution, we have
\begin{equation}
	C^\mathrm{diel}_\kappa = \frac{A_\kappa h^3}{2(2\pi m_\mathrm{e} k_\mathrm{B}T)^{3/2}} \frac{g_j}{g_i}
		       \frac{A^\mathrm{a}}{\left (1+\frac{\Delta E_{ji}}{(\kappa-1.5)k_\mathrm{B}T}\right)^{\kappa+1}}\,,
	\label{Eq:C_diel_kappa}
\end{equation}
which leads to
\begin{equation}
	C^\mathrm{diel}_\kappa = C^\mathrm{diel}_\mathrm{Maxw} \frac{A_\kappa \mathrm{e}^{\frac{\Delta E_{ji}}{k_\mathrm{B}T}}}{\left (1+\frac{\Delta E_{ji}}{(\kappa-1.5)k_\mathrm{B}T}\right)^{\kappa+1}}\,.
	\label{Eq:C_diel_kappa2}
\end{equation}
%

%
\section{The non-Maxwellian Continuum}
\label{Sect:4}

The continuum for the non-Maxwellian $\kappa$-distributions is treated here using the approach of \citet{Dudik12}. Contributions from the free-free and free-bound continua are considered. The two-photon continuum is not considered, as its emissivity for $\kappa$-distributions is not known, and its contribution is usually weak for the Maxwellian distribution  \citep{Young03,Phillips08} especially at higher densities. Nevertheless, at least for the Maxwellian distribution and a limited wavelength range, the two-photon continuum may not be a negligible contribution to the total continuum. We plan to implement it in the future. 

\subsection{The Free-Free Continuum}
\label{Sect:4.1}

The total emissivity of the free-free continuum arising due to electron-ion bremsstrahlung is given by \citep{Dudik11,Dudik12} 
   \begin{eqnarray}
	\nonumber \varepsilon_\mathrm{ff}(&&\lambda,\kappa,T)  = \frac{A_\kappa T^{1/2}}{\lambda^2} n_\mathrm{e} n_\mathrm{H}  \times \\
	&& \times \sum_{Z}  K_X(\kappa,T)A_X \int\limits_{0}^{\infty} \frac{g_\mathrm{ff}(y,w)}{\left(1+\frac{y+w}{\kappa-3/2}\right)^{\kappa+1}} \mathrm{d}y \,,
       \label{Eq:ff}
   \end{eqnarray}
where $A_X$ is the element abundance relative to hydrogen, $w = hc/\lambda k_\mathrm{B}T = {\cal E} / k_\mathrm{B}T$ is the scaled photon energy, the $g_\mathrm{ff}$ is the free-free Gaunt factor, and the  $K_X(\kappa,T)$ is a function of $\kappa$ and $T$ through the dependence on the ionization balance $n_k/n_X$ (see Sect. \ref{Sect:3.1})
   \begin{equation}
       K_Z(\kappa,T) = \frac{1}{4 \pi} \frac{32\pi}{3} \frac{e^6}{m_\mathrm{e} c^2} \sqrt{\frac{2\pi k_\mathrm{B}}{3 m_\mathrm{e}}} \sum_{k} k^2 \frac{n_k}{n_Z} \,,
       \label{Eq:ff_constant}
   \end{equation}
where $k$ is the ionization degree. The units of $\varepsilon_\mathrm{ff}$ are ergs\,cm$^{-3}$s$^{-1}$sr$^{-1}$\AA$^{-1}$.

The bremsstrahlung spectrum is strongly dependent on $\kappa$ mainly at short wavelengths \citep{Dudik12}, where the tail of the $\kappa$-distribution strongly enhances the bremsstrahlung emission. Near the wavelength where the $\varepsilon_\mathrm{ff}$ peaks for the Maxwellian distribution, the free-free emission drops with $\kappa$. At larger wavelengths it is enhanced again (see Figs. 2 and 3 in \cite{Dudik12}).

\subsection{Free-Bound Continuum}
\label{Sect:4.2}

The emissivity of the recombination processes resulting in $k$-times ionized ions of element $X$ with an electron on an excited level $j$ is for the $\kappa$-distributed incident electrons given by \citep{Dudik12}
   \begin{eqnarray}
       \nonumber \varepsilon_\mathrm{fb}(&& \lambda,\kappa,T) = \frac{1}{4\pi} \sqrt{\frac{2}{\pi}} \frac{{\cal E}^5}{hc^3 \left(m_\mathrm{e} k_\mathrm{B}T\right)^{3/2}}  n_\mathrm{e} n_\mathrm{H} \times \\
                    && \times \sum_{k,X}{ \frac{n_\mathrm{k+1}}{n_X} A_X \frac{g_j}{g_0} \sigma_j^\mathrm{bf} A_\kappa \frac{1}{\left(1 +\frac{{\cal E}-I_j}{\left(\kappa -3/2\right) k_\mathrm{B}T}\right)^{\kappa+1}}}\,,\hspace{0.7cm}
       \label{Eq:fb}
   \end{eqnarray}
where ${\cal E}$\,=\,$hc/\lambda$\,=\,$E+I_j$ is the photon energy, $I_j$ is the ionization potential from the level $j$ with statistical weight $g_j$, and $\sigma_j^\mathrm{bf}$ is the ionization cross-section from the level $j$. 

A conspicuous feature of the free-bound spectra for the $\kappa$-distributions are the greatly enhanced ionization edges \citep[see Fig. 5 in][]{Dudik12}. Generally, this increase comes from Eq. (\ref{Eq:fb}) through the increase of low-energy electrons in a $\kappa$-distribution with respect to the Maxwellian distribution at the same $T$. However, details also depend on the ionization equilibrium together with $T$ and $\kappa$ \citep{Dudik12}.

%
\section{The KAPPA Package}
\label{Sect:5}

The KAPPA package\footnote{http://kappa.asu.cas.cz} currently allows for calculation of the synthetic spectra for integer values of $\kappa$\,=\,2, 3, 4, 5, 7, 10, 15, 25, and 33, for which the ionization equilibria are tabulated. These values should cover the parameter space with sufficient density. The database and software for the KAPPA package is based on the IDL version of the freely available CHIANTI database and software\footnote{www.chiantidatabase.org} \citep{Dere97,Landi13}. The routines and database of the KAPPA package are contained in a standalone folder. It cannot be contained within the CHIANTI itself in order to prevent its automatic removal by CHIANTI updates. The path to the folder can be set by an IDL system variable in the \textit{idl\_startup.pro} file

$defsysv,\;'!data\_ pth',\;'path\; to\; package' $.

The KAPPA folder contains the modified CHIANTI routines for calculation of spectra for the $\kappa$-distributions, with the ``data\_k'' subfolder having the same structure as the CHIANTI's ``dbase`` subfolder. The modified CHIANTI routines follow the original CHIANTI routines as closely as possible. Their names end with an extra ''\textit{\_k}`` before the \textit{.pro} extension. The calling parameters of these routines are kept the same, except that the first parameter is always the value of $\kappa$.

The subdirectories within the database contain datas for ionization and recombination rates (Sect. \ref{Sect:5.1.2}) together with the tabulated $\Upsilon_{ij}(T,\kappa)$ and \rotatebox[origin=c]{180}{$\Upsilon$}$_{ji}(T,\kappa)$ files in the ASCII format. Previous versions of the modification corresponding to CHIANTI v5.2 \citep{Dzifcakova06b} contained the coefficients for the approximation of $\Omega$. Then, the calculations for the $\kappa$-distributions were aproximately ten times longer compared to the CHIANTI for the Maxwellian distribution. Therefore, we decided to pre-calculate $\Upsilon(\kappa,T)$ for a grid temperatures and $\kappa$. These pre-calculated values of $\Upsilon(\kappa,T)$ are contained in files names according to the ion and the value of $\kappa$ with the extension \textit{.ups}, e.g., \textit{c\_5\_k2.ups} for \ion{C}{5} and $\kappa$\,=\,2. IDL savefiles containing the $\Upsilon_{ij}(T,\kappa)$ and \rotatebox[origin=c]{180}{$\Upsilon$}$_{ji}(T,\kappa)$ are also provided.

At present, the KAPPA package fully corresponds to the atomic data contained in the CHIANTI version 7.1. Similarly, the routines provided in the KAPPA package are based on CHIANTI 7.1 routines, with the exception of routines for free-free continuum (see Sect. \ref{Sect:5.3.1}).


\subsection{Ionization Equilibrium}
\label{Sect:5.1}

\subsubsection{Ionization Equilibrium Files}
\label{Sect:5.1.1}

The ionization equilibrium \textit{.ioneq} and similar files were originally provided by \citet{Dzifcakova13}. A minor software bug in the calculation of radiative recombination rates for the $\kappa$-distributions was found and corrected. This problem affected the ionization equilibria at log$(T/K$)\,$<$\,5 with the error being much smaller than the effect of $\kappa$-distributions on the ionization equilibrium.

These \textit{.ioneq} files are produced in the same format as the original \textit{chianti.ioneq} file. Therefore, these can be read by the CHIANTI routine \textit{read\_ioneq.pro} directly. The names of these files are \textit{kappa\_02.ioneq} and similar, where the numbers give the integer value of $\kappa$. For more details on the \textit{.ioneq} file format, see \citet{Dzifcakova13}, Appendix A therein. 

\subsubsection{Ionization and Recombination Rates}
\label{Sect:5.1.2}

In addition to the ionization equilibria, total ionization and recombination rates are provided for each ion and a range of temperatures. Here, the total ionization rate is a sum of the direct collisional ionization rate and the autoionization rate. Similarly, the total recombination rate is given by the sum of the radiative recombination rate and the total dielectronic recombination rate \citep{Dzifcakova13}. These rates are stored in the respective database folder for each ion, e.g., the \textit{dbase/c/c\_5/c\_5\_k25.tionizr} is the total ionization rate file for \ion{C}{5} and $\kappa$\,=\,25. The file format is ASCII. The total recombination rate file has the same name except the \textit{.trecombr} extension. The routines \textit{read\_rate\_ioniz\_k.pro} and \textit{read\_rate\_recomb\_k.pro} are provided for reading these files.

%
\begin{table*}[!ht]
\begin{center}
\caption{List of routines within the KAPPA package.
\label{Table:1}}
\begin{tabular}{ll}
\tableline
\tableline
Routine name		& Function \\
\tableline
kappa.pro		& interactive widget for calculation of synthetic spectra, based on ch\_ss.pro				\\ 
ch\_synthetic\_k.pro	& calculates line intensities as a function of $\kappa$, $n_\mathrm{e}$ and $T$				\\
descale\_diel\_k.pro	& converts $\Upsilon_{ij}(T,\kappa)$  and \rotatebox[origin=c]{180}{$\Upsilon$}$_{ji}(T,\kappa)$from the scaled domain 	\\
			& for dielectronic satellite lines and performs correction in Eq. (\ref{Eq:C_diel_kappa2}) 		\\
emiss\_calc\_k.pro	& calculates $hc/\lambda$~$A_{ji} n(X_j^{+k})$								\\
freebound\_ion\_k.pro	& calculates the free-bound continuum arising from a single ion						\\
freebound\_k.pro	& calculates the free-bound continuum									\\
freefree\_k.pro		& free-free continuum interpolated from pre-calculated data						\\
freefree\_k\_integral.pro & calculates the free-free continuum directly								\\
isothermal\_k.pro	& calculates isothermal spectra as a function of $\lambda$						\\
make\_kappa\_spec\_k.pro & routine for calculating the synthetic spectra							\\
plot\_populations\_k.pro & calculates and plots relative level populations							\\
pop\_solver\_k.pro	& calculates the relative level population								\\
read\_ff\_k.pro 	& reads the pre-calculated free-free continuum as a function of $Z$ and $T$				\\
read\_rate\_ioniz\_k.pro & reads the total ionization and recombination rates							\\
read\_rate\_recomb\_k.pro & reads the total ionization and recombination rates							\\
ups\_kappa\_interp.pro	& routine for interpolating the $\Upsilon_{ij}(T,\kappa)$  and \rotatebox[origin=c]{180}{$\Upsilon$}$_{ji}(T,\kappa)$ \\
\tableline
\tableline
\end{tabular}
\end{center}
\end{table*}
%
%
\subsection{Tools for calculation of line spectra}
\label{Sect:5.2}

The KAPPA package provides several routines for calculation of line intensities. These are listed in Table \ref{Table:1}. As already mentioned, these routines are based on CHIANTI routines, version 7.1. They can be used in the same manner as the CHIANTI routines, with the exception that the value of $\kappa$ is always the first parameter.

The most important of these routines is the \textit{pop\_solver\_k.pro} routine that calculates the relative level population based on the distribution-averaged collision strengths $\Upsilon_{ij}(T,\kappa)$ and \rotatebox[origin=c]{180}{$\Upsilon$}$_{ji}(T,\kappa)$ calculated using the method described in Sect. \ref{Sect:3.3}. Other routines for calculating line intensities (Table \ref{Table:1}) rely on this routine. Examples of synthetic spectra calculated for $\kappa$\,=\,2 and their comparison to the Maxwellian spectra at the same $T$ are given in Sect. \ref{Sect:6.1}.

We note here that the method for calculation of the collisional electron excitation and deexcitation rates described in Sect. \ref{Sect:3.3} cannot be applied to the collisional excitation by protons due to unavailability of the proton excitation cross sections. The proton excitation is typically negligible, but may be important for some transitions. In the synthesis of line spectra, the proton excitation rate for $\kappa$-distribution is assumed to be the same as for the Maxwellian distribution at the same temperature. It is currently unknown if this assumption is justified. Because of this, we advocate caution in using such lines for e.g. diagnostics of $\kappa$ from observations.

An interactive widget for calculating the synthetic spectra is provided in the \textit{kappa.pro} routine, based on the CHIANTI's \textit{ch\_ss.pro}. The value of $\kappa$ is selected by the choice of the ionization equilibrium. Subsequently, the excitation and line intensities are calculated for the same value of $\kappa$. All other functionality of the \textit{ch\_ss.pro} routine is retained.


\subsection{Tools for calculation of continuum}
\label{Sect:5.3}

\subsubsection{Free-free continuum}
\label{Sect:5.3.1}

The CHIANTI database relies on the approximations to the Maxwellian bremsstrahlung calculated by \citet{Itoh00} and \citet{Sutherland98} and incorporated in the \textit{freefree.pro} routine together with the \textit{itoh.pro} and \textit{sutherland.pro} routines. This approach cannot be followed in the modified CHIANTI, since no fitting formulae exist for the free-free continuum for $\kappa$-distributions. Instead, we provide two options to calculate the free-free continuum for $\kappa$-distributions:
\begin{enumerate}
 \item Direct integration using the Eqs. (\ref{Eq:ff}) and (\ref{Eq:ff_constant}). This approach is implemented in the \textit{freefree\_k\_integral.pro} routine and requires the \textit{data\_k/continuum/gffew.dat} file containing the scaled $g_\mathrm{ff}(y,w)$ values provided by \citet{Sutherland98}. These $g_\mathrm{ff}$ values are then de-scaled and numerically integrated. Since the scaling depends on the ionization energy (and thus on the ionization stage $k$ and the proton number $Z$), it has to be carried out for each ion separately \citep{Dudik12}. Therefore, the direct integration using the \textit{freefree\_k\_integral.pro} is time-consuming and impractical. We note that, in practice, restricting the integration to elements with relative abundance of $A_Z$\,$\geqq$\,10$^{-6}$ introduces a relative error smaller than $10^{-4}$ and speeds up the calculations by a factor of $\approx$2. Note that the Maxwellian-integrated $g_\mathrm{ff}(y,w)$ are a part of the CHIANTI database.

 \item To overcome the long calculation times, the free-free continuum has been pre-calculated as a function of $Z$ for 101 logaritmically spaced temperatures spanning log$(T/$K)\,=\,$\left<4,9\right>$ with a step of $\Delta$log$(T/$K)\,=\,0.05, together with 29 logarithmically spaced points in $\lambda$\,=\,0.1\AA\,--3\,$\times$10$^{4}$\AA~with a step of log$(\lambda/$\AA)\,=\,0.2. These calculations are contained in the \textit{data\_k/continuum/ff\_kappa\_02.dat} file and analogous files for other values of $\kappa$, each file for a single value of $\kappa$. The files are in the ASCII format. The routine \textit{freefree\_k.pro} reads these files, folds and sums them over the abundances to produce a semi-final free-free continuum. The final free-free continuum is then calculated for the user-input ranges of $T$ and $\lambda$ within in the ranges specified above. This is achieved first by linearly interpolating in log$(T/$K) and then by spline-interpolating in log$(\lambda/$\AA). In this way, an accuracy of few per cent is achieved in the $\lambda$\,=\,1\AA\,--2\,$\times$10$^{4}$\AA~range, with the calculation time of few seconds. We note that this method should not be used for calculation of free-free continua below 1\AA~and 2\,$\times$10$^{4}$\AA, where the spline interpolation results in errors of several $\times$\,10\,\% or more.
\end{enumerate}

\subsubsection{Free-bound continuum}
\label{Sect:5.3.2}

Using Eq. (\ref{Eq:fb}), the free-bound continuum is straightforward to calculate. The \textit{freebound\_k.pro} and \textit{freebound\_ion\_k.pro} routines can be used in the same manner as the CHIANTI's \textit{freebound.pro} and \textit{freebound\_ion.pro} routines. The only change is that these routines require a value of $\kappa$ as an extra input. Ionization equilibrium files (Sect. \ref{Sect:5.1}) are read together with the cross-sections. The speed of the calculation is the same as using the original CHIANTI.

%
   \begin{figure*}[!t]
	\centering
	\includegraphics[width=8.6cm]{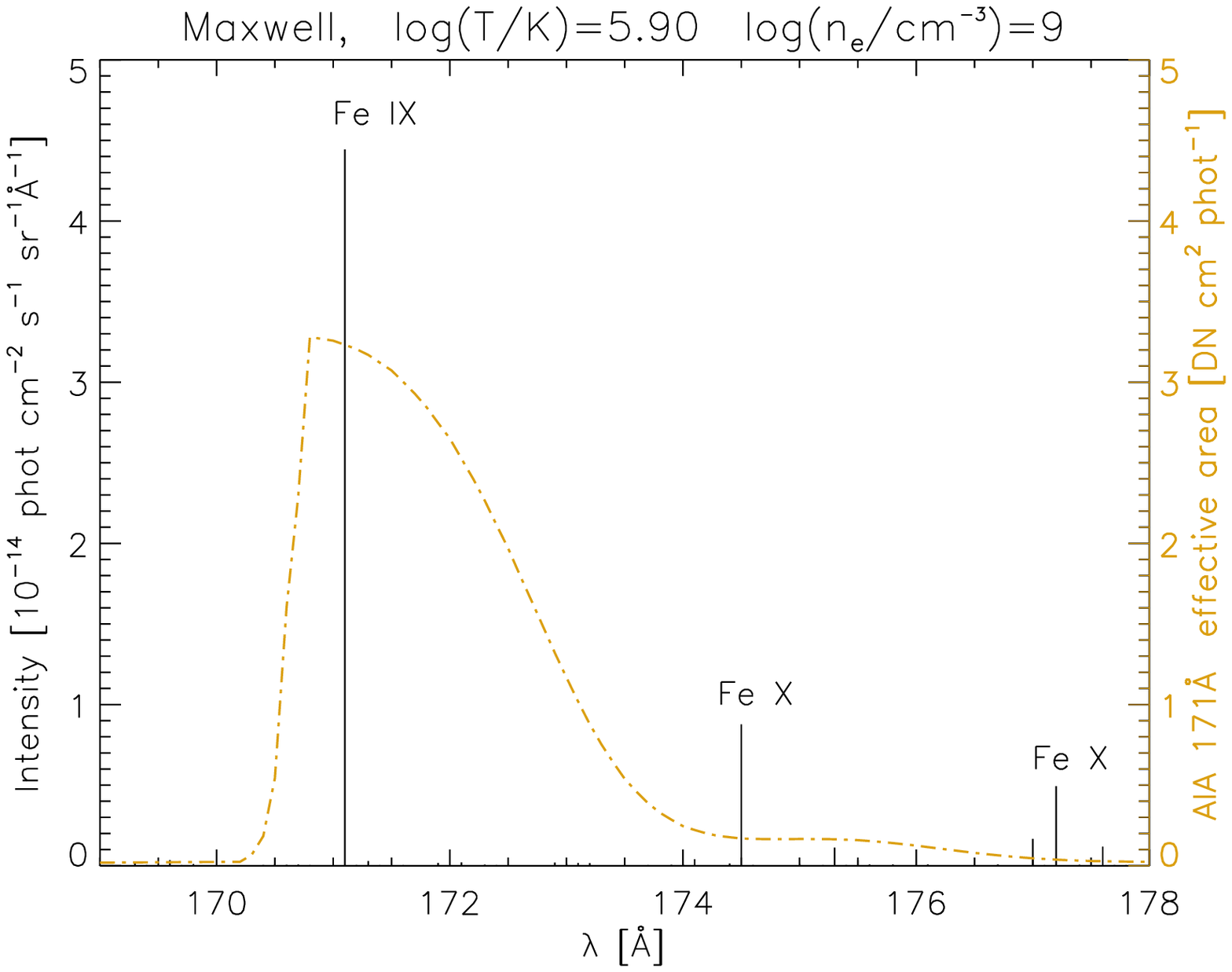}
	\includegraphics[width=8.6cm]{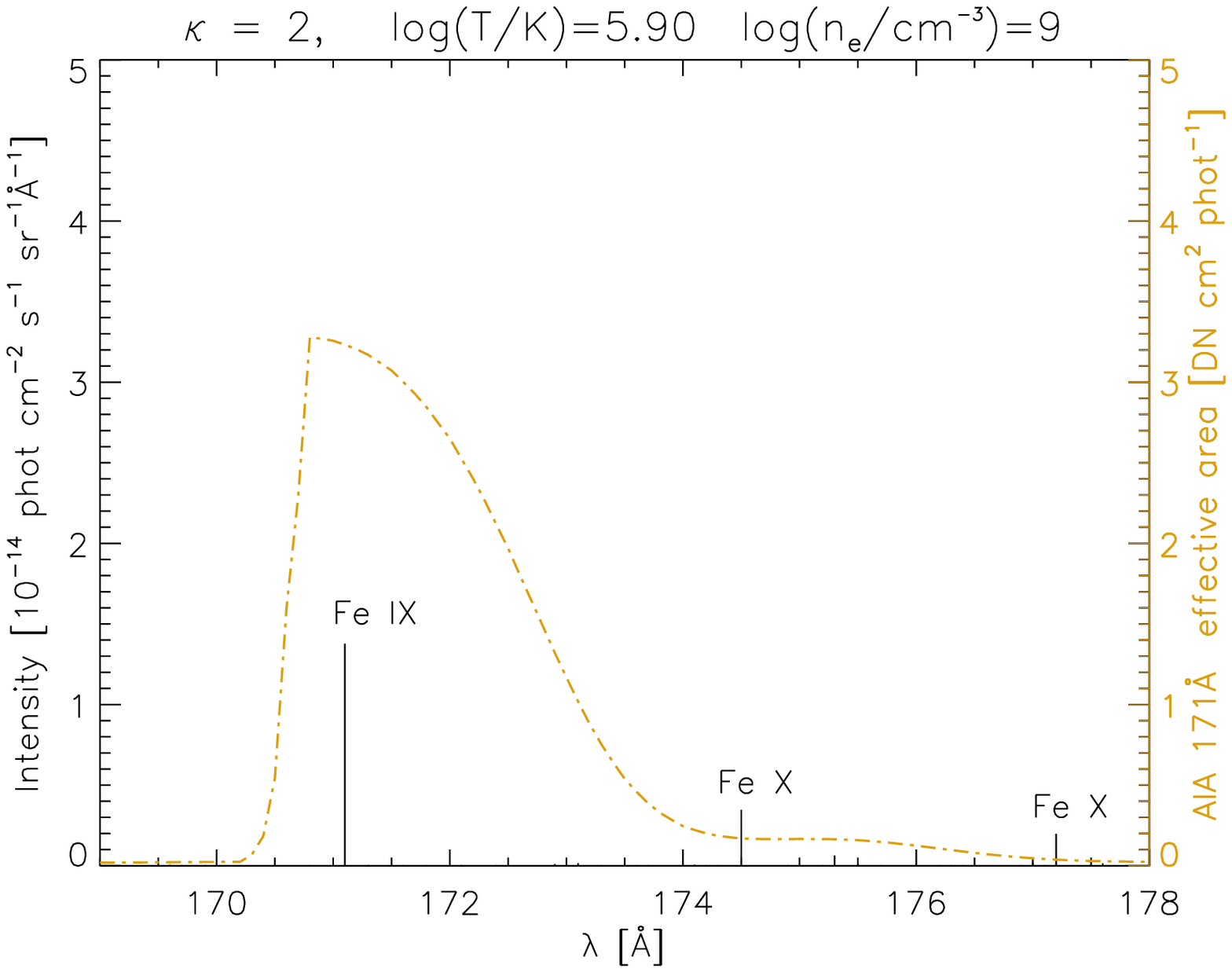}
	\includegraphics[width=8.6cm]{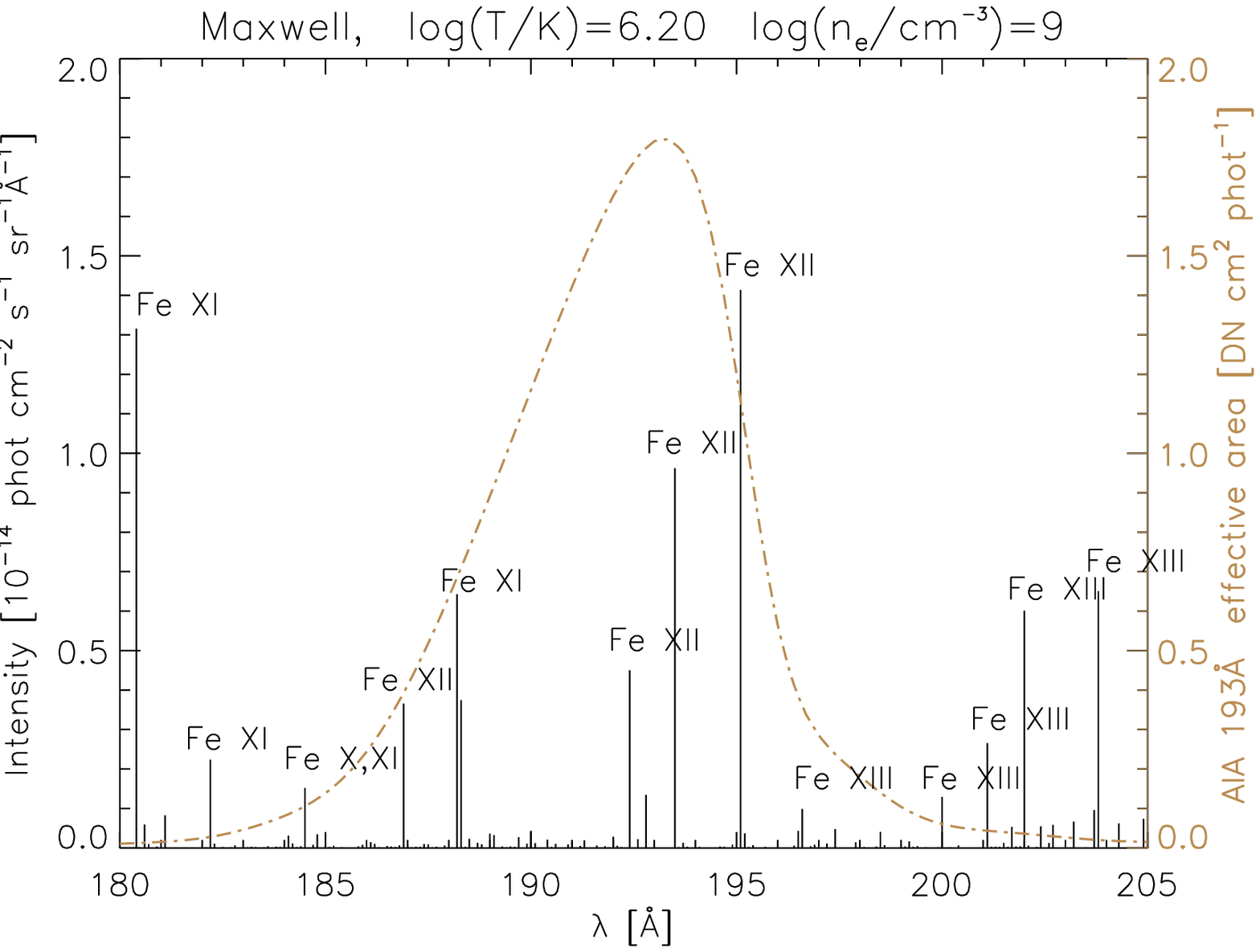}
	\includegraphics[width=8.6cm]{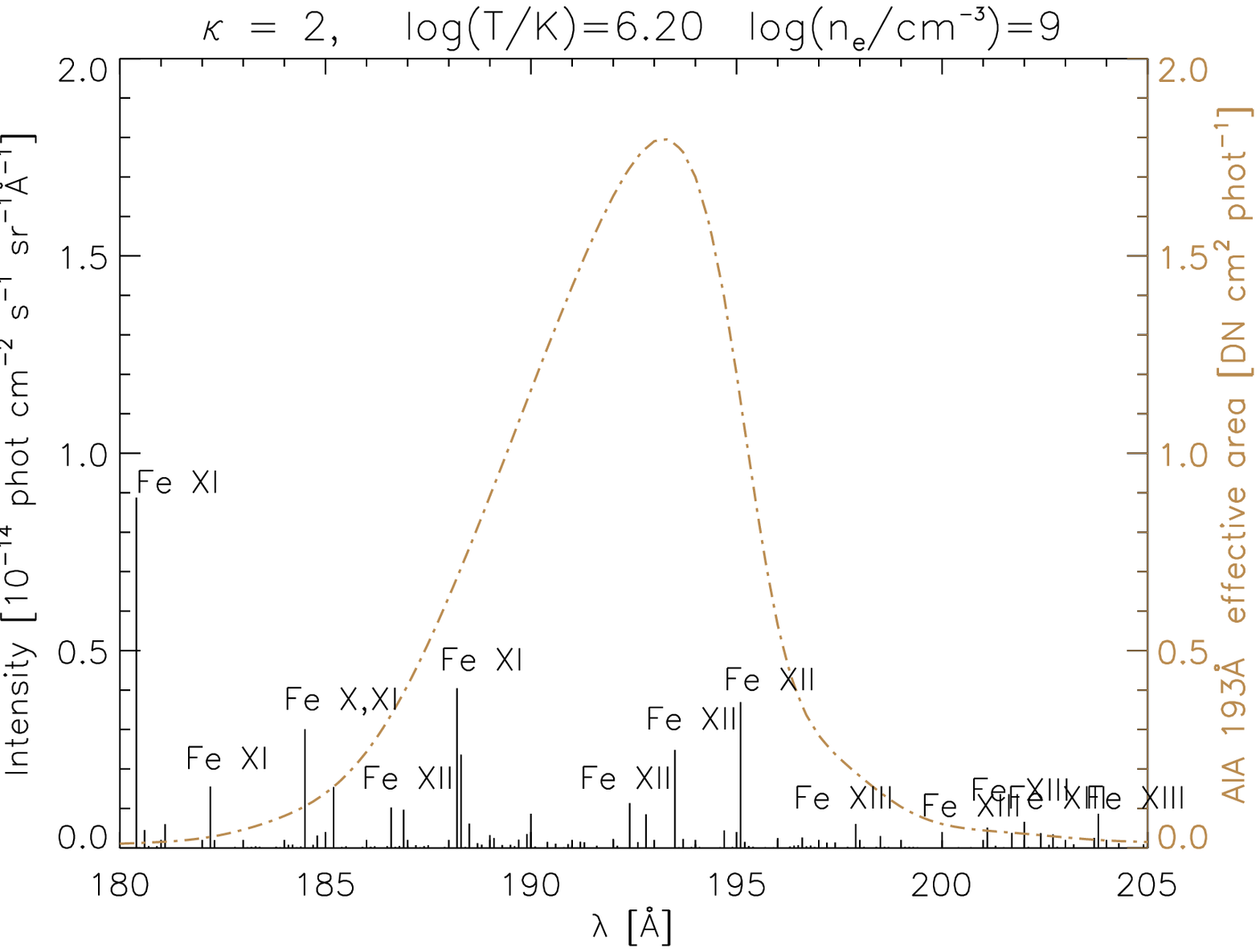}
	\caption{Example isothermal spectra at log$(T/K$)\,=\,5.9 near the peak of the AIA 171\AA~wavelength response (\textit{top}) and at log$(T/$K)\,=\,6.2 near the peak of the AIA 193\AA~filter (\textit{bottom}). The electron density assumed is log$(n_\mathrm{e}$/cm$^{-3}$)\,=\,9.0. \\ A color version of this image is available in the online journal.}
       \label{Fig:AIA_spectra}
   \end{figure*}
%
%
   \begin{figure*}[!t]
	\centering
	\includegraphics[width=8.6cm]{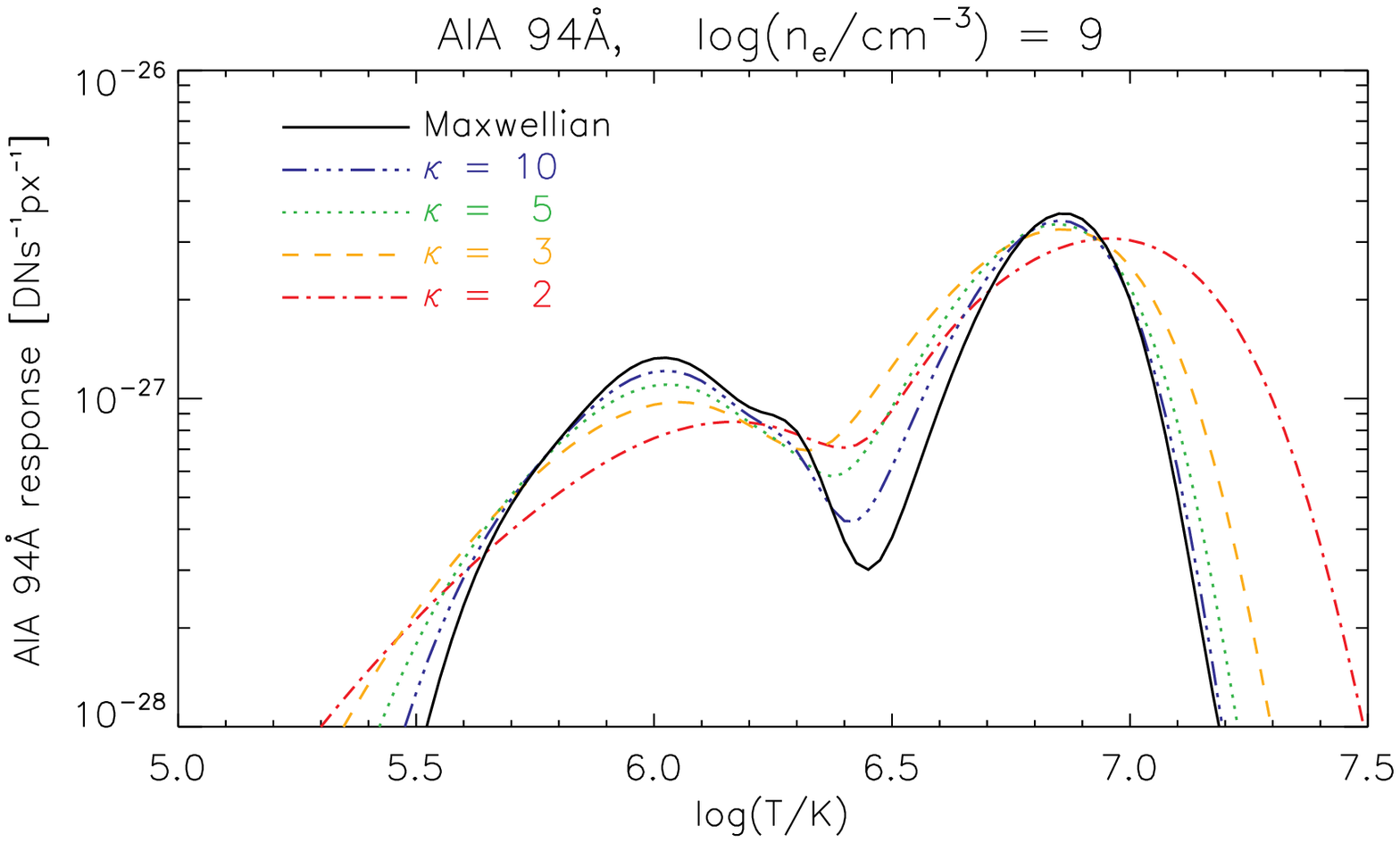}
	\includegraphics[width=8.6cm]{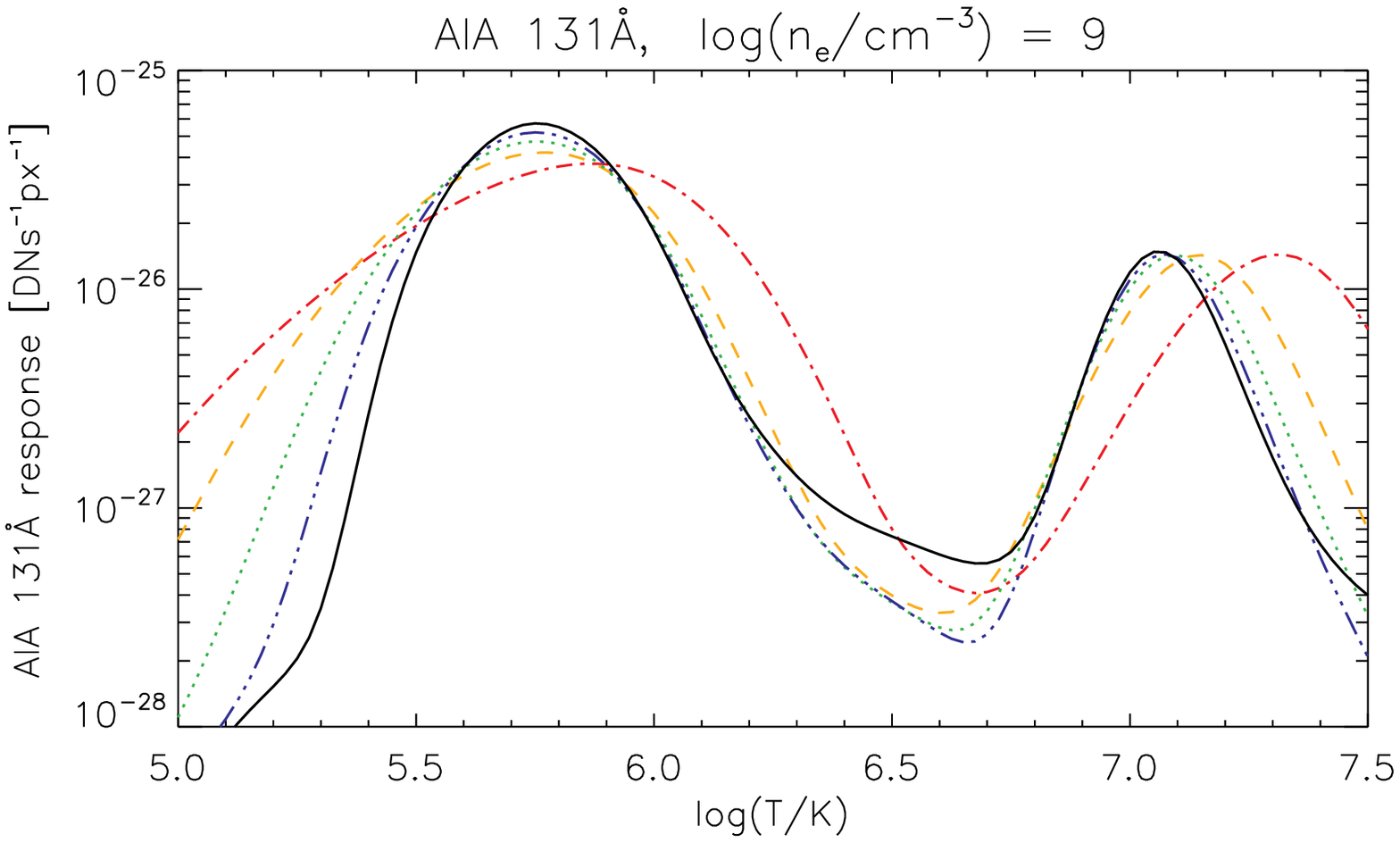}
	\includegraphics[width=8.6cm]{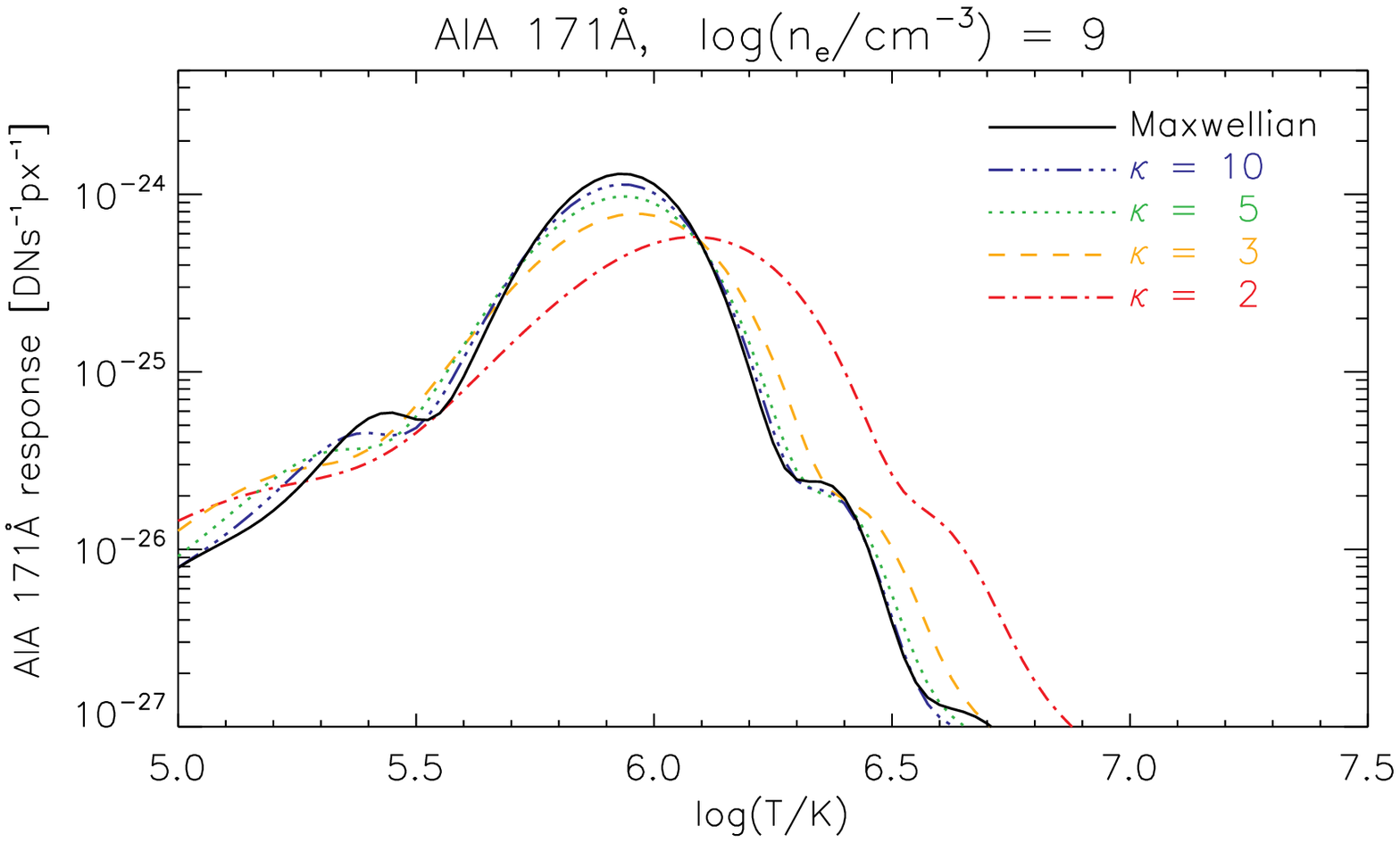}
	\includegraphics[width=8.6cm]{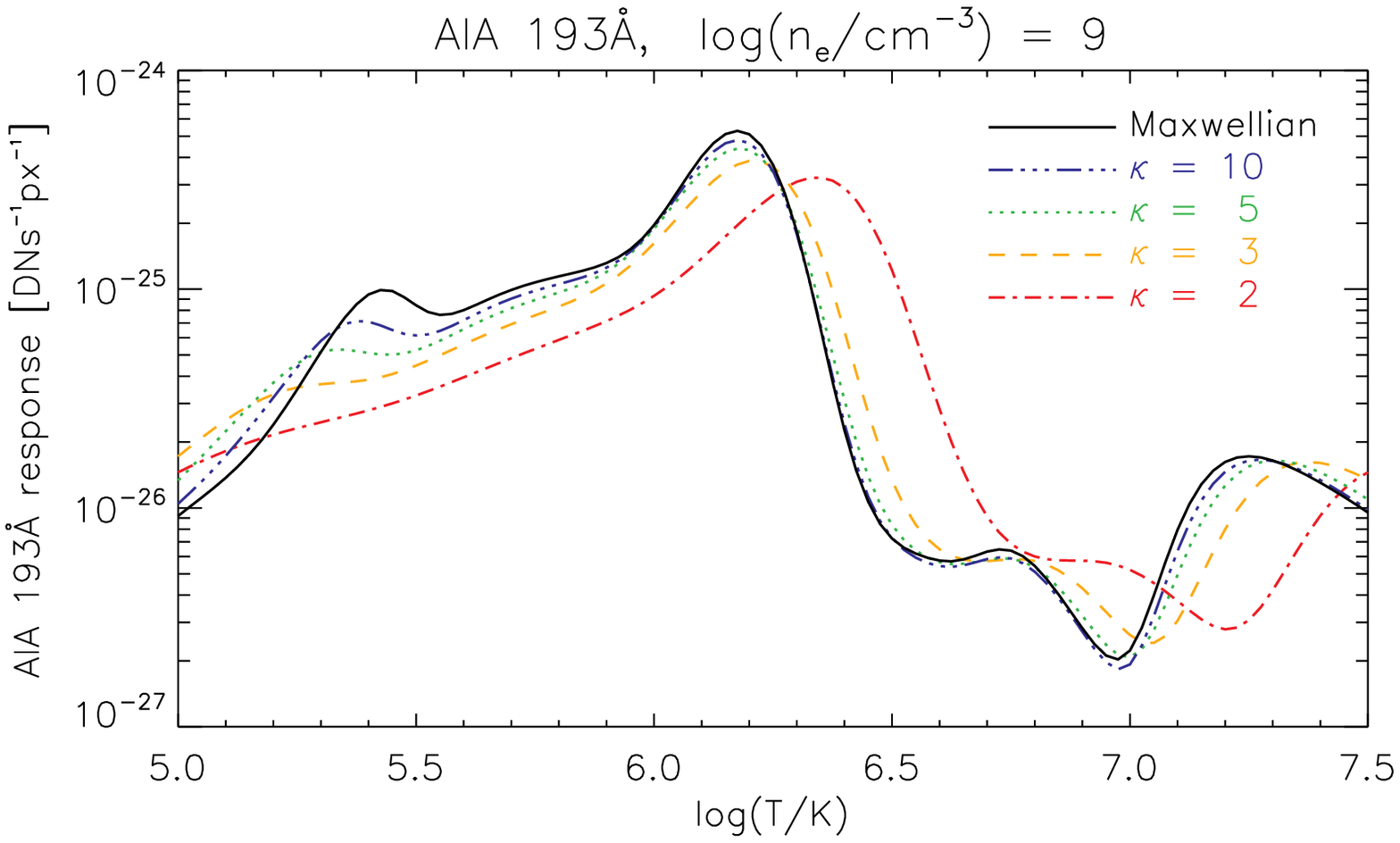}
	\includegraphics[width=8.6cm]{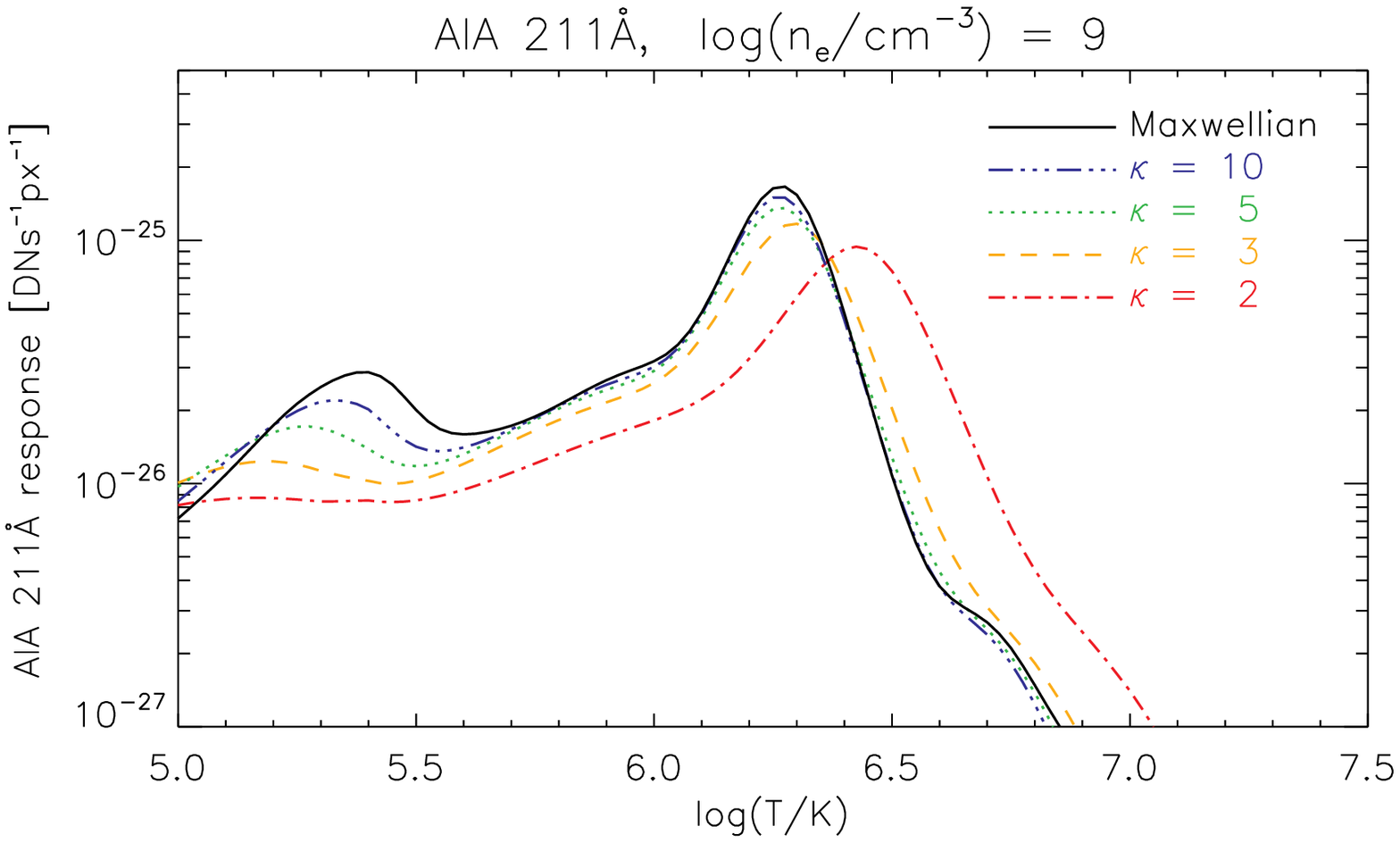}
	\includegraphics[width=8.6cm]{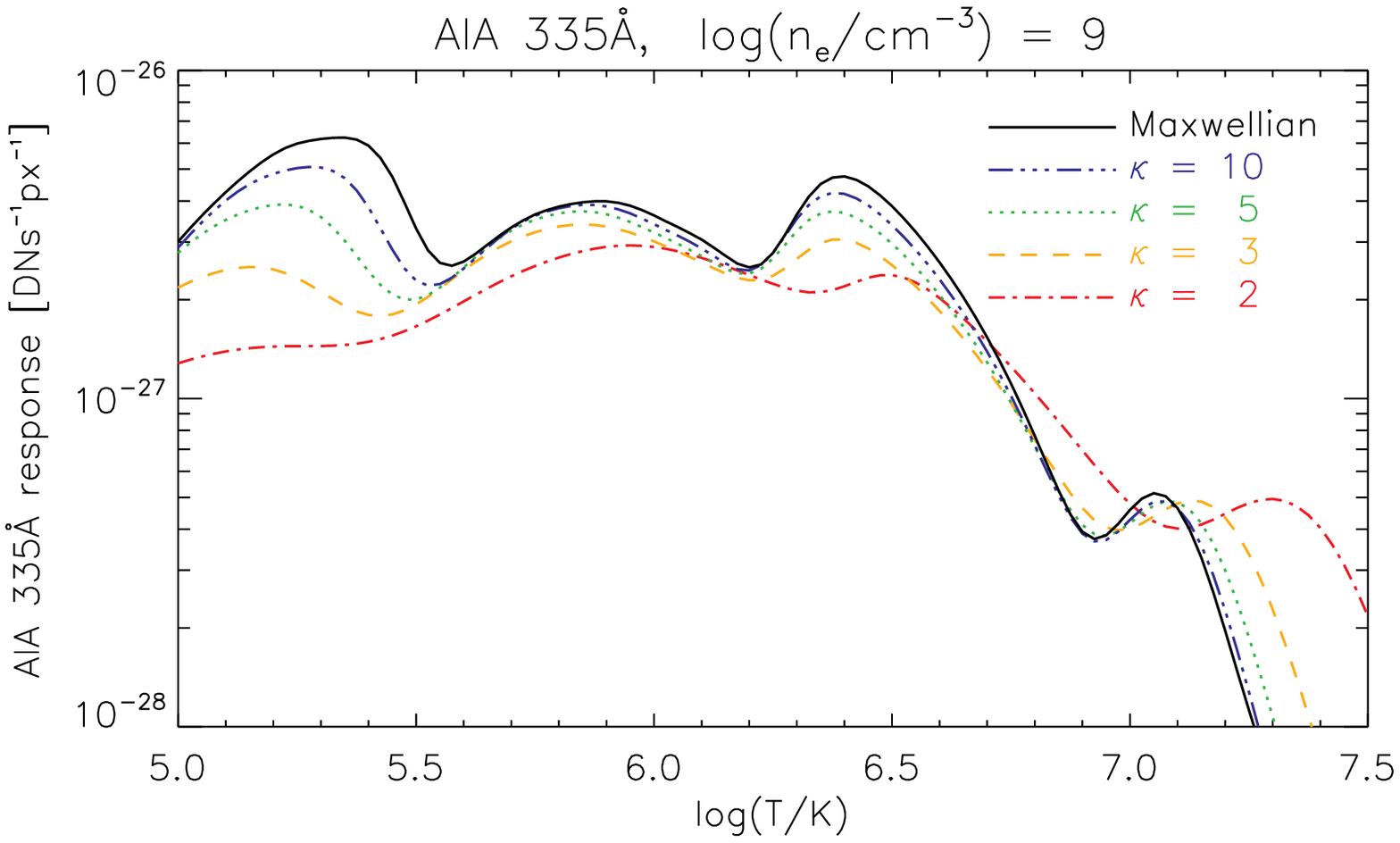}
	\caption{Responses of the AIA EUV filters for the $\kappa$-distributions. Individual colors and linestyles stand for different values of $\kappa$, as indicated. \\ A color version of this image is available in the online journal.}
       \label{Fig:AIA_responses}
   \end{figure*}
%
%
\section{Synthetic Spectra and \textit{SDO}/AIA Responses}
\label{Sect:6}

In this section, we provide some examples of the calculated spectra that are of interest to the physics of the solar corona. Note that the behaviour of individual lines observed by the \textit{Hinode}/EIS spectrometer \citep{Culhane07} and the possible observational diagnostics with and without the effect of the ionization equilibrium are described elsewhere \citep{Dzifcakova10,Mackovjak13,Dzifcakova13,Dudik14b}, as is the application for DEM diagnostics \citep{Mackovjak14}.

\subsection{Synthetic EUV Spectra}
\label{Sect:6.1}

An example of the synthetic isothermal spectra calculated using the \textit{isothermal\_k.pro} for the Maxwellian and $\kappa$\,=\,2 are shown in Fig. \ref{Fig:AIA_spectra}. These examples show synthetic line and continuum spectra within the AIA 171\AA~and 193\AA~channels. The spectra are calculated for the electron density of $n_\mathrm{e}$\,=\,10$^9$\,cm$^{-3}$ and temperatures of log$(T/K$)\,=\,5.9 for AIA 171\AA~and 6.2 for AIA 193\AA~channels, respectively. Note that the temperature is kept the same for both distributions shown. These temperatures correspond to the maximum of the relative ion abundance of \ion{Fe}{9} and \ion{Fe}{12} under the Maxwellian distribution, respectively (see Fig. \ref{Fig:Ioneq}). The line intensities for $\kappa$\,=\,2 are decreased by a factor of several compared to the Maxwellian distribution. This is mainly an effect of the ionization equilibrium, with the maximum of the relative ion abundance for $\kappa$\,=\,2 being shifted to higher log$(T/K$) (Fig. \ref{Fig:Ioneq}, \textit{bottom}). Note that the intensities of the continuum is several orders of magnitude smaller than the line intensities; therefore, the continuum is not visible in the linear scale on Fig. \ref{Fig:AIA_spectra}.

From Fig. \ref{Fig:AIA_spectra} we see that at log$(T/K$)\,=\,5.9, the AIA 171\AA~channel is dominated by \ion{Fe}{9} independently of the value of $\kappa$. Contrary to that, the situation for the AIA 193\AA~channel and log$(T/K$)\,=\,6.2 is more complex. This filter is dominated by \ion{Fe}{12} transitions between energy levels 1--30 at 192.394\AA, 1--29 at 193.509\AA~and 1--27 at 195.119\AA~\citep[][Table B.4]{Dudik14b}. However, contributions from the 1--38 and 1--37 transitions in \ion{Fe}{11} at 188.216\AA~and 188.299\AA~are present as well. The relative contribution of these transitions to the total filter response to emission at log$(T/K$)\,=\,6.2 increases from 4.4\% for the Maxwellian distribution to 6.5\% for $\kappa$\,=\,2. This is because this temperature is closer to the ionization peak of \ion{Fe}{11} than \ion{Fe}{12} for $\kappa$\,=\,2 (Fig. \ref{Fig:Ioneq}).

%
\subsection{AIA responses for the $\kappa$-distributions}
\label{Sect:6.2}

As an example of the usage of the synthetic line and continuum spectra we calculated the responses of the Atmospheric Imaging Assembly \citep[AIA][]{Boerner12,Lemen12} onboard the \textit{Solar Dynamics Observatory (SDO)} \citep{Pesnell12} for the $\kappa$-distributions. Note that even though the continuum intensities are weak compared to the line intensities (Sect. \ref{Sect:6.1}) at a particular wavelength, the continuum is a significant contributor to some of the AIA bands \citep{ODwyer10,DelZanna13b}. This is because the filter response to plasma emission is given by the wavelength integral of the filter and instrument transmissivity times the emitted spectrum \citep[e.g., Eq. (6) in][]{Dudik09}.

The \textit{SDO}/AIA responses calculated for $\kappa$-distributions and log($n_\mathrm{e}/$cm$^{-3}$)\,=\,9 are shown in Fig. \ref{Fig:AIA_responses}. The peaks of the responses are typically flatter and wider, and can be shifted to higher log$(T/$K) for low $\kappa$. This behaviour is typical since it is given mainly by the ionization equilibrium (Sect. \ref{Sect:3.1}). It has been reported for the TRACE filter responses by \citet{Dudik09} and for \textit{Hinode}/XRT responses by \citet{Dzifcakova12}, where the AIA responses were also calculated for an earlier set of atomic data corresponding to CHIANTI v5.2 \citep{Landi06}. The AIA responses calculated here represent a significant improvement over the \citet{Dzifcakova12} ones due to advances in the atomic data for AIA bands \citep{DelZanna13b}.

We note that some of the secondary maxima, such as those at log$(T/$K)\,$\approx$\,5.4 for AIA 171\AA, 193\AA, 211\AA,~and 335\AA~ disappear for low $\kappa$. This is again mostly because of the wider ionization peaks and their relative contributions to individual filter responses \citep{Dudik09}, which gradually smooth out these secondary maxima with decreasing $\kappa$. We also note that the contribution from \ion{Fe}{10} and \ion{Fe}{14} to the AIA 94\AA~response \citep[][Fig. 3 therein]{DelZanna13b} form a single, smooth secondary peak of the response for $\kappa$\,$\lesssim$\,5.

These AIA responses for $\kappa$-distributions can be used to obtain the DEMs using the regularized DEM inversion developed by \citet{Hannah12} and \citet{Hannah13}, as well as for diagnostics of the distribution from combination of imaging and spectroscopic observations (Dud\'ik et al. 2014, in preparation).

%
\section{Summary}
\label{Sect:7}

We have developed tools for calculation of synthetic optically thin line and continuum spectra arising from collisionally dominated astrophysical plasmas characterized by a $\kappa$-distribution. These tools constitute the KAPPA package, which is based on the freely available CHIANTI database and software. At present, the KAPPA package can handle only values of $\kappa$\,=\,2, 3, 4, 5, 7, 10, 15, 25, and 33, which should provide sufficient coverage for most spectroscopic purposes. Ionization and recombination rates are provided together with ionization equilibrium calculations. Approximations to the distribution-averaged collision strengths are provided. These are based on the reverse-engineered collision strengths obtained from the Maxwellian-averaged collision strengths available within CHIANTI. This is done for all transitions in all ions available within CHIANTI, version 7.1. We have tested the validity of this approximate method by comparison with the directly integrated collision strengths. For temperatures typical of the formation of individual ions, typical errors of less than 5\% were found. It was also found that the errors are \textit{always} less than 10\%. The errors are typically of the order of a few per cent for strong transitions, but the precision decreases for weaker transitions and low values of $\kappa$. Considering the uncertainties in the atomic data calculations themselves, these errors are considered acceptable.

Several routines for calculation of the synthetic line spectra, free-free and free-bound continua are provided. These routines are based on the CHIANTI routines and can be used in the same manner, except that the first input parameter is always the value of $\kappa$. The calculation of the free-free continuum is based on interpolation from pre-calculated values; however, an option of direct integration of the free-free gaunt factors is also provided. We aim to keep the database updated to reflect the newer releases of CHIANTI.

\acknowledgements
The authors thank G. Del Zanna, P. R. Young and H. E. Mason for useful discussions. The collision strengths $\Omega$ used to validate the approximate method in Sect. \ref{Sect:3.4} were provided by G. Del Zanna and are gratefully acknowledged. EDZ, PK, FF and AK acknowledges the support by Grant Agency of the Czech Republic, Grant No. P209/12/1652. JD acknowledges support from the Royal Society via the Newton Fellowships Programme. The authors also acknowledge the support from the International Space Science Institute through its International Teams program. CHIANTI is a collaborative project involving the NRL (USA), RAL (UK), MSSL (UK), the Universities of Florence (Italy) and Cambridge (UK), and George Mason University (USA). It is a great spectroscopic database and software and the authors are very grateful for its existence and availability.

\bibliographystyle{apj}
\bibliography{kappa}

\begin{thebibliography}{95}
\expandafter\ifx\csname natexlab\endcsname\relax\def\natexlab#1{#1}\fi

\bibitem[{{Abramowitz} \& {Stegun}(1965)}]{Abramowitz65}
{Abramowitz}, M., \& {Stegun}, I.~A. 1965, {Handbook of mathematical functions
  with formulas, graphs, and mathematical tables}

\bibitem[{{Anderson} {et~al.}(1996){Anderson}, {Raymond}, \& {van
  Ballegooijen}}]{Anderson96}
{Anderson}, S.~W., {Raymond}, J.~C., \& {van Ballegooijen}, A. 1996, \apj, 457,
  939

\bibitem[{{Battaglia} \& {Kontar}(2013)}]{Battaglia13}
{Battaglia}, M., \& {Kontar}, E.~P. 2013, \apj, 779, 107

\bibitem[{{Bian} {et~al.}(2014){Bian}, {Emslie}, {Stackhouse}, \&
  {Kontar}}]{Bian14}
{Bian}, N.~H., {Emslie}, A.~G., {Stackhouse}, D.~J., \& {Kontar}, E.~P. 2014,
  ArXiv e-prints

\bibitem[{{Binette} {et~al.}(2012){Binette}, {Matadamas}, {H{\"a}gele},
  {Nicholls}, {Magris C.}, {Pe{\~n}a-Guerrero}, {Morisset}, \&
  {Rodr{\'{\i}}guez-Gonz{\'a}lez}}]{Binette12}
{Binette}, L., {Matadamas}, R., {H{\"a}gele}, G.~F., {et~al.} 2012, \aap, 547,
  A29

\bibitem[{{Boerner} {et~al.}(2012){Boerner}, {Edwards}, {Lemen}, {Rausch},
  {Schrijver}, {Shine}, {Shing}, {Stern}, {Tarbell}, {Title}, {Wolfson},
  {Soufli}, {Spiller}, {Gullikson}, {McKenzie}, {Windt}, {Golub}, {Podgorski},
  {Testa}, \& {Weber}}]{Boerner12}
{Boerner}, P., {Edwards}, C., {Lemen}, J., {et~al.} 2012, \solphys, 275, 41

\bibitem[{{Bradshaw}(2009)}]{Bradshaw09}
{Bradshaw}, S.~J. 2009, \aap, 502, 409

\bibitem[{{Bradshaw} {et~al.}(2004){Bradshaw}, {Del Zanna}, \&
  {Mason}}]{Bradshaw04}
{Bradshaw}, S.~J., {Del Zanna}, G., \& {Mason}, H.~E. 2004, \aap, 425, 287

\bibitem[{{Bradshaw} \& {Mason}(2003)}]{Bradshaw03}
{Bradshaw}, S.~J., \& {Mason}, H.~E. 2003, \aap, 401, 699

\bibitem[{{Bradshaw} \& {Raymond}(2013)}]{Bradshaw13}
{Bradshaw}, S.~J., \& {Raymond}, J. 2013, \ssr, 178, 271

\bibitem[{{Bryans}(2006)}]{Bryans06}
{Bryans}, P. 2006, {On the spectral emission of non-Maxwellian plasmas, Ph.D.
  thesis} (University of Strathclyde)

\bibitem[{{Burgess} \& {Tully}(1992)}]{Burgess92}
{Burgess}, A., \& {Tully}, J.~A. 1992, \aap, 254, 436

\bibitem[{{Che} \& {Goldstein}(2014)}]{Che14}
{Che}, H., \& {Goldstein}, M.~L. 2014, \apjl, 795, L38

\bibitem[{{Collier}(2004)}]{Collier04}
{Collier}, M.~R. 2004, Advances in Space Research, 33, 2108

\bibitem[{{Collier} {et~al.}(1996){Collier}, {Hamilton}, {Gloeckler},
  {Bochsler}, \& {Sheldon}}]{Collier96}
{Collier}, M.~R., {Hamilton}, D.~C., {Gloeckler}, G., {Bochsler}, P., \&
  {Sheldon}, R.~B. 1996, \grl, 23, 1191

\bibitem[{{Culhane} {et~al.}(2007){Culhane}, {Harra}, {James}, {Al-Janabi},
  {Bradley}, {Chaudry}, {Rees}, {Tandy}, {Thomas}, {Whillock}, {Winter},
  {Doschek}, {Korendyke}, {Brown}, {Myers}, {Mariska}, {Seely}, {Lang}, {Kent},
  {Shaughnessy}, {Young}, {Simnett}, {Castelli}, {Mahmoud}, {Mapson-Menard},
  {Probyn}, {Thomas}, {Davila}, {Dere}, {Windt}, {Shea}, {Hagood}, {Moye},
  {Hara}, {Watanabe}, {Matsuzaki}, {Kosugi}, {Hansteen}, \&
  {Wikstol}}]{Culhane07}
{Culhane}, J.~L., {Harra}, L.~K., {James}, A.~M., {et~al.} 2007, \solphys, 243,
  19

\bibitem[{{De Pontieu} {et~al.}(2014){De Pontieu}, {Title}, {Lemen}, {Kushner},
  {Akin}, {Allard}, {Berger}, {Boerner}, {Cheung}, {Chou}, {Drake}, {Duncan},
  {Freeland}, {Heyman}, {Hoffman}, {Hurlburt}, {Lindgren}, {Mathur}, {Rehse},
  {Sabolish}, {Seguin}, {Schrijver}, {Tarbell}, {W{\"u}lser}, {Wolfson},
  {Yanari}, {Mudge}, {Nguyen-Phuc}, {Timmons}, {van Bezooijen}, {Weingrod},
  {Brookner}, {Butcher}, {Dougherty}, {Eder}, {Knagenhjelm}, {Larsen},
  {Mansir}, {Phan}, {Boyle}, {Cheimets}, {DeLuca}, {Golub}, {Gates}, {Hertz},
  {McKillop}, {Park}, {Perry}, {Podgorski}, {Reeves}, {Saar}, {Testa}, {Tian},
  {Weber}, {Dunn}, {Eccles}, {Jaeggli}, {Kankelborg}, {Mashburn}, {Pust},
  {Springer}, {Carvalho}, {Kleint}, {Marmie}, {Mazmanian}, {Pereira}, {Sawyer},
  {Strong}, {Worden}, {Carlsson}, {Hansteen}, {Leenaarts}, {Wiesmann},
  {Aloise}, {Chu}, {Bush}, {Scherrer}, {Brekke}, {Martinez-Sykora}, {Lites},
  {McIntosh}, {Uitenbroek}, {Okamoto}, {Gummin}, {Auker}, {Jerram}, {Pool}, \&
  {Waltham}}]{DePontieu14}
{De Pontieu}, B., {Title}, A.~M., {Lemen}, J.~R., {et~al.} 2014, \solphys

\bibitem[{{Del Zanna}(2011)}]{DelZanna11b}
{Del Zanna}, G. 2011, \aap, 536, A59

\bibitem[{{Del Zanna}(2013)}]{DelZanna13b}
---. 2013, \aap, 558, A73

\bibitem[{{Del Zanna} {et~al.}(2014){Del Zanna}, {Storey}, {Badnell}, \&
  {Mason}}]{DelZanna14}
{Del Zanna}, G., {Storey}, P.~J., {Badnell}, N.~R., \& {Mason}, H.~E. 2014,
  \aap, in press

\bibitem[{{Del Zanna} {et~al.}(2010){Del Zanna}, {Storey}, \&
  {Mason}}]{DelZanna10a}
{Del Zanna}, G., {Storey}, P.~J., \& {Mason}, H.~E. 2010, \aap, 514, A40

\bibitem[{{Dere}(2007)}]{Dere07}
{Dere}, K.~P. 2007, \aap, 466, 771

\bibitem[{{Dere} {et~al.}(1997){Dere}, {Landi}, {Mason}, {Monsignori Fossi}, \&
  {Young}}]{Dere97}
{Dere}, K.~P., {Landi}, E., {Mason}, H.~E., {Monsignori Fossi}, B.~C., \&
  {Young}, P.~R. 1997, \aaps, 125, 149

\bibitem[{{Dopita} {et~al.}(2013){Dopita}, {Sutherland}, {Nicholls}, {Kewley},
  \& {Vogt}}]{Dopita13}
{Dopita}, M.~A., {Sutherland}, R.~S., {Nicholls}, D.~C., {Kewley}, L.~J., \&
  {Vogt}, F.~P.~A. 2013, \apjs, 208, 10

\bibitem[{{Drake} {et~al.}(2006){Drake}, {Swisdak}, {Che}, \& {Shay}}]{Drake06}
{Drake}, J.~F., {Swisdak}, M., {Che}, H., \& {Shay}, M.~A. 2006, \nat, 443, 553

\bibitem[{{Dud{\'{\i}}k} {et~al.}(2014{\natexlab{a}}){Dud{\'{\i}}k}, {Del
  Zanna}, {Dzif{\v c}{\'a}kov{\'a}}, {Mason}, \& {Golub}}]{Dudik14a}
{Dud{\'{\i}}k}, J., {Del Zanna}, G., {Dzif{\v c}{\'a}kov{\'a}}, E., {Mason},
  H.~E., \& {Golub}, L. 2014{\natexlab{a}}, \apjl, 780, L12

\bibitem[{{Dud{\'{\i}}k} {et~al.}(2014{\natexlab{b}}){Dud{\'{\i}}k}, {Del
  Zanna}, {Mason}, \& {Dzif{\v c}{\'a}kov{\'a}}}]{Dudik14b}
{Dud{\'{\i}}k}, J., {Del Zanna}, G., {Mason}, H.~E., \& {Dzif{\v
  c}{\'a}kov{\'a}}, E. 2014{\natexlab{b}}, \aap, 570, A124

\bibitem[{{Dud{\'{\i}}k} {et~al.}(2011){Dud{\'{\i}}k}, {Dzif{\v
  c}{\'a}kov{\'a}}, {Karlick{\'y}}, \& {Kulinov{\'a}}}]{Dudik11}
{Dud{\'{\i}}k}, J., {Dzif{\v c}{\'a}kov{\'a}}, E., {Karlick{\'y}}, M., \&
  {Kulinov{\'a}}, A. 2011, \aap, 529, A103

\bibitem[{{Dud{\'{\i}}k} {et~al.}(2012){Dud{\'{\i}}k}, {Ka{\v s}parov{\'a}},
  {Dzif{\v c}{\'a}kov{\'a}}, {Karlick{\'y}}, \& {Mackovjak}}]{Dudik12}
{Dud{\'{\i}}k}, J., {Ka{\v s}parov{\'a}}, J., {Dzif{\v c}{\'a}kov{\'a}}, E.,
  {Karlick{\'y}}, M., \& {Mackovjak}, {\v S}. 2012, \aap, 539, A107

\bibitem[{{Dud{\'{\i}}k} {et~al.}(2009){Dud{\'{\i}}k}, {Kulinov{\'a}}, {Dzif{\v
  c}{\'a}kov{\'a}}, \& {Karlick{\'y}}}]{Dudik09}
{Dud{\'{\i}}k}, J., {Kulinov{\'a}}, A., {Dzif{\v c}{\'a}kov{\'a}}, E., \&
  {Karlick{\'y}}, M. 2009, \aap, 505, 1255

\bibitem[{{Dzif{\v c}{\'a}kov{\'a}}(2002)}]{Dzifcakova02}
{Dzif{\v c}{\'a}kov{\'a}}, E. 2002, \solphys, 208, 91

\bibitem[{{Dzif{\v c}{\'a}kov{\'a}}(2006{\natexlab{a}})}]{Dzifcakova06}
---. 2006{\natexlab{a}}, \solphys, 234, 243

\bibitem[{{Dzif{\v c}{\'a}kov{\'a}}(2006{\natexlab{b}})}]{Dzifcakova06b}
{Dzif{\v c}{\'a}kov{\'a}}, E. 2006{\natexlab{b}}, in ESA Special Publication,
  Vol. 617, SOHO-17. 10 Years of SOHO and Beyond

\bibitem[{{Dzif{\v c}{\'a}kov{\'a}} \& {Dud{\'{\i}}k}(2013)}]{Dzifcakova13}
{Dzif{\v c}{\'a}kov{\'a}}, E., \& {Dud{\'{\i}}k}, J. 2013, \apjs, 206, 6

\bibitem[{{Dzif{\v c}{\'a}kov{\'a}} {et~al.}(2012){Dzif{\v c}{\'a}kov{\'a}},
  {Dud{\'{\i}}k}, \& {Karlick{\'y}}}]{Dzifcakova12}
{Dzif{\v c}{\'a}kov{\'a}}, E., {Dud{\'{\i}}k}, J., \& {Karlick{\'y}}, M. 2012,
  in Astronomical Society of the Pacific Conference Series, Vol. 456, Fifth
  Hinode Science Meeting, ed. L.~{Golub}, I.~{De Moortel}, \& T.~{Shimizu}, 135

\bibitem[{{Dzif{\v c}{\'a}kov{\'a}} {et~al.}(2011){Dzif{\v c}{\'a}kov{\'a}},
  {Homola}, \& {Dud{\'{\i}}k}}]{Dzifcakova11}
{Dzif{\v c}{\'a}kov{\'a}}, E., {Homola}, M., \& {Dud{\'{\i}}k}, J. 2011, \aap,
  531, A111

\bibitem[{{Dzif{\v c}{\'a}kov{\'a}} \& {Kulinov{\'a}}(2010)}]{Dzifcakova10}
{Dzif{\v c}{\'a}kov{\'a}}, E., \& {Kulinov{\'a}}, A. 2010, \solphys, 263, 25

\bibitem[{{Dzif{\v c}{\'a}kov{\'a}} \& {Mason}(2008)}]{DzifcakovaMason08}
{Dzif{\v c}{\'a}kov{\'a}}, E., \& {Mason}, H.~E. 2008, \solphys, 247, 301

\bibitem[{{Dzif\v{c}{\'a}kov{\'a}}(1992)}]{Dzifcakova92}
{Dzif\v{c}{\'a}kov{\'a}}, E. 1992, \solphys, 140, 247

\bibitem[{{Feldman} {et~al.}(2007){Feldman}, {Landi}, \& {Doschek}}]{Feldman07}
{Feldman}, U., {Landi}, E., \& {Doschek}, G.~A. 2007, \apj, 660, 1674

\bibitem[{{Gontikakis} {et~al.}(2013){Gontikakis}, {Patsourakos},
  {Efthymiopoulos}, {Anastasiadis}, \& {Georgoulis}}]{Gontikakis13}
{Gontikakis}, C., {Patsourakos}, S., {Efthymiopoulos}, C., {Anastasiadis}, A.,
  \& {Georgoulis}, M.~K. 2013, \apj, 771, 126

\bibitem[{{Hannah} {et~al.}(2010){Hannah}, {Hudson}, {Hurford}, \&
  {Lin}}]{Hannah10}
{Hannah}, I.~G., {Hudson}, H.~S., {Hurford}, G.~J., \& {Lin}, R.~P. 2010, \apj,
  724, 487

\bibitem[{{Hannah} \& {Kontar}(2012)}]{Hannah12}
{Hannah}, I.~G., \& {Kontar}, E.~P. 2012, \aap, 539, A146

\bibitem[{{Hannah} \& {Kontar}(2013)}]{Hannah13}
---. 2013, \aap, 553, A10

\bibitem[{{Hasegawa} {et~al.}(1985){Hasegawa}, {Mima}, \&
  {Duong-van}}]{Hasegawa85}
{Hasegawa}, A., {Mima}, K., \& {Duong-van}, M. 1985, Physical Review Letters,
  54, 2608

\bibitem[{{Itoh} {et~al.}(2000){Itoh}, {Sakamoto}, {Kusano}, {Nozawa}, \&
  {Kohyama}}]{Itoh00}
{Itoh}, N., {Sakamoto}, T., {Kusano}, S., {Nozawa}, S., \& {Kohyama}, Y. 2000,
  \apjs, 128, 125

\bibitem[{{Ka{\v s}parov{\'a}} \& {Karlick{\'y}}(2009)}]{Kasparova09}
{Ka{\v s}parov{\'a}}, J., \& {Karlick{\'y}}, M. 2009, \aap, 497, L13

\bibitem[{{Laming} \& {Lepri}(2007)}]{Laming07}
{Laming}, J.~M., \& {Lepri}, S.~T. 2007, \apj, 660, 1642

\bibitem[{{Laming} {et~al.}(2013){Laming}, {Moses}, {Ko}, {Ng}, {Rakowski}, \&
  {Tylka}}]{Laming13}
{Laming}, J.~M., {Moses}, J.~D., {Ko}, Y.-K., {et~al.} 2013, \apj, 770, 73

\bibitem[{{Landi} {et~al.}(2006){Landi}, {Del Zanna}, {Young}, {Dere}, {Mason},
  \& {Landini}}]{Landi06}
{Landi}, E., {Del Zanna}, G., {Young}, P.~R., {et~al.} 2006, \apjs, 162, 261

\bibitem[{{Landi} {et~al.}(2013){Landi}, {Young}, {Dere}, {Del Zanna}, \&
  {Mason}}]{Landi13}
{Landi}, E., {Young}, P.~R., {Dere}, K.~P., {Del Zanna}, G., \& {Mason}, H.~E.
  2013, \apj, 763, 86

\bibitem[{{Le Chat} {et~al.}(2011){Le Chat}, {Issautier}, {Meyer-Vernet}, \&
  {Hoang}}]{LeChat11}
{Le Chat}, G., {Issautier}, K., {Meyer-Vernet}, N., \& {Hoang}, S. 2011,
  \solphys, 271, 141

\bibitem[{{Lemen} {et~al.}(2012){Lemen}, {Title}, {Akin}, {Boerner}, {Chou},
  {Drake}, {Duncan}, {Edwards}, {Friedlaender}, {Heyman}, {Hurlburt}, {Katz},
  {Kushner}, {Levay}, {Lindgren}, {Mathur}, {McFeaters}, {Mitchell}, {Rehse},
  {Schrijver}, {Springer}, {Stern}, {Tarbell}, {Wuelser}, {Wolfson}, {Yanari},
  {Bookbinder}, {Cheimets}, {Caldwell}, {Deluca}, {Gates}, {Golub}, {Park},
  {Podgorski}, {Bush}, {Scherrer}, {Gummin}, {Smith}, {Auker}, {Jerram},
  {Pool}, {Soufli}, {Windt}, {Beardsley}, {Clapp}, {Lang}, \&
  {Waltham}}]{Lemen12}
{Lemen}, J.~R., {Title}, A.~M., {Akin}, D.~J., {et~al.} 2012, \solphys, 275, 17

\bibitem[{{Leubner}(2002)}]{Leubner02}
{Leubner}, M.~P. 2002, \apss, 282, 573

\bibitem[{{Leubner}(2004)}]{Leubner04a}
---. 2004, \apj, 604, 469

\bibitem[{{Liang} {et~al.}(2012){Liang}, {Badnell}, \& {Zhao}}]{Liang12}
{Liang}, G.~Y., {Badnell}, N.~R., \& {Zhao}, G. 2012, \aap, 547, A87

\bibitem[{{Lin} {et~al.}(2002){Lin}, {Dennis}, {Hurford}, {Smith}, {Zehnder},
  {Harvey}, {Curtis}, {Pankow}, {Turin}, {Bester}, {Csillaghy}, {Lewis},
  {Madden}, {van Beek}, {Appleby}, {Raudorf}, {McTiernan}, {Ramaty}, {Schmahl},
  {Schwartz}, {Krucker}, {Abiad}, {Quinn}, {Berg}, {Hashii}, {Sterling},
  {Jackson}, {Pratt}, {Campbell}, {Malone}, {Landis}, {Barrington-Leigh},
  {Slassi-Sennou}, {Cork}, {Clark}, {Amato}, {Orwig}, {Boyle}, {Banks},
  {Shirey}, {Tolbert}, {Zarro}, {Snow}, {Thomsen}, {Henneck}, {McHedlishvili},
  {Ming}, {Fivian}, {Jordan}, {Wanner}, {Crubb}, {Preble}, {Matranga}, {Benz},
  {Hudson}, {Canfield}, {Holman}, {Crannell}, {Kosugi}, {Emslie}, {Vilmer},
  {Brown}, {Johns-Krull}, {Aschwanden}, {Metcalf}, \& {Conway}}]{Lin02}
{Lin}, R.~P., {Dennis}, B.~R., {Hurford}, G.~J., {et~al.} 2002, \solphys, 210,
  3

\bibitem[{{Livadiotis} \& {McComas}(2009)}]{Livadiotis09}
{Livadiotis}, G., \& {McComas}, D.~J. 2009, \jgr, 114, A11105

\bibitem[{{Livadiotis} \& {McComas}(2010)}]{Livadiotis10}
---. 2010, \apj, 714, 971

\bibitem[{{Livadiotis} \& {McComas}(2011)}]{Livadiotis11a}
---. 2011, \apj, 741, 88

\bibitem[{{Livadiotis} \& {McComas}(2013)}]{Livadiotis13}
---. 2013, \ssr, 175, 183

\bibitem[{{Mackovjak} {et~al.}(2013){Mackovjak}, {Dzif{\v c}{\'a}kov{\'a}}, \&
  {Dud{\'{\i}}k}}]{Mackovjak13}
{Mackovjak}, {\v S}., {Dzif{\v c}{\'a}kov{\'a}}, E., \& {Dud{\'{\i}}k}, J.
  2013, \solphys, 282, 263

\bibitem[{{Mackovjak} {et~al.}(2014){Mackovjak}, {Dzif{\v c}{\'a}kov{\'a}}, \&
  {Dud{\'{\i}}k}}]{Mackovjak14}
---. 2014, \aap, 564, A130

\bibitem[{{Maksimovic} {et~al.}(1997{\natexlab{a}}){Maksimovic}, {Pierrard}, \&
  {Lemaire}}]{Maksimovic97a}
{Maksimovic}, M., {Pierrard}, V., \& {Lemaire}, J.~F. 1997{\natexlab{a}}, \aap,
  324, 725

\bibitem[{{Maksimovic} {et~al.}(1997{\natexlab{b}}){Maksimovic}, {Pierrard}, \&
  {Riley}}]{Maksimovic97b}
{Maksimovic}, M., {Pierrard}, V., \& {Riley}, P. 1997{\natexlab{b}}, \grl, 24,
  1151

\bibitem[{{Mason} \& {Monsignori Fossi}(1994)}]{Mason94}
{Mason}, H.~E., \& {Monsignori Fossi}, B.~C. 1994, \aapr, 6, 123

\bibitem[{{Mewe}(1972)}]{Mewe72}
{Mewe}, R. 1972, \aap, 20, 215

\bibitem[{{Meyer-Vernet}(2007)}]{Meyer-Vernet07}
{Meyer-Vernet}, N. 2007, {Basics of the Solar Wind} (Cambridge University
  Press)

\bibitem[{{Meyer-Vernet} {et~al.}(1995){Meyer-Vernet}, {Moncuquet}, \&
  {Hoang}}]{Meyer-Vernet95}
{Meyer-Vernet}, N., {Moncuquet}, M., \& {Hoang}, S. 1995, \icarus, 116, 202

\bibitem[{{Nicholls} {et~al.}(2012){Nicholls}, {Dopita}, \&
  {Sutherland}}]{Nicholls12}
{Nicholls}, D.~C., {Dopita}, M.~A., \& {Sutherland}, R.~S. 2012, \apj, 752, 148

\bibitem[{{Nicholls} {et~al.}(2013){Nicholls}, {Dopita}, {Sutherland},
  {Kewley}, \& {Palay}}]{Nicholls13}
{Nicholls}, D.~C., {Dopita}, M.~A., {Sutherland}, R.~S., {Kewley}, L.~J., \&
  {Palay}, E. 2013, \apjs, 207, 21

\bibitem[{{O'Dwyer} {et~al.}(2010){O'Dwyer}, {Del Zanna}, {Mason}, {Weber}, \&
  {Tripathi}}]{ODwyer10}
{O'Dwyer}, B., {Del Zanna}, G., {Mason}, H.~E., {Weber}, M.~A., \& {Tripathi},
  D. 2010, \aap, 521, A21

\bibitem[{{Oka} {et~al.}(2013){Oka}, {Ishikawa}, {Saint-Hilaire}, {Krucker}, \&
  {Lin}}]{Oka13}
{Oka}, M., {Ishikawa}, S., {Saint-Hilaire}, P., {Krucker}, S., \& {Lin}, R.~P.
  2013, \apj, 764, 6

\bibitem[{{Owocki} \& {Scudder}(1983)}]{Owocki83}
{Owocki}, S.~P., \& {Scudder}, J.~D. 1983, \apj, 270, 758

\bibitem[{{Pesnell} {et~al.}(2012){Pesnell}, {Thompson}, \&
  {Chamberlin}}]{Pesnell12}
{Pesnell}, W.~D., {Thompson}, B.~J., \& {Chamberlin}, P.~C. 2012, \solphys,
  275, 3

\bibitem[{{Phillips} {et~al.}(2008){Phillips}, {Feldman}, \&
  {Landi}}]{Phillips08}
{Phillips}, K.~J.~H., {Feldman}, U., \& {Landi}, E. 2008, {Ultraviolet and
  X-ray Spectroscopy of the Solar Atmosphere} (Cambridge University Press)

\bibitem[{{Pierrard} \& {Lazar}(2010)}]{Pierrard10}
{Pierrard}, V., \& {Lazar}, M. 2010, \solphys, 267, 153

\bibitem[{{Raymond} {et~al.}(2010){Raymond}, {Winkler}, {Blair}, {Lee}, \&
  {Park}}]{Raymond10}
{Raymond}, J.~C., {Winkler}, P.~F., {Blair}, W.~P., {Lee}, J.-J., \& {Park}, S.
  2010, \apj, 712, 901

\bibitem[{{Scudder} \& {Karimabadi}(2013)}]{Scudder13}
{Scudder}, J.~D., \& {Karimabadi}, H. 2013, \apj, 770, 26

\bibitem[{{Seaton}(1953)}]{Seaton53}
{Seaton}, M.~J. 1953, Royal Society of London Proceedings Series A, 218, 400

\bibitem[{{Seely} {et~al.}(1987){Seely}, {Feldman}, \& {Doschek}}]{Seely87}
{Seely}, J.~F., {Feldman}, U., \& {Doschek}, G.~A. 1987, \apj, 319, 541

\bibitem[{{Storey} \& {Sochi}(2014)}]{Storey14}
{Storey}, P.~J., \& {Sochi}, T. 2014, ArXiv e-prints

\bibitem[{{Storey} {et~al.}(2013){Storey}, {Sochi}, \& {Badnell}}]{Storey13}
{Storey}, P.~J., {Sochi}, T., \& {Badnell}, N.~R. 2013, ArXiv e-prints

\bibitem[{{Sutherland}(1998)}]{Sutherland98}
{Sutherland}, R.~S. 1998, \mnras, 300, 321

\bibitem[{{Teriaca} {et~al.}(2012){Teriaca}, {Warren}, \& {Curdt}}]{Teriaca12b}
{Teriaca}, L., {Warren}, H.~P., \& {Curdt}, W. 2012, \apjl, 754, L40

\bibitem[{{Testa} {et~al.}(2014){Testa}, {De Pontieu}, {Allred}, {Carlsson},
  {Reale}, {Daw}, {Hansteen}, {Martinez-Sykora}, {Liu}, {DeLuca}, {Golub},
  {McKillop}, {Reeves}, {Saar}, {Tian}, {Lemen}, {Title}, {Boerner}, {Hulburt},
  {Tarbell}, {Wuelser}, {Kleint}, {Kankelborg}, \& {Jaeggli}}]{Testa14}
{Testa}, P., {De Pontieu}, B., {Allred}, J., {et~al.} 2014, Science, 346,
  1255724

\bibitem[{{Tsallis}(1988)}]{Tsallis88}
{Tsallis}, C. 1988, Journal of Statistical Physics, 52, 479

\bibitem[{{Tsallis}(2009)}]{Tsallis09}
---. 2009, {Introduction to Nonextensive Statistical Mechanics} ({Springer New
  York, 2009})

\bibitem[{{Vasyliunas}(1968)}]{Vasyliunas68}
{Vasyliunas}, V.~M. 1968, in Astrophysics and Space Science Library, Vol.~10,
  Physics of the Magnetosphere, ed. {R.~D.~L.~Carovillano \& J.~F.~McClay}, 622

\bibitem[{{Vocks} \& {Mann}(2003)}]{Vocks03}
{Vocks}, C., \& {Mann}, G. 2003, \apj, 593, 1134

\bibitem[{{Vocks} {et~al.}(2008){Vocks}, {Mann}, \& {Rausche}}]{Vocks08}
{Vocks}, C., {Mann}, G., \& {Rausche}, G. 2008, \aap, 480, 527

\bibitem[{{Wannawichian} {et~al.}(2003){Wannawichian}, {Ruffolo}, \&
  {Kartavykh}}]{Wannawichian03}
{Wannawichian}, S., {Ruffolo}, D., \& {Kartavykh}, Y.~Y. 2003, \apjs, 146, 443

\bibitem[{{Warren} {et~al.}(2012){Warren}, {Winebarger}, \&
  {Brooks}}]{Warren12}
{Warren}, H.~P., {Winebarger}, A.~R., \& {Brooks}, D.~H. 2012, \apj, 759, 141

\bibitem[{{Wilhelm} {et~al.}(1995){Wilhelm}, {Curdt}, {Marsch}, {Sch{\"u}hle},
  {Lemaire}, {Gabriel}, {Vial}, {Grewing}, {Huber}, {Jordan}, {Poland},
  {Thomas}, {K{\"u}hne}, {Timothy}, {Hassler}, \& {Siegmund}}]{Wilhelm95}
{Wilhelm}, K., {Curdt}, W., {Marsch}, E., {et~al.} 1995, \solphys, 162, 189

\bibitem[{{Young} {et~al.}(2003){Young}, {Del Zanna}, {Landi}, {Dere}, {Mason},
  \& {Landini}}]{Young03}
{Young}, P.~R., {Del Zanna}, G., {Landi}, E., {et~al.} 2003, \apjs, 144, 135

\end{thebibliography}



\end{document}